\documentclass[iop]{emulateapj}
\usepackage{natbib}
\usepackage{multirow}
\usepackage{booktabs}
\usepackage{amsmath}
\usepackage{graphicx}
\usepackage{xspace} 
\usepackage[colorlinks,linkcolor=blue,citecolor=blue,urlcolor=blue]{hyperref}
\newcommand{\myemail}{jfarias@ufl.edu}
\graphicspath{ {Figures/} }

\def\Rcl{R_{\rm cl}}
\def\krho{k_{\rho}}
\def\Sigmacl{\Sigma_{\rm cloud}}
\def\sigmas{\sigma_{\rm cl,s}}
\def\Msun{{\rm M}_{\odot}}
\def\MSun{{\rm M}_{\odot}}
\def\phipbar{\phi_{\bar{P}}}
\def\phipcl{\phi_{P,\rm cl}}
\def\phib{\phi_{\rm B}}
\def\fbin{f_{\rm bin}}
\def\fb{f_{\rm bound}}
\def\Qi{Q_{i}}
\def\dr{{\rm d} r}

\def\tcross{ t_{\rm cr}}
\def\rc{r_{\rm c}} 
\def\rh{r_{\rm h}}

\def\Mcl{M_{\rm cl}}
\def\Gammab{ \Gamma_{\rm b}}

\def\meanms{ \langle m_{\rm s} \rangle}
\def\Gammaeff{\Gamma_{\rm b,eff}}
\def\foc{{\cal F}}

\def\equalmass{\texttt{equal\_mass}\xspace}          
\def\nobinaries{\texttt{single\_imf}\xspace}        
\def\nose{\texttt{binaries\_50}\xspace}               
\def\fiducial{\texttt{fiducial}\xspace}            
\def\binariesun{\texttt{binaries\_un}\xspace}
\def\binariesth{\texttt{binaries\_th}\xspace}
\def\segregated{\texttt{segregated}\xspace}
\def\fullbinaries{\texttt{binaries\_100}\xspace}

\defcitealias{mt03}{MT03}
\defcitealias{t13}{T13}

\slugcomment{Draft version \today}
\shorttitle{Star Cluster Formation from Turbulent Clumps}
\shortauthors{Farias, Tan \& Chatterjee}

\begin{document}

\title{Star Cluster Formation from Turbulent Clumps.\\
      I. The Fast Formation Limit}

\author{Juan P. Farias\altaffilmark{1,$\star$}, Jonathan C. Tan\altaffilmark{1} and Sourav Chatterjee\altaffilmark{2}}
\affil{$^{1}$Astronomy Department, University of Florida, Gainesville, FL 32611, USA\\
      $^{2}$Center for Interdisciplinary Exploration \& Research in Astrophysics (CIERA)\\ 
      Physics \& Astronomy, Northwestern University, Evanston, IL 60202, USA}
\email[$^\star$E-mail:]{ \myemail}

\begin{abstract}
We investigate the formation and early evolution of star clusters
assuming that they form from a turbulent starless clump of given mass
bounded inside a parent self-gravitating molecular cloud characterized
by a particular mass surface density.  As a first step we assume
instantaneous star cluster formation and gas expulsion. We draw our
initial conditions from observed properties of starless clumps.  We
follow the early evolution of the clusters up to 20~Myr, investigating
effects of different star formation efficiencies, primordial binary
fractions and eccentricities and primordial mass segregation
levels. We investigate clumps with initial masses of $M_{\rm
  cl}=3000\:\Msun$ embedded in ambient cloud environments with mass
surface densities, $\Sigmacl=0.1$ and $1\:{\rm g\:cm^{-2}}$. We show
that these models of fast star cluster formation result, in the
fiducial case, in clusters that expand rapidly, even considering only
the bound members. Clusters formed from higher $\Sigmacl$ environments
tend to expand more quickly, so are soon larger than clusters born
from lower $\Sigmacl$ conditions. To form a young cluster of a given
age, stellar mass and mass surface density, these models need to
assume a parent molecular clump that is many times denser, which is
unrealistic compared to observed systems. We also show that in these
models the initial binary properties are only slightly modified by
interactions, meaning that binary properties, e.g., at 20~Myr, are
very similar to those at birth. With this study we set up the basis of
future work where we will investigate more realistic models of star
formation compared to this instantaneous, baseline case.
\end{abstract}


\section{Introduction}

Most stars tend to form together in clusters \citep[e.g.,][]{Gutermut2009}, 
which are created from over dense gas {\it clumps}, typically found in
giant molecular clouds (GMCs) \citep[e.g.,][]{Mckee2007}.  Thus
understanding how stars form comes with the direct need to understand
how and where star clusters form. It is still a matter of debate if
star cluster formation depends on properties of the GMC environment or
not.  While theoretical studies have taught us the essential physical
processes that determine a star cluster's evolution after gas is
dispersed, the transition from the dense star-forming clump to the
star cluster that emerges from the gas is not yet well understood
\citep[see, e.g.,][for a review]{Banerjee2015}.
In particular, it is debated whether the process is slow with the
clump evolving in a quasi-equilibrium state
\citep{Tan2006,Nakamura2007}, or very rapid with star cluster
formation occurring in just a crossing time of the system
\citep{Elmegreen2000,Elmegreen2007,Hartmann2007}.

There are numerous physical processes potentially involved in such a
transition from gas clump to star cluster, including: fragmentation of
a magnetized, turbulent and self-gravitating medium to a population of
pre-stellar cores; collapse of these cores via rotationally supported
disks to single or multiple star systems; feedback from the forming
stars, especially protostellar outflows that can maintain turbulence
in the clump \citep{Nakamura2007,Nakamura2014}, and eventually
radiative feedback processes from more massive stars
\citep[e.g.,][]{Dale2015}; continued infall of gas to the clump;
dispersal of clump gas by feedback; dynamical interactions among the
forming and recently formed stars as they orbit in the protocluster
potential \citep[e.g.,][]{ Chatterjee2012}. 
This is a complicated, multi-scale problem, the full solution of which
is beyond current computational capabilities. Thus approximate models
are necessary. By investigating how the outcome of star cluster
formation depends on the adopted approximations we can learn which
processes are most important.

Our approach in this paper and subsequent papers in this series is to
follow the dynamics of formed stars, including binary properties,
accurately via direct $N$-body integrations, but approximate various
models for how individual stars are born within the star-forming
clump. Our initial conditions are based on the Turbulent Core/Clump
model of \cite{mt03} (hereafter MT03), which approximates clumps as
singular polytropic virialized spheres that are in pressure
equilibrium with their surrounding cloud medium. This surrounding
cloud is also assumed to be self-gravitating so its ambient pressure
is $\bar{P}\sim G \Sigma^2$, where $\Sigma$ is the cloud mass surface
density---the main variable describing different environmental
conditions.

In this first paper, we start with the simplest approximation for star
formation, i.e., instantaneous formation of the stellar population
from the initial gas clump along with simultaneous, instantaneous
expulsion of the remaining gas that is not incorporated into
stars. This approximation has often been adopted by previous $N$-body
studies
\citep[e.g.,][]{Bastian2006,Parker2014a,Pfalzner2015}. However, in
comparison to these previous studies our work is distinguished by (1)
adopting initial conditions that have been explicitly developed for
self-gravitating gas clumps (i.e., singular polytropic spheres as
approximations for turbulent, magnetized clumps); (2) following the
full evolution of binary systems.

A number of authors have studied the dynamics of binaries in star
clusters using numerical models
\citep[e.g.,][]{Kroupa1999,Kroupa2001b,Parker2009,Kouwenhoven2010,Parker2011,Kaczmarek2011},
focusing on various aspects of their dynamics. In our work we follow
the evolution of binary properties due to stellar interactions and
stellar evolution in clusters forming in different environments. We
use and will use these results to constrain assumptions about the star
cluster process, e.g., its duration, and individual star formation
processes, e.g., how binaries form, i.e., the primordial binary
properties. We also examine the role of binaries in producing
dynamical and binary supernova ejections, especially fast runaway
stars.

We anticipate that, in reality, star cluster formation is likely to
proceed in a relatively gradual manner, i.e., taking at least several
and perhaps many local free-fall timescales of the gas clump. Modeling
this process of gradual star cluster formation will be the topic of
the second paper in this series. One aspect that is needed in such a
model is the gradual evolution of the potential of the natal gas
clump, which can be approximated via a simple analytic relation. Such
an approach of an evolving approximate background gas potential has
been adopted previously by
\cite[e.g.,][]{Tutukov1978,Lada1984,Geyer2001,Boily2003,Chen2009,Smith2011,Farias2015,Brinkmann2016}.
However, in all these studies the full stellar population was
introduced instantaneously at the beginning of the simulation. None of
these studies included a full treatment of binaries, 
i.e., including a significant fraction of primordial binaries.

The second aspect is the gradual formation of the stellar population
in the cluster. This will be the main focus of Paper II in our series.
There have so far been relatively few studies using such an approach.
\cite{Proszkow2009} presented a study involving gradual star formation
and then the early evolution of young clusters, focusing on low to
intermediate mass clusters (100 to 3000 members), without primordial
binaries.
However, the stellar densities of their models were relatively low,
limiting the effects of stellar interactions. They also did not
include stellar evolution and only simulated up to 10~Myr.

These methods of $N$-body modeling can be contrasted with other
approaches to simulating star cluster formation, e.g., simulations
that follow the (magneto-)hydrodynamics of the collapsing clump
\citep[e.g.,][]{Price2009,Padoan2012,Myers2014,Padoan2014}.  Such
simulations must still implement sub-grid models for how stars form
and inject feedback into the gas. Typically they do not have the
resolution to accurately follow binary orbital evolution. Still, these
are complementary approaches to those based on pure $N$-body
approaches and comparison of the results of the different methods will
be instructive.

\section{Pre-Cluster Clumps as Initial Conditions}\label{sec:ic}

The initial conditions for star clusters are constrained by the
observed properties of dense gas clumps within GMCs.  These locations
are also expected to be the sites of future massive star formation.
The properties of these clumps have been summarized by \cite{Tan2014},
based on Galactic observational studies of Infrared Dark Clouds
(IRDCs) \citep[e.g.,][]{Rathborne2006,Butler2009,Butler2012} and
mm/sub-mm dust continuum emission and molecular line surveys of clumps
\citep[e.g.,][]{Schuller2009,Ginsburg2012,Ma2013}.  There are a range
of clump masses observed from $\sim100\:M_\odot$ to $\sim
10^5\:M_\odot$.

In the fiducial Turbulent Clump model, the mass surface density of the
clump of interest is only a factor of 1.22 times higher than that of
its surrounding cloud \citepalias{mt03}. Therefore the observed
surface densities of clumps give a reasonable estimate of the mass
surface densities of the ambient clouds, $\Sigmacl$,
which is the other main parameter needed to set up the initial
conditions of the models. The observed values of mass surface density
of Galactic clumps and protoclusters are in the range from
$\sim0.03$--$1\:{\rm g\:cm^{-2}}$.

In the setup of our initial conditions in this paper we make several
simplifying assumptions as first steps in describing the complexity of
star cluster formation: i) the parent clump is isolated (i.e., no
external tidal fields); ii) the clump is in hydrostatic and virial
equilibrium with the structure of a singular polytropic sphere
\citepalias{mt03}; iii) stars are born with the same velocities as
their parent gas, so that their velocity dispersion profile is the
same as that of the initial gas; iv) all stars form instantaneously
and the remaining gas is also expelled instantaneously at this time;
iv) the star formation efficiency (SFE) is spatially constant, which
means that stars follow the same spatial distribution as the initial
gas; v) there is no initial spatial or kinematic substructure given to
the stars, except that which results from random, Poisson sampling;
vi) following an initial test model of equal mass stars, a standard
Kroupa IMF for the stars \citep{Kroupa2001} is adopted, then with
various binary properties investigated. It should be remembered that
these are starting assumptions and that many of these will be relaxed
in subsequent investigations. However, first the behavior of this
simplest, idealized model needs to be understood.

\subsection{Initial stellar phase-space distributions}

First we define the physical and kinematic properties of the
pre-cluster clump, i.e., mass, size, density profile and velocity
dispersion profile. Stars are born from this clump and initially
follow the same phase-space (position, velocity)
distribution. \citetalias{mt03} characterize pre-cluster clumps and
pre-stellar cores as singular polytropic spheres in virial and
hydrostatic equilibrium. The density profile of such clumps is then:
\begin{eqnarray}
        \label{eq:dens}
        \rho_{\rm cl} (r) &=& \rho_{\rm s,cl} \left(\frac{r}{\Rcl} \right)^{-\krho},
\end{eqnarray}
where $\rho_{\rm s,cl}$ is the density at the surface of the clump,
i.e., at radius $\Rcl$. We adopt $\krho=1.5$ as a fiducial value,
i.e., the same as that of \citetalias{mt03} who made their choice
based on observations of clumps reported at the time. No significant
difference has been found in later measurements performed by
\cite{Butler2012} in IRDCs where they found $\krho\simeq1.6$. We thus
keep $\krho=1.5$ as our fiducial value for simplicity and consistency
with the previous analysis of \citetalias{mt03}. The density at the
surface of the clump can be expressed as
\begin{eqnarray}
        \rho_{\rm s,cl}&=& \frac{(3-\krho)M_{\rm cl}}{4\pi\Rcl^3},
\end{eqnarray}
where $M_{\rm cl} = M(r<\Rcl)$ is the total mass of the clump. 

The radius of a clump in virial equilibrium and pressure equilibrium
with its surroundings, i.e., a larger self-gravitating cloud of given
mass surface density, $\Sigmacl$, is (\citetalias{mt03};
\citealt{t13}, hereafter T13)
\begin{eqnarray}
        \Rcl &=& 0.50 \left( \frac{A}{k_{p} \phipcl \phipbar} \right)^{1/4} 
         \left( \frac{M_{\rm cl}}{3000\:\Msun}  \right)^{1/2} \nonumber \\
         & & \times \left( \frac{\Sigmacl}{1 {\rm ~g~cm^{-2}}} \right)^{-1/2}\:{\rm pc}\\
         \label{eq:rcl}
 & \rightarrow & 0.37 M_{\rm cl,3000}^{1/2} \Sigma_{\rm cloud,1}^{-1/2}\:{\rm pc}
\end{eqnarray}
where $k_p=2(\krho -1)$ is the power law exponent of the pressure
($P$) within the clump; $\phipcl$ is the ratio between the pressure at
the surface of the clump ($P_{\rm s,cl}$) and the mean pressure inside
the cloud, $\bar{P}_{\rm cloud}$; $\phipbar$ is a normalization
constant, $\sim{\cal O}(1)$, in the relation $\bar{P}_{\rm
  cloud}\equiv\phipbar G\Sigmacl^2$; and $A=(3-\krho)(\krho-1)f_g
\rightarrow 3/4$, since we assume the clump is initially starless so
$f_g=1$.  As fiducial values we choose $\phi_{P,\rm cl}=2$ and
$\phipbar=1.32$ \citepalias[see][]{mt03,t13}.  Thus the structural
properties of the fiducial clump are specified by two parameters:
$M_{\rm cl}$ and $\Sigma_{\rm cloud}$. In this study we will focus on
the case of $M_{\rm cl}=3000\:\Msun$ and investigate $\Sigmacl=0.1$
and $1\:{\rm g\:cm}^{-2}$, which are representative of the range of
values observed in real clumps \citep{Tan2014}.

The total mass of stars, $M_*$, that form from the clump is given
\begin{eqnarray}
M_* = \epsilon M_{\rm cl},
\end{eqnarray}
where $\epsilon$ is the overall SFE. We will consider a range of
values from $\epsilon=0.1$ to 1, with a fiducial value of 0.5.  The
stars are assumed to have the same structural profile as the clump,
i.e., $\epsilon$ is independent of radial location. Then, the density
profile of the stars is simply: $\rho_*(r)=\epsilon\rho_{\rm cl}(r)$.
The cumulative radial mass distribution of the stars is
\begin{eqnarray}
        \label{eq:mstars}
        M_*(r<\Rcl) &=& M_* \left( \frac{r}{\Rcl} \right)^{3-\krho}.
\end{eqnarray}
To set up the positions of the stars, we first create the mass sample
from a given IMF (including binary companions), i.e., in the fiducial
case a standard Kroupa IMF \citep{Kroupa2001} with individual masses
in the range $0.01\:\MSun <m_{i}< 100\:\MSun$, in random order. Next,
we create a cumulative mass array from the previous sample to then
choose the individual distance from the center $r_{i}$ according to
Eq.~\ref{eq:mstars} (in the case of 100\% mass segregation, see below,
masses are sorted from the most massive to the less massive before
this step). Finally, we place the star randomly on the surface of the
sphere of radius $r_{i}$.  In this way we ensure that the clusters
always match the desired initial density profile, no matter the nature
of the stellar IMF.

The kinematic properties of the clump are specified by the condition
of virial equilibrium. In a virialized clump, the velocity dispersion
$\sigma$ scales in the same manner as the effective sound speed $c
\equiv (P/\rho)^{1/2}$. Therefore, the velocity dispersion profile of
the clump is:
\begin{eqnarray}
\sigma_{\rm cl}(r) &=& \sigmas \left( \frac{r}{\Rcl} \right)^{(2-\krho)/2},
\label{eq:sigmar}
\end{eqnarray}
where $\sigmas$ is the velocity dispersion at the surface of the
clump. The presence of large-scale magnetic fields can provide some
support to the clump so that a smaller turbulent velocity dispersion
is needed to achieve virial equilibrium. The effect of magnetic fields
on the stability of the clump can be expressed as $\phib \equiv
\langle c^2 \rangle / \langle \sigma^2 \rangle$. Then, the velocity
dispersion at the surface is:
\begin{eqnarray}
\sigmas &=& 5.08 \left( \frac{\phipcl \phipbar}{Ak_{P}^2 \phib^4} \right)^{1/8} 
\left(\frac{M_{\rm cl}}{3000~\Msun} \right)^{1/4} \nonumber \\
& & \times \left( \frac{\Sigmacl }{1~{\rm g~cm^{-2}}} \right)^{1/4}\:{\rm km~s^{-1}}
\label{eq:sigmas}
\end{eqnarray}
where we use $\phib=2.8$ as a fiducial value, which is the value for
regions with an Alfv\'en Mach number of 1 \citepalias[see Appendix A2
  of][]{mt03}. The initial velocity dispersion profile of the stars
then follows Eq.~\ref{eq:sigmar}. The individual stellar velocities
are then assigned velocities in each of the $x$, $y$ and $z$
directions independently by drawing from a Gaussian centered at zero with width
$\sigma(r)$. Note that the mass averaged velocity dispersion of the clump/cluster is
\citepalias{t13}
\begin{eqnarray}
\label{eq:sigma}
\sigma_{\rm cl}&=& \frac{2(3-\krho)}{8-3\krho} \sigmas \rightarrow \frac{6}{7}\sigmas.
\end{eqnarray}
The resulting velocity distribution has the form of a
Maxwell-Boltzmann distribution with $\sigma_{3D} = \sqrt{3} \sigma_{\rm cl} $ 
but with a one dimensional velocity dispersion
profile as in equation \ref{eq:sigmar}.
The properties of the clumps, i.e., the low and high $\Sigma$ cases,
are summarized in Table~\ref{tab:clumps}.

\begin{table*}
\centering
\caption{Parent clump parameters}
\begin{tabular}{ccccccccc} \toprule
 & $\Sigmacl$ & $M_{\rm cl}$($\Msun$)&$\Rcl$ (pc)     & $\krho$ & $\phipcl$ & $\phipbar$
 & $\phib$ & $\sigma_{\rm cl}$ (km/s)  \\[0.1cm] \hline
  ``Low-$\Sigma$'' Clump & 0.1 & 3000 & 1.159 & 1.5 & 2 & 1.31 & 2.8 & 1.71 \\ 
  ``High-$\Sigma$'' Clump & 1   & 3000 & 0.367 & 1.5 & 2 & 1.31 & 2.8 & 3.04 \\ \bottomrule
\end{tabular}
\label{tab:clumps}
\end{table*}

\begin{table*}
\centering
\caption{Initial conditions for simulation sets}
\label{tab:ic}
\begin{tabular}{rcccccccl} \toprule
Set name & $\epsilon$ & $\langle N_* \rangle$      & $\fbin$ & $f(e)$    & IMF            &IMS& S.E. & Comment in plots\\ \hline
  \equalmass          & 0.5 & 1500        & 0 & -- & single mass   & N &N   & Single equal mass particles  \\
  \nobinaries         & 0.5 & $4000$ & 0   & --         & \cite{Kroupa2001}& N &N   & Single stars with IMF (No SE)  \\
  \nose               & 0.5 & $4000$ & 0.5 &$\delta(e) $& \cite{Kroupa2001} & N &N   & 50\% binaries (No SE)\\
  \fiducial           & 0.5 & $4000$ & 0.5 &$\delta(e) $& \cite{Kroupa2001} & N &Y   & Fiducial Case  \\
  \binariesun         & 0.5 & $4000$ & 0.5 & uniform    & \cite{Kroupa2001} & N &Y   & $e$ uniform distribution \\
  \binariesth         & 0.5 & $4000$ & 0.5 &$ 2e$       & \cite{Kroupa2001} & N &Y   & $e$ thermal distribution  \\
  \segregated         & 0.5 & $4000$ & 0.5 &$\delta(e) $& \cite{Kroupa2001} & Y &Y   & Mass segregated\\
  \fullbinaries       & 0.5 & $4000$ & 1 & $\delta(e) $ & \cite{Kroupa2001} & N &Y   & 100\% binaries  \\
  \texttt{sfe\_10}    & 0.1 & $850 $ & 0.5 &$\delta(e) $& \cite{Kroupa2001} & N &Y   & SFE = 10\%\\
  \texttt{sfe\_30}    & 0.3 & $2500$ & 0.5 &$\delta(e) $& \cite{Kroupa2001} & N &Y   & SFE = 30\%\\
  \texttt{sfe\_80}    & 0.8 & $6500$ & 0.5 &$\delta(e) $& \cite{Kroupa2001} & N &Y   & SFE = 80\%\\
  \texttt{sfe\_100}   & 1.0 & $7300$ & 0.5 &$\delta(e) $& \cite{Kroupa2001} & N &Y   & SFE = 100\%\\ \bottomrule
\end{tabular}
\tablecomments{For each of the sets named in the first column 20 simulations were
performed for each of the clumps parameters listed in Table \ref{tab:clumps}. Second
column shows the assumed SFE, third column shows the average number of stars per
simulation, fourth column the primordial binary fraction, fifth column the eccentricity
distribution function, column six shows the assumed IMF, column seven stands for whether
the cluster is initially mass segregated (IMS). Column eight shows if Stellar evolution
is included in the set and last column is a comment from which we will referring to the
set in the graphs for clarity. }
\end{table*}

\subsection{The primordial binary population}
\label{sec:binary}

Observational evidence shows that about half of star systems in the
field are binaries
\citep[e.g.,][]{Duquennoy1991,Fischer1992,Mason1998,Preibisch1999,Close2003,Basri2006,Raghavan2010}.
Given the densities of star-forming clumps and young star clusters, it
is likely that most of these binaries where born together inside
individual cores, rather then forming via subsequent dynamical
interactions \citep[e.g.,][]{Parker2014b}. However, this is a question
that our simulations will be able to address.

Theoretically, a full understanding of binary formation from a
collapsing gas core is likely to require a full non-ideal MHD
treatment to resolve formation of the accretion disk and then later
its potential fragmentation. The difficulty of this problem means that
essentially the statistical properties of primordial binaries are very
uncertain, and so we will investigate several different choices based
on observations.

For most of our simulations that include binaries, we assume a binary
system fraction, $f_{\rm bin}=0.5$. We adopt a period distribution
from the survey \citep{Raghavan2010} using a log-normal period
distribution with mean of $P=293.3$ years and standard deviation of
$\sigma_{ {\rm log} P } = 2.28$ (with $P$ in days).  We use a
companion mass ratio distribution (CMRD) of the form $dN/dq \propto
q^{0.7}$, based on observations in young star clusters
\citep{Reggiani2011}.
The eccentricity distribution remains less well constrained. While
\cite{Duquennoy1991} found a thermal distribution, i.e., $f(e)=2e$ for
solar-type stars in the solar neighborhood, a similar more recent
study \citep{Raghavan2010} found a flat eccentricity distribution for
the same kind of stars. However, if binaries form mainly via disk
fragmentation we would expect that they are born with near circular
orbits. In order to measure how much binaries are affected by
dynamical interactions in the different models we adopt initially
circular orbits for the eccentricities in our fiducial model. We also
investigate cases with initially thermal and uniform distributions of
eccentricities.

\subsection{Overview of the Cluster Models}

For each of the low and high $\Sigma$ clumps (see
Table~\ref{tab:clumps}), we set up a stellar cluster as described
above, i.e., assuming constant SFE$(r)$ and a velocity dispersion
profile equal to that of the parent clump. Thus, the initial crossing
time (i.e., dynamical time) is defined by the properties of the parent
clump to be $t_{\rm cr}\equiv R_{\rm cl}/\sigma_{\rm cl}$, i.e.,
$0.663$ and $0.118\:{\rm Myr}$ for the $\Sigmacl=0.1$ and $1\:{\rm
  g\:cm^{-2}}$ cases, respectively.

We run 20 realizations for each set of initial conditions, summarized
in Table \ref{tab:ic}. The simulations are run for 20 Myr, varying
only the random seed between simulations in the same set, which
affects the initial positions and velocities of the stars as well as
the IMF sampling.

We construct these sets of simulations starting from the simplest case
to the one that defines our fiducial case. We start using only single
mass particles of $m_{i}=1\:{\rm M_\odot}$ with no primordial binaries
and with a SFE of 50\% in the set \equalmass. Next, we include an IMF
assuming the \cite{Kroupa2001} distribution with a mass range from
0.01 to $100\:\MSun$ defining the set \nobinaries, again with no
initial binaries. We then include 50\% primordial binaries with
circular orbits and with other properties described in
\S\ref{sec:binary}, defining the set \nose. The above three simulation
sets do not include stellar evolution (SE).  We define the \fiducial
simulation set by assuming a Kroupa IMF, SFE of 50\%, with 50\% of
stars as primordial binary systems with initial circular orbits and
with stellar evolution included.

Next, we test other choices and parameters of the fiducial model. We
test two other eccentricity distributions, a thermal eccentricity
distribution, i.e., $f(e)=2e$ in the set \binariesth, and a uniform
eccentricity distribution between 0 and 1 in the set \binariesun.  An
extreme scenario of mass segregation is tested in the set \segregated
in which stars are sorted in descending order of individual stellar
mass from the center of the cluster.  We also test the extreme case in
which all stars are binary systems ($\fbin=1$) in the set
\fullbinaries.

We also carry out simulations with different SFE. These simulations
only differ from the \fiducial set in their SFE, i.e., the average
number of stars per simulation on each set increases with the SFE
since we use the same parent clump of $\Mcl = 3000\:\Msun$. The SFEs
investigated are ${\rm SFE}= 10\%$, 30\%, 80\% and 100\% and the sets
are named \texttt{sfe\_10}, \texttt{sfe\_30}, \texttt{sfe\_80} and
\texttt{sfe\_100}, respectively.

\subsection{Numerical Methods}

We follow the evolution of the star clusters for 20 Myr utilizing the
direct $N$-body integrator \textsc{Nbody6++} \citep{Wang2015} which is
a GPU/MPI optimized version of the classical and widely used direct
integrator \textsc{Nbody6} \citep{Aarseth2003}. \textsc{Nbody6++} has
implemented special regularizations to accurately follow the evolution
of binaries and high order systems in the cluster being able to
efficiently simulate star clusters with high binary fractions with no
loose of accuracy. For cases with stellar evolution we used the recipe
included in \textsc{Nbody6++} based on the analytical models for single and binary
stellar evolution developed by \cite{Hurley2000,Hurley2002}.
The code also has implemented artificial velocity kicks to emulate
asymmetrical supernovae ejections. The magnitude of the kicks are
drawn from a Maxwell distribution with $\sigma=265$~km/s following the
observations of \cite{Hobbs2005} on pulsar proper motions.

\section{Results}

\subsection{Initial dynamical state of the clusters}
\label{sec:qi}

\begin{figure}
\includegraphics[width=\columnwidth]{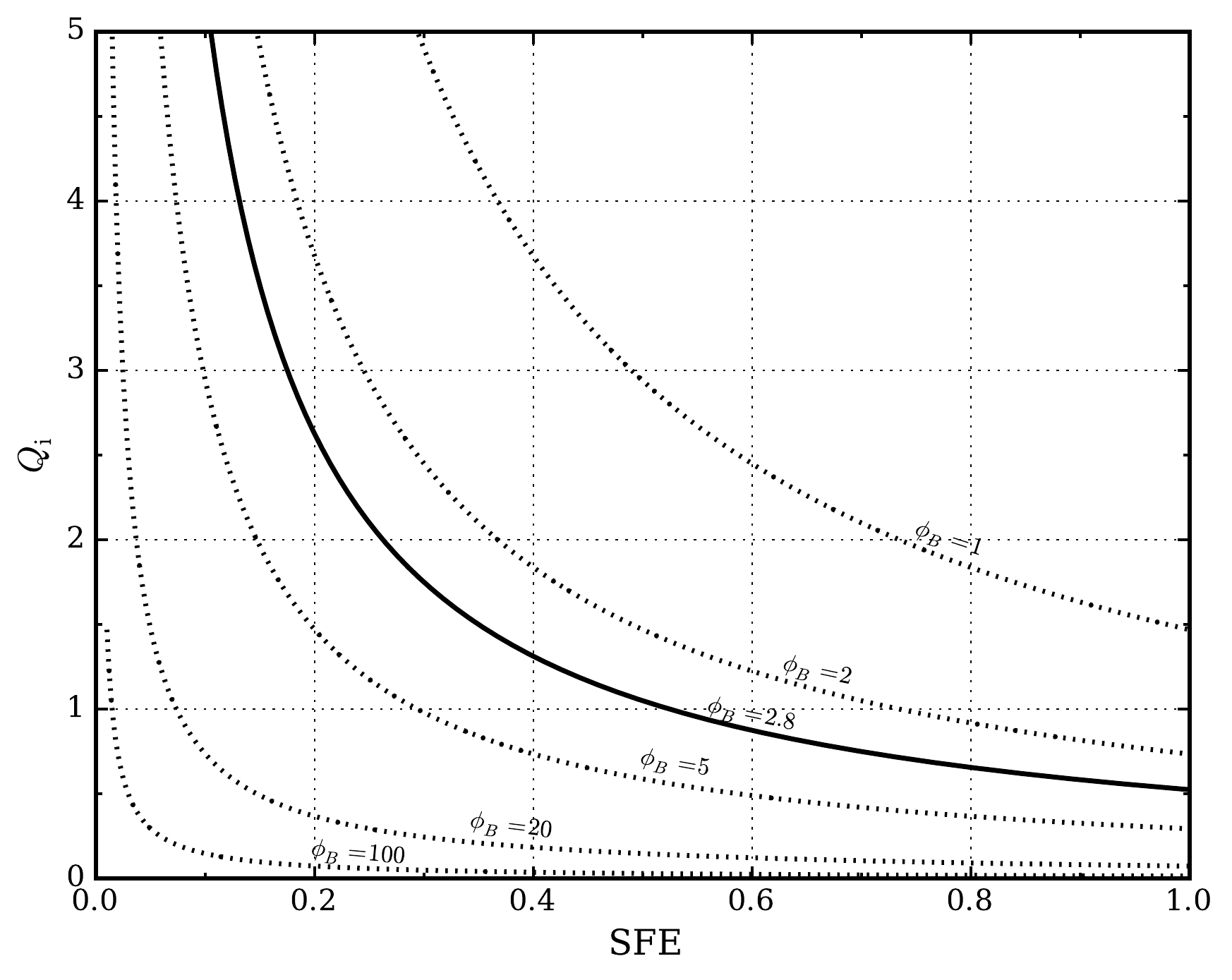}
\caption{
The initial virial ratio $Q_{i}$ as a function of the SFE from the
clump. Solid line shows the relation for our fiducial value and dotted
lines the relation for different values of $\phib$.}\label{fig:Qsfe}
\end{figure}

Before performing any simulation, from the assumptions described in
\S\ref{sec:ic}, we first derive the initial dynamical state of the
clusters by characterizing their virial ratio, i.e.,
\begin{eqnarray}
\Qi &=& - \frac{T_*}{\Omega}
\label{eq:qi}
\end{eqnarray}
where $T_*$ is the total kinetic energy of the stars and $\Omega$ is
their total gravitational potential energy. A cluster with $\Qi<1$ is
bound and $\Qi=0.5$ is the value for a state of virial equilibrium. We
assumed the gas was expelled immediately after the stars formed, thus
$\Omega$ only depends on the stars in the cluster, i.e.,
$\Omega=\Omega_*$. The potential of the stars is then
\begin{eqnarray}
\Omega_* &=& -\frac{G}{2} \int_0^{\Rcl} \left[ \frac{M(r<\Rcl)}{r}\right]^2 \dr  \nonumber \\
 & & -\frac{G}{2} \int_{\Rcl}^{\infty} \left(  \frac{M_*}{r} \right)^2 \dr \\
        &=& -\left(\frac{3-\krho}{5-2\krho}\right)\frac{GM_*^2}{\Rcl}.
        \label{eq:omega}
\end{eqnarray}
The kinetic energy of the stars is given by
\begin{eqnarray}
\label{eq:t*}
T_*&=& \frac{3}{2}M_*\sigma^2,
\end{eqnarray}
where $\sigma$ is the one dimensional mass averaged velocity
dispersion. Assuming that the stars are born from the gas following
the same dispersion profile, then $\sigma$ is related to that at the
surface by Eq.~\ref{eq:sigma}.

Replacing equations~\ref{eq:sigma},~\ref{eq:t*}
and~\ref{eq:omega} in eq.~\ref{eq:qi}, and also replacing $\Rcl$ and
$\sigmas$ by their expressions in eq.~\ref{eq:rcl} and \ref{eq:sigmas}
respectively, we obtain
\begin{eqnarray}
\label{eq:qi2}
\Qi&=& \frac{3(5-2\krho)(3-\krho)}{(8-3\krho)^2(\krho-1)}\frac{1}{\epsilon\phib}\\
\Qi&\rightarrow& \frac{0.51}{\epsilon},
\label{eq:qi3}
\end{eqnarray}
where the arrow shows the relation using the fiducial values for the
clump. Values of $Q_i$ versus SFE are shown for different models in
Figure~\ref{fig:Qsfe}.

The dynamical state of the clusters also depends on the presence of
magnetic fields in the initial clump, i.e., $\phi_B$.
In the absence of magnetic field support ($\phi_B=1$) the velocities
needed for virial equilibrium are higher and stars formed from such
gas will have higher values of $\Qi$, e.g., even in the best case
scenario with a SFE of a 100\% we have $\Qi\simeq1.6$ (and $\simeq3$
for SFE of 50\%). However, in the fiducial case with approximate
equipartition of energy density from large scale magnetic fields and
turbulence ($\phi_B\simeq2.8$) a SFE of about 50\% leads to a cluster
that is marginally gravitationally bound ($\Qi\simeq1$). 
Note that this variation of $\phi_B$ also corresponds to a variation
in the virial parameter of the gas clump, $\alpha_{\rm vir} = 5
\langle \sigma^2\rangle R/(GM)$, since for the fiducial case with
$k_\rho=1.5$ we have $\alpha_{\rm vir}=15/(4\phi_B)\rightarrow 1.34$
(see Appendix A of \citetalias{mt03}).

\subsection{The bound stellar cluster}
\label{sec:bound}

\begin{figure}
\includegraphics[width=\columnwidth]{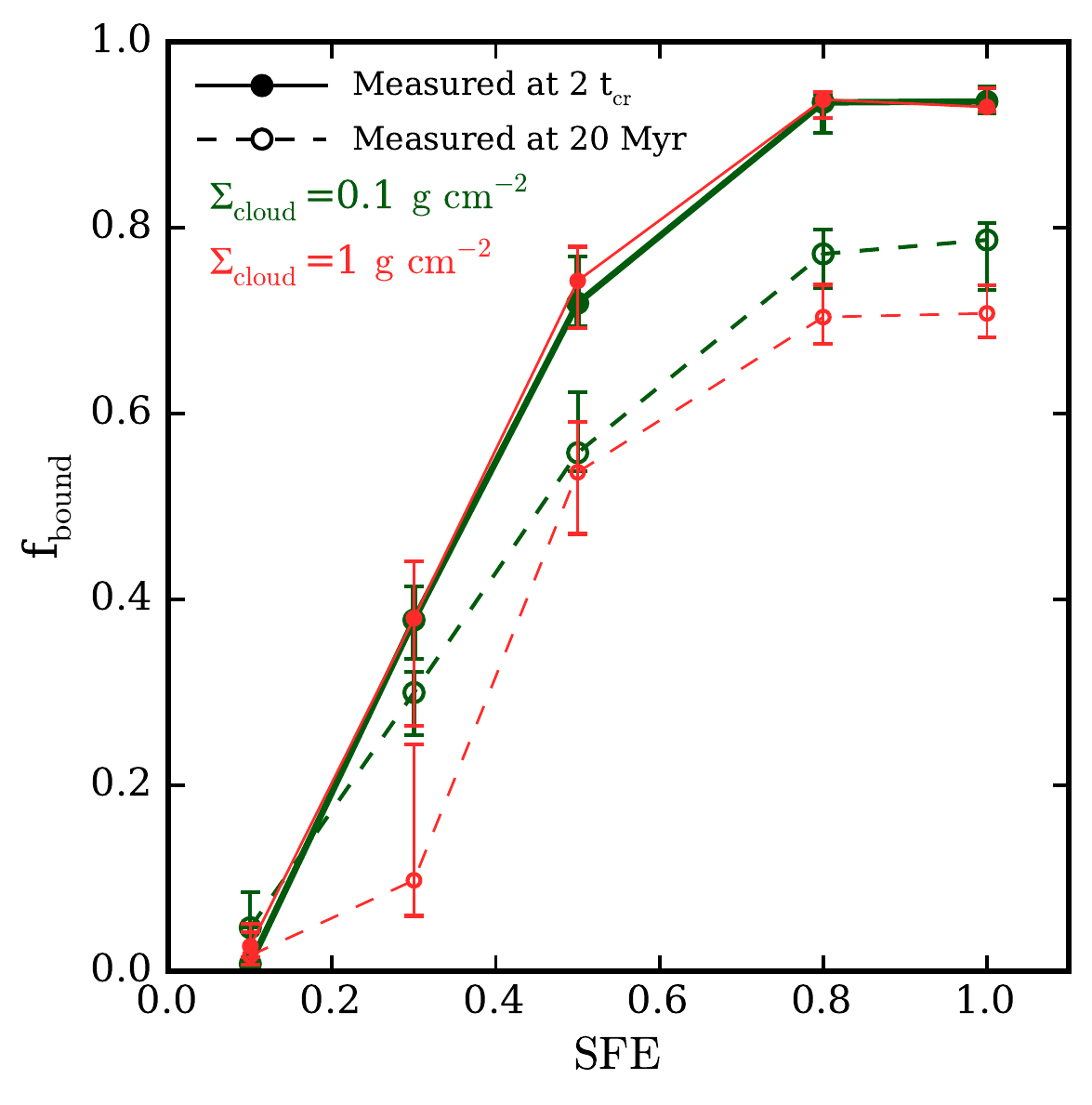}
\caption{
Bound mass fraction, $\fb$, as a function of SFE. Solid lines and
points show results at early times ($2t_{\rm cr}$); dashed lines and
open points show results at the end of the simulations (20~Myr).  The
values are the mean averages for the simulation sets (each of 20
realizations) with cases shown for $\Sigmacl=0.1$ (green) and $1\:{\rm
  g\:cm^{-2}}$ (red). Errorbars show the values between the 25th and
75th percentiles. Dynamical evolution leads to a general decrease of
$\fb$ during the first 20~Myrs of these clusters.}\label{fig:sfe}
\end{figure}

As discussed in \S\ref{sec:qi}, star clusters born from turbulent
clumps bounded by high pressure ambient environments can start with
relatively high velocity dispersions.  Their virial ratio after gas
expulsion will depend on the global SFE and the contribution of
magnetic fields to the support of the parent clump, i.e $\phib$. There
will be significant initial mass loss of the stars that are born
unbound, occurring on a timescale of a few crossing times.  However,
the gaseous clump is assumed to have a positive power law for the
velocity dispersion with radius (see Eq. \ref{eq:sigma}) and so is
more likely to be left with a central gravitationally bound core. In
contrast, a relaxed star cluster (e.g., with a Plummer profile) has a
velocity dispersion profile that decreases with radius. We thus expect
differences in the early evolution of our clusters compared to those
modeled with initial Plummer profiles by,
e.g. \cite{Goodwin2006,Baumgardt2007,Pfalzner2013}.

To measure the bound mass fraction, $\fb$, at each timestep of cluster
evolution. We construct the bound entity based on an accurate measure
of the mean velocity of the bound stars and we select all stars with
negative total energy in the frame of reference of the bound
cluster. The velocity of the bound cluster is not known {\it a priori}
(although it is expected to be close to the zero velocity of the
reference frame), and thus this is solved in an iterative
calculation. We start by selecting all stars with negative energy
inside the half mass radius of the full cluster and then iterate until
the members between iterations do not change by more than two members.
This method, called ``snowballing'', is described in \cite{Smith2013}.

Figure \ref{fig:sfe} shows the bound fractions measured at 2 crossing
times and after 20~Myr
for clusters with different SFE for simulations with a parent clump
with $\Sigmacl=0.1$ and $1.0\:{\rm g\:cm^{-2}}$.
The initially positive radial gradient of velocity dispersion of the
star clusters causes stars in the outer parts to leave first, while
the central core can remain bound. The mass fraction of this remnant
bound core depends on the initial global SFE. Later dynamical
evolution and internal mass loss of its members due to stellar
evolution leads to a slower decrease in mass of the bound core over
time.

The results shown in Figure \ref{fig:sfe} are those for the fiducial
case, i.e., with $\phib=2.8$. As discussed in \S\ref{sec:qi}, the
initial virial ratio of the star clusters depends sensitively on this
value: a higher value of $\phib$ shifts the trend shown in this figure
upwards so that even clusters with $\rm SFE=10\%$ may retain a
significant bound core if $\phib$ is sufficiently high.

\subsection{Global evolution}

\begin{figure*}
        \centering
        \includegraphics[width=\textwidth]{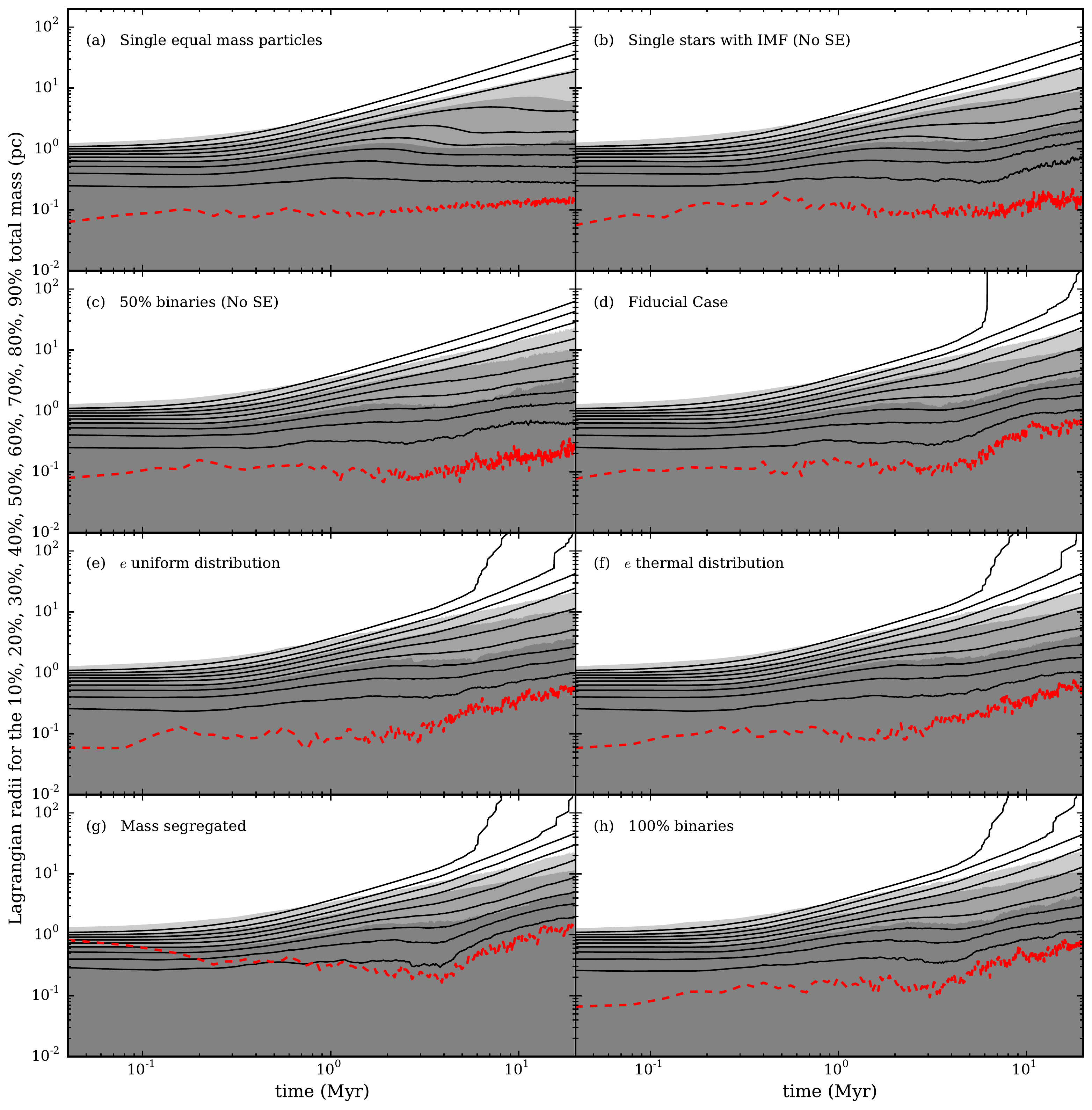}
        \caption{
Average Lagrangian radii evolution for different set of simulations
for star clusters born from a parent clump with $\Sigmacl=0.1\:{\rm
  g\:cm^{-2}}$.  We show the Lagrangian radii for the 10\%, 20\%,
30\%, 40\%, 50\%, 60\%, 70\%, 80\% and 90\% masses (black solid
lines). Red dashed lines are the core radii defined in
\cite{Aarseth2003}.  Gray shaded areas represent the regions below the
50\%, 95\% and 100\% mass radius of the bound cluster.}\label{fig:lrad}
 \end{figure*}

 \begin{figure*}
	  \centering
        \includegraphics[width=\textwidth]{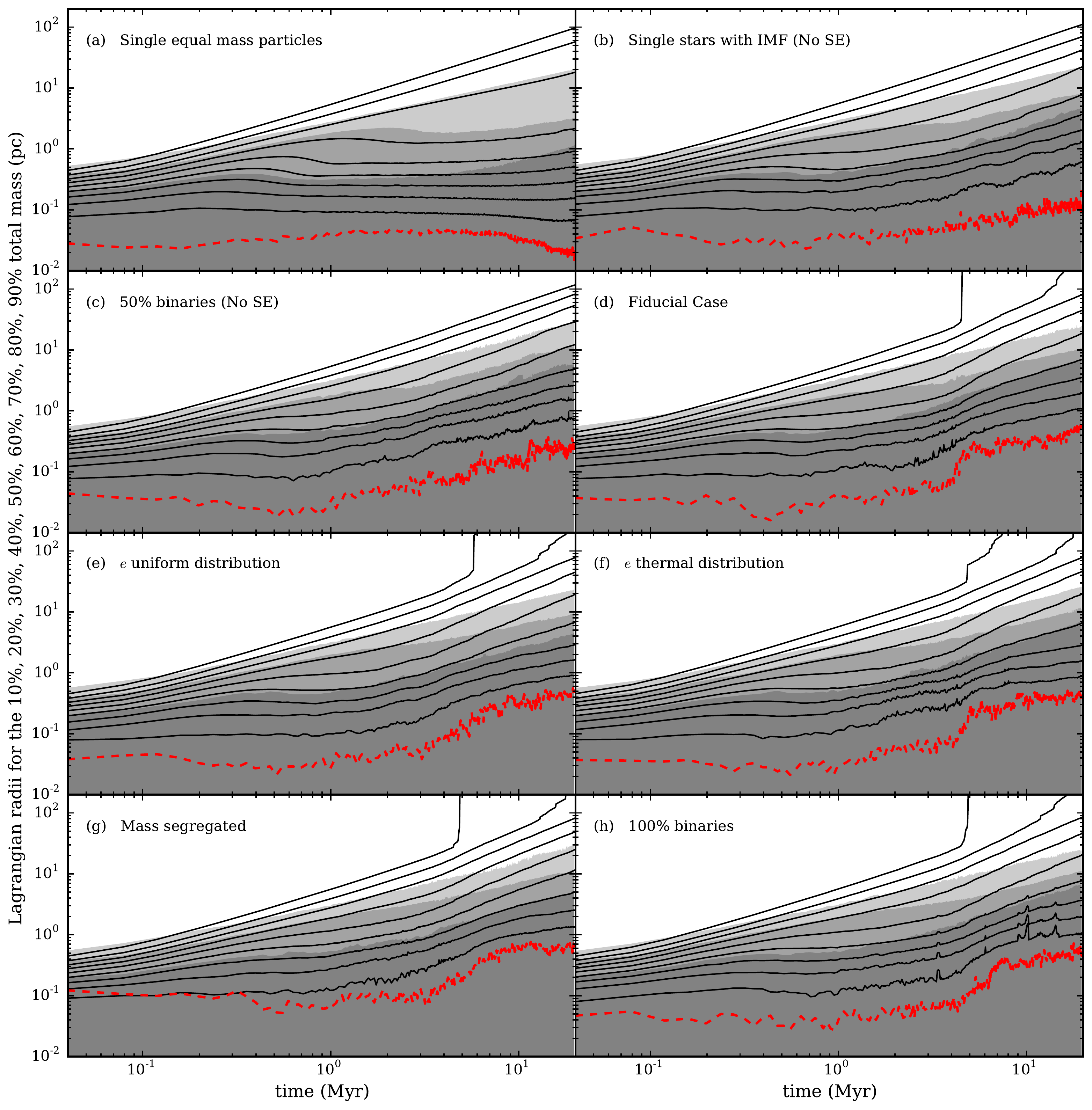}
        \caption{
Same as Figure \ref{fig:lrad} for star clusters born from a clump with
$\Sigmacl=1\:{\rm g\:cm^{-2}}$.}
        \label{fig:lradss1}
 \end{figure*}

Here we explore the evolution of the structure of the clusters with
time. The fiducial case has SFE of 50\% and $\Qi = 1.02$, slightly
above the criterion for global boundedness. Therefore, initial
expansion and some initial mass loss is expected. We show the
evolution of the Lagrangian radii with time in
Figures \ref{fig:lrad} and \ref{fig:lradss1} for $\Sigmacl=0.1$ and
$1\:{\rm g\:cm^{-2}}$, respectively, where the values presented are
the average of the 20 simulations performed for each set. In each
figure, the top four panels, (a) to (d), show the effects of gradually
adding greater degrees of realism to make the fiducial model. Then the
lower four panels show the effects of different choices of initial
binary properties and degree of initial mass segregation.  These
figures also show the evolution of the core radius $\rc$ (red dashed
line), which is the density averaged distance from the density center
of the cluster \citep[see \S15.2 in][]{Aarseth2003}.

As expected, the clusters expand with time. The expansion rate of the
outer Lagrangian radii of the clusters, i.e., of the unbound stars, is
determined by the initial velocity dispersion of the parent clump,
which is higher at higher mass surface densities. Thus star clusters
starting from a clump with $\Sigmacl=1\:{\rm g\:cm}^{-2}$ are soon
more extended than the clusters forming from clumps with
$\Sigmacl=0.1\:{\rm g\:cm}^{-2}$ of the same mass and SFE, i.e., the
half-mass radius at 20~Myr of the first group is $\sim 20\:$pc
compared to $\sim10\:$pc for the lower $\Sigma$ case.

Initial expansion of the bound portion of the cluster happens early
within a few crossing times as the clusters relax to a virialized
state. Then the later evolution is affected by dynamical interactions
between the stars (i.e., mass segregation, evolution of binaries and
dynamical ejection of stars from unstable multiple systems) and mass
loss resulting from stellar evolution. The relative importance of
these effects can be gauged by examining the sequence of panels from
(a) to (d) in Figures \ref{fig:lrad} and \ref{fig:lradss1}. The later
stage expansion of the bound cluster is negligible in the case of
equal mass stars. The model with an IMF but only single stars
undergoes mass segregation that leads to noticeable expansion after
about 6 Myr in the case of $\Sigmacl=0.1\:{\rm g\:cm^{-2}}$ and after
about 1~Myr in the case of $\Sigmacl=1.0\:{\rm g\:cm^{-2}}$.

Note that before adding binaries and stellar evolution in our models,
evolution of the star clusters with high $\Sigmacl$ would be exactly
the same as those with low $\Sigmacl$ after properly scaling for the
initial size and crossing time \cite[see ][]{Aarseth1998}. However,
the characteristic timescales introduced by binaries (e.g., at their
typical orbital separation) and by stellar evolution break this
self-similarity.

When binaries are added, their presence leads to another potential
source of expansion, since their binding energy
couples with the internal energy of the stellar cluster, leading to a
change of kinetic energy in each interaction
\citep{Heggie1975,Hills1975}. However little difference appears when
moving from single stars to 50\% binaries, even in simulations with
the high $\Sigma$ initial condition that can suffer more
interactions. As we will see in \S\ref{sec:binaries}, the initial
densities of these models are not high enough and/or do not last long
enough to give binaries, on average, the chance to interact
significantly with other stars.

The inclusion of stellar evolution causes the cluster to expand even
more. Stellar evolution starts becoming important after a few Myr,
when the first massive stars lose mass by stellar winds and then
explode as supernovae. The supernova explosions may cause stars to be
ejected (e.g., as fast runaway stars) either by the destruction of
tight binaries, velocity kicks caused by the asymmetrical explosion,
or both combined effects (see the outer two lines in Figures
\ref{fig:lrad} and \ref{fig:lradss1}). We focus on the ejection of
runaway stars below in \S\ref{sec:ejections}.  For now we see that
cluster expansion is increased by this effect and also by the fact
that the potential well of the cluster is made more shallow with the
loss of mass through stellar winds and supernovae. However, the loss
of runaway stars does not affect the global evolution of the cluster
too much, even in the most extreme case with $\fbin=1$. Finally, the
lower four panels in these figures show that variations of binary
orbital properties, degree of initial mass segregation or primordial
binary fractions have relatively minor effects.

\begin{figure*}
        \centering
\includegraphics[width=\textwidth]{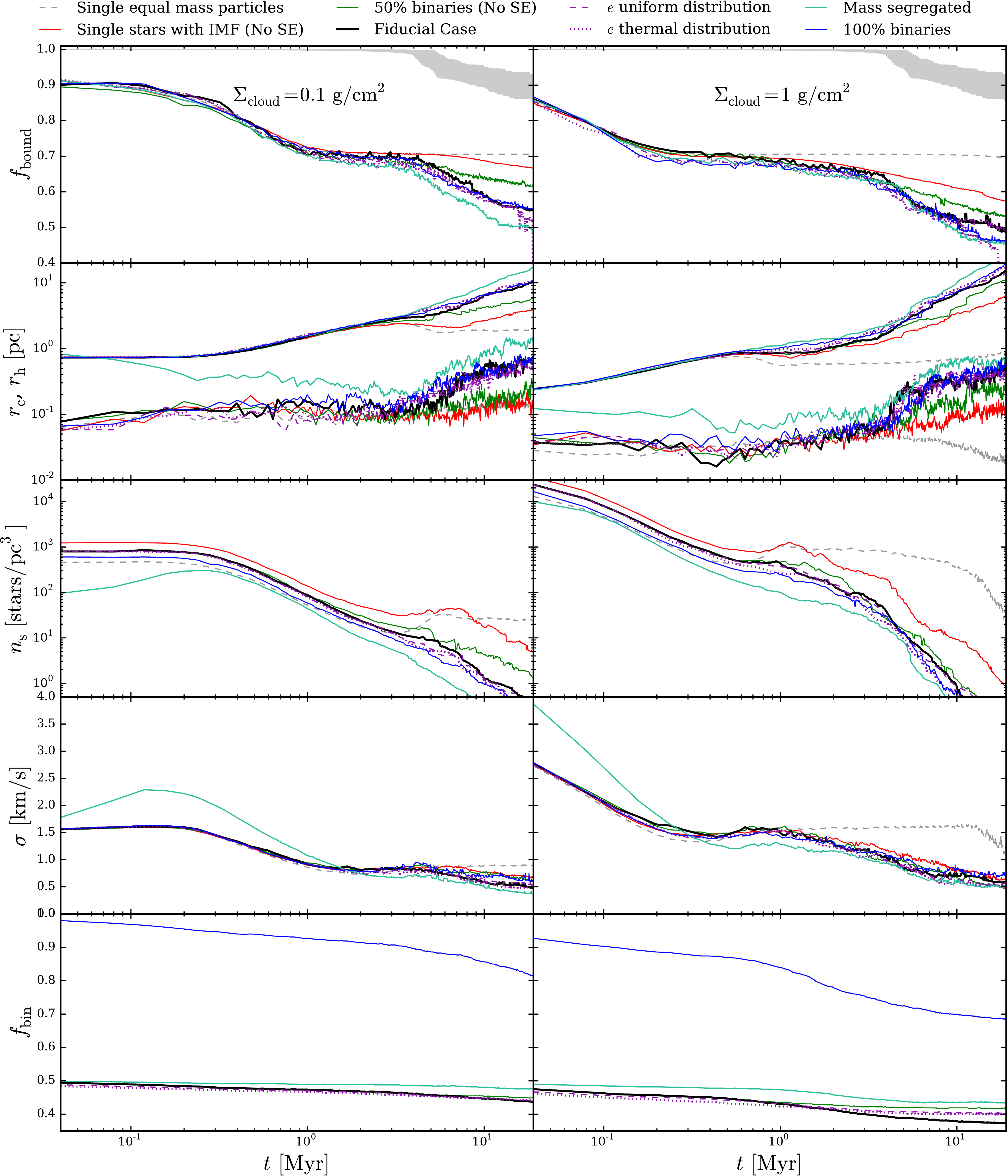}
\caption{
Time evolution of properties of fiducial clusters forming with
SFE=50\%. Left column shows clusters forming from a $\Sigmacl=0.1$ g
cm$^{-2}$ environment; right column from a $\Sigmacl=1$ g cm$^{-2}$
environment. The lines in each panel show median values calculated
from the 20 simulations performed for each set. Top row shows the
fraction of bound mass in the cluster relative to the initial mass, 
where in this figure unbound stars inside the 95\% mass radius of the
bound cluster are kept to show the timescale of their escape.  Here we
also show the fraction of total stellar mass in the \fiducial
simulations that remains after accounting for stellar evolution mass
loss: gray shaded region shows the area between 25th and 75th
percentiles.  Second row shows the evolution of core radius ($\rc$)
and half mass radius $\rh$ for all the stars in the simulation. Third
row shows the average number density of systems ($n_{s}$), where per
systems we refeer to singles and binaries, measured inside the volume
defined by $\rh$. Fourth row show the evolution of the velocity
dispersion measured inside $\rh$, and bottom row shows the evolution
of the global binary fraction.}\label{fig:evol}
\end{figure*}

In Figure \ref{fig:evol} we compare the evolution of several
parameters of the different sets of simulations for clusters born with
different initial mass surface densities: $\Sigmacl=0.1\:{\rm
  g\:cm^{-2}}$ on the left; $\Sigmacl=1\:{\rm g\:cm^{-2}}$ on the
right.  In the first row we show evolution of the bound mass fraction
$\fb$.
Only for the purposes of this figure, to show the timescale on which
initially unbound stars leave the cluster, we count all stars inside
the 95\% radius of the bound cluster as also being bound. 
Here, and in all panels in this figure, values are the medians of each
set of simulations with parameters given in Table \ref{tab:ic}. Also
shown in the top row is a shaded area representing the loss of mass
due to stellar evolution for all stars in the fiducial simulations
(including unbound stars). The second row shows the evolution of the
core radii, $\rc$, and the half-mass radii, $\rh$. The third row shows
the evolution of the effective number density, $n_{\rm s}$, i.e., the number
of systems (a binary is counted as one system) inside the volume
defined by $\rh$. The fourth row shows the evolution of the velocity
dispersion measured inside $\rh$, while the fifth row shows the
evolution of the total binary fraction.

It takes about 1.5 $\tcross$ for initially unbound stars to leave the
bound cluster, leading to $\fb$ decreasing to about 0.7. The velocity
profile of the clusters have a positive slope, i.e., higher speeds in
the outskirts, this causes outer stars to be more likely to be
unbound, with practically no chance of interacting with others. These
stars leave the cluster with a velocity dispersion determined by the
parent clump, i.e. $\sigmas$. We refer to these as \emph{unbound
  stars}, distinguishing from the \emph{ejected stars} that escape
later due to dynamical ejections. After the first $\sim1.5\:\tcross$,
all initially unbound stars leave the cluster and later evolution is
determined by dynamical interactions and stellar evolution.

Simulations with equal mass stars essentially do not lose further
members. With an IMF, then mass segregation does lead to some
additional mass loss. When including 50\% primordial binaries mass
loss at later times is moderately enhanced. Adding in stellar
evolution, i.e., in the fiducial model, continues this trend, with a
final value of $\fb\simeq 0.5$. These trends are mirrored in the
expansion of the clusters. Variations of binary orbital properties or
primordial binary fractions are seen to have relatively minor effects.

Models with full initial mass segregation show some differences. In
the case with $\Sigma_{\rm cloud}=0.1\:{\rm g\:cm^{-2}}$, the number
densities at the center are quite low initially and the core of the
cluster contracts significantly.
After this contraction, the number density is raised in the core,
which then later expands quite rapidly. Even though number densities
of these clusters are never too high, the few interactions that do
occur are enough to expand the cluster and the evolution of the 50\%
mass radius is determined by these interactions.

\begin{figure}
        \centering
        \includegraphics[width=\columnwidth]{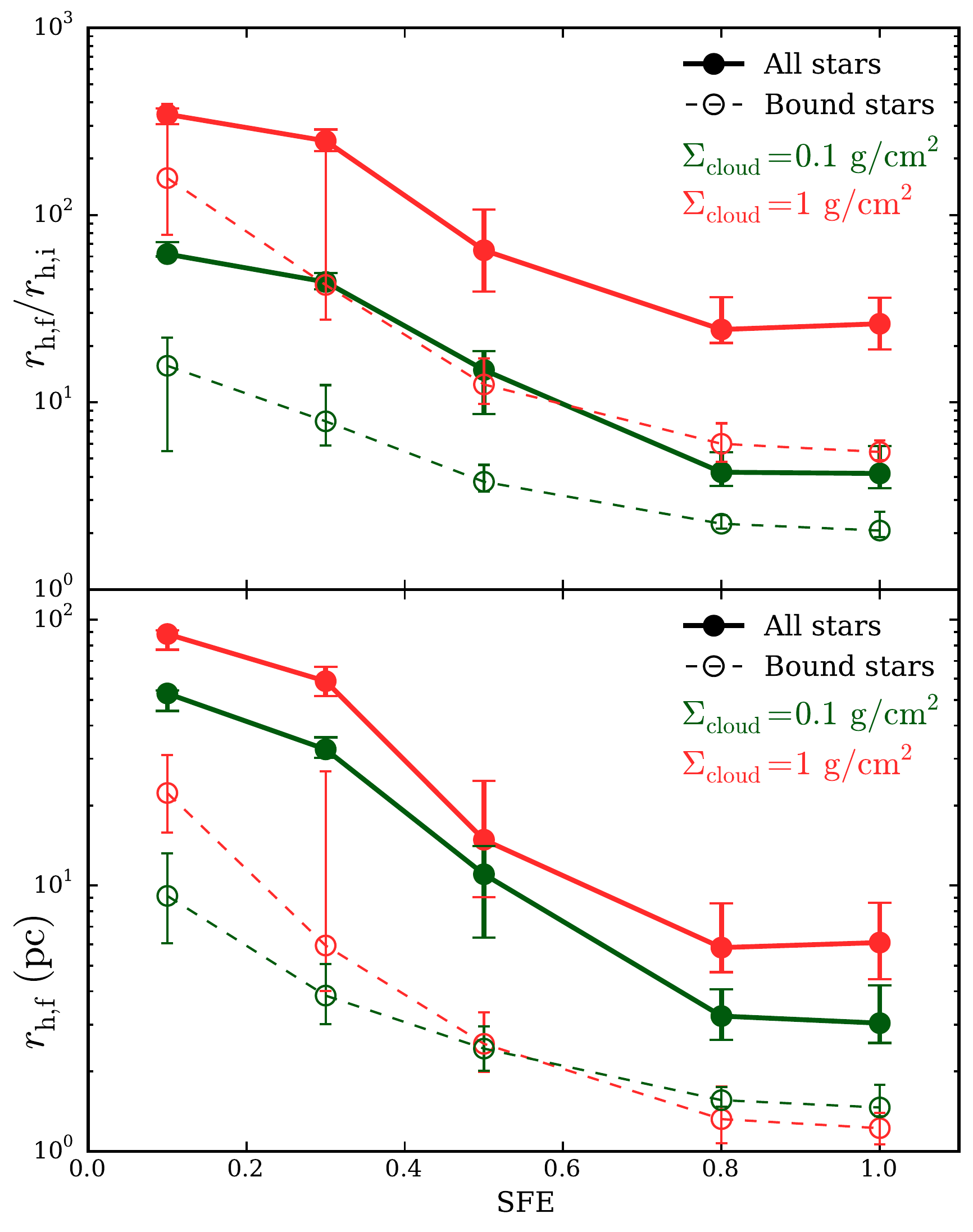}
        \caption{
The size of star clusters at 20 Myr as a function of the SFE.  Top
pannel shows the final half-mass radii, $r_{\rm h,f}$, compared to the
initial half-mass radii, $r_{\rm h,i}$. Bottom panel shows $r_{\rm
  h,f}$ in physical units.  Filled circles show measurements using all
the stars; open circles using only the bound cluster.  Values are
medians over the 20 simulations performed for each set and errorbars
shows the region between the 25th and 75th percentiles.  
}
        \label{fig:sizes}
\end{figure}

In general, the star clusters presented here expand considerably
regardless of the different parameters of the simulations. The amount
of expansion depends on the SFE and the initial cluster density. The
top panel of Figure \ref{fig:sizes} shows the ratio between the
half-mass radius at the end of the simulations, $r_{\rm h,f}$, and the
half-mass radius at the start, $r_{\rm h,i}$.  The difference between
the models with high and low initial densities are explained mainly by
the initial velocity dispersion of the parent clump. Stars that are
born unbound in the denser case escape with a higher typical velocity
than the low dense case, causing the differences in the
expansion. However, when considering the bound part of the cluster,
the actual size of star clusters at 20 Myr is similar regardless the
initial density, as shown in the bottom panel of
Figure~\ref{fig:sizes}.  Differences between the sizes of the bound
clusters arise when the SFEs are low. 
This is due to the fact that the crossing times of these bound systems
are large ($t_{\rm cr} \approx 30\:\rm Myr$) and they have not yet achieved
an equilibrium distribution by 20 Myr. Thus low SFE clusters are still
in the first phase of their expansion.
Regardless of the initial density, the final size of the bound systems
depends mainly on the initial SFE: low SFE results in a more extended
bound system.

\subsection{The effects and evolution of binaries}
\label{sec:binaries}

\begin{figure*}
        $\begin{array}{c}
        \includegraphics[width=\textwidth]{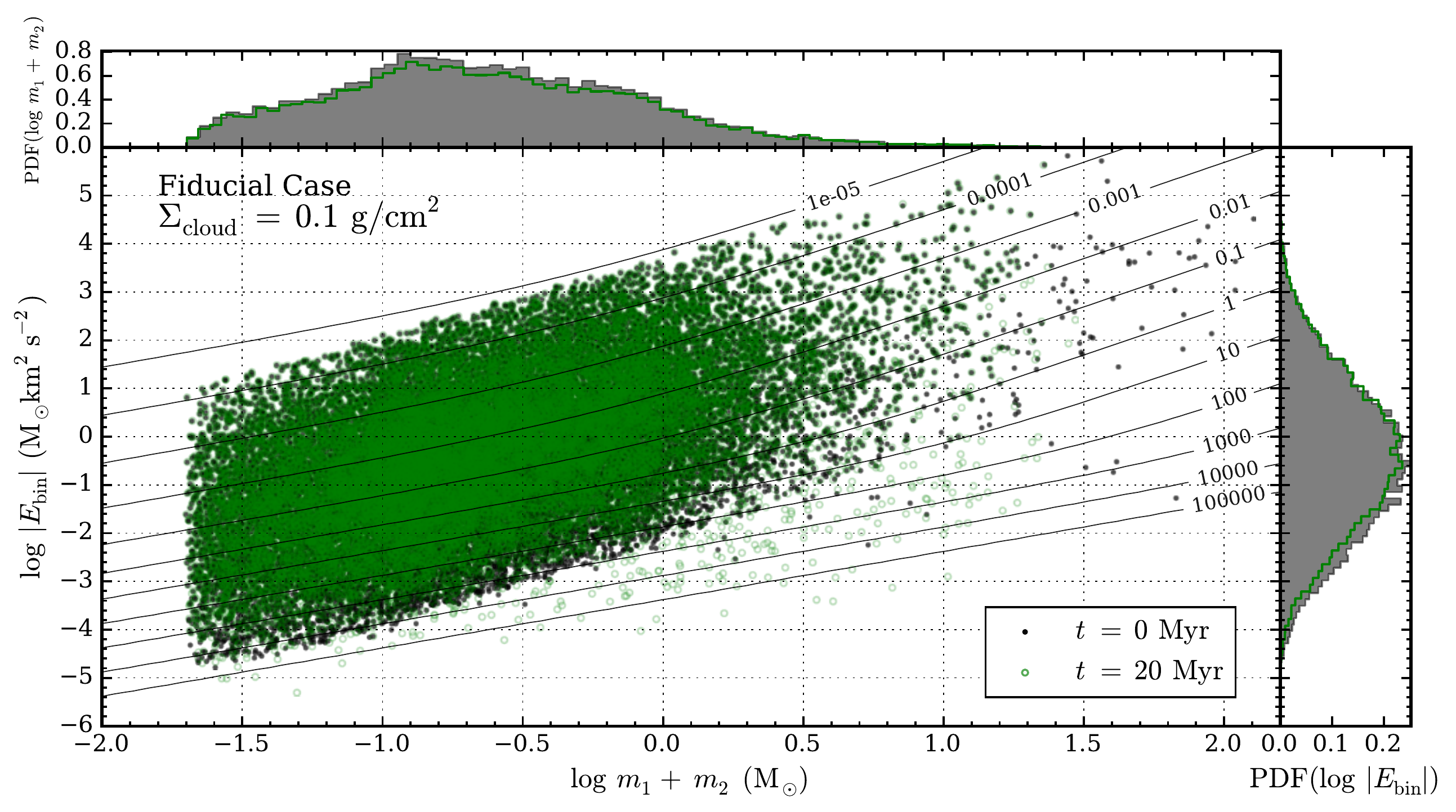} \\
        \includegraphics[width=\textwidth]{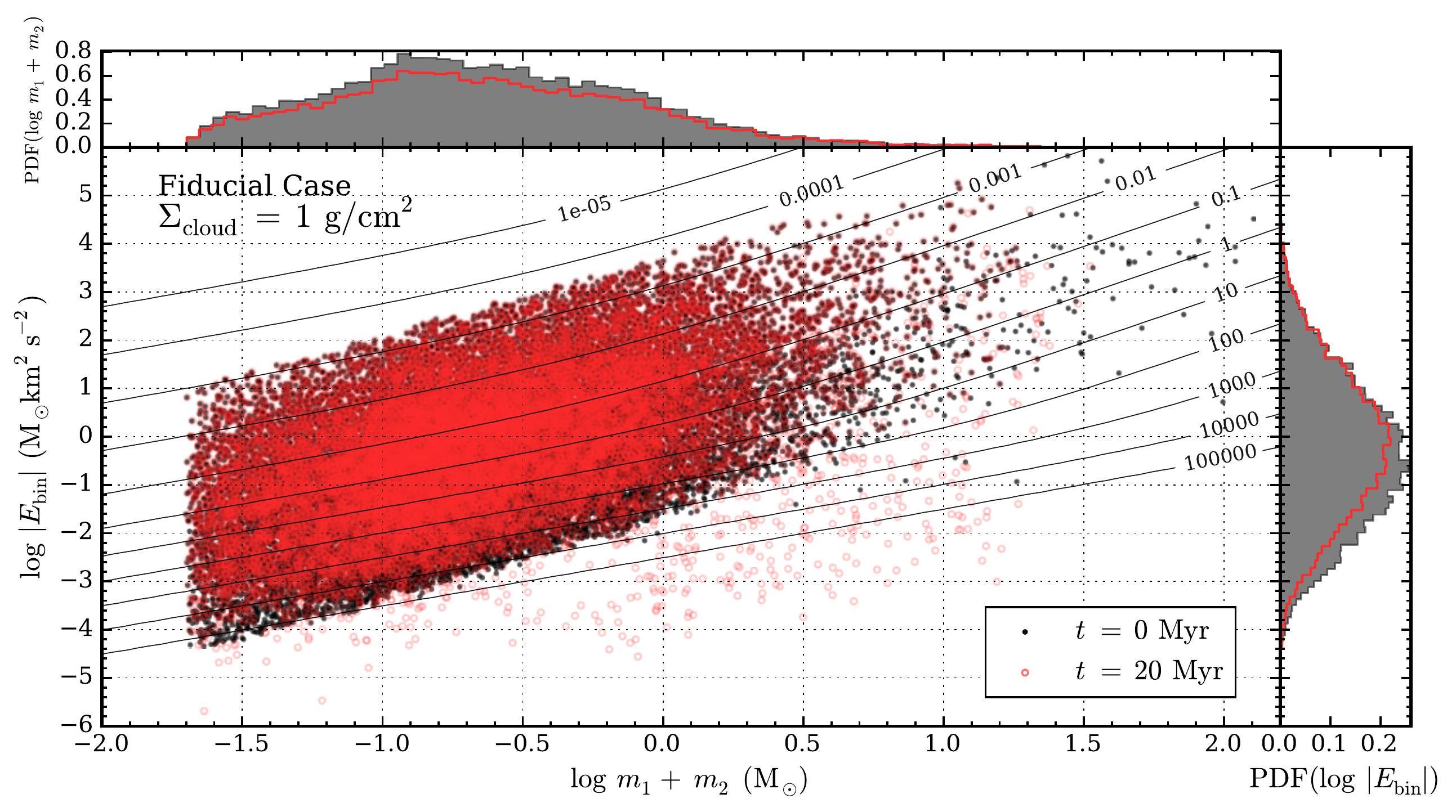}
        \end{array}$
                \caption{
Binding energies as a function of total mass for primordial binaries
at the start (black points) and at 20~Myr (red open circles) in the
set \fiducial for simulations with $\Sigmacl = 0.1\:{\rm g\:cm^{-2}}$
(top panel) and $\Sigmacl = 1\:{\rm g\:cm^{-2}}$ (bottom panel). Side
panels show the corresponding probability distribution functions
(PDFs), where the case at 20~Myr (red lines) is normalized by the
initial number of binaries. Labeled lines represent contours of
$\Gammaeff$ (Eq.~\ref{eq:gammaeff}) at the beginning of the
simulations in units of Myr$^{-1}$, calculated using values for $n_s$,
$m_s$ and $\sigma$ inside the half mass radius of the cluster.  Few
binary stars are modified in these models and typically only those
with small binding energies interchange energy with the
cluster.}\label{fig:massebin}
\end{figure*}

\begin{figure}
        \centering
        \includegraphics[width=\columnwidth]{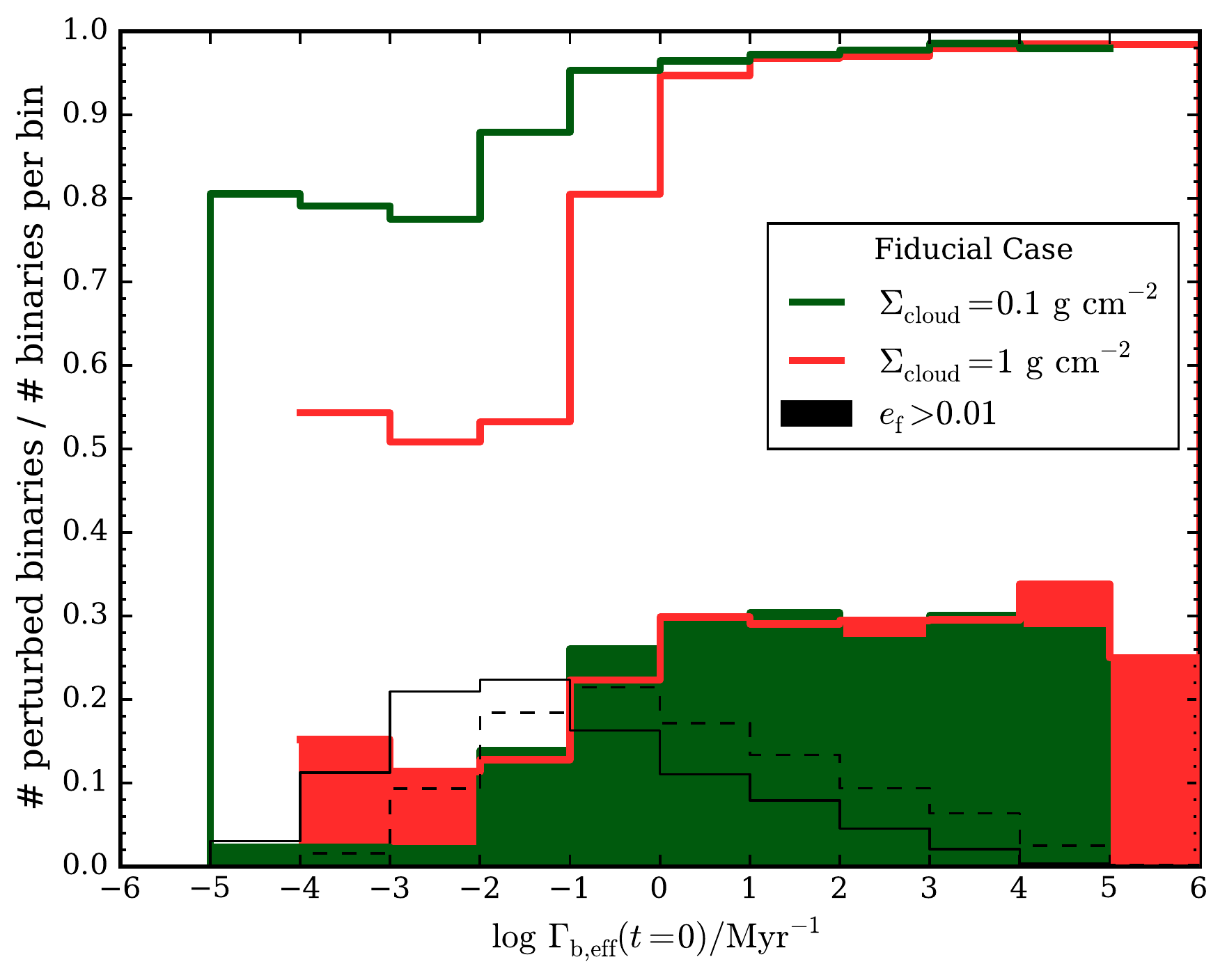}
\caption{
Number of binaries with variations in their binding energies per
initial $\Gammaeff$ bin divided by the total number of binaries on
each bin for all binaries in the set \fiducial, i.e. the probability
of a given binary of being modified as a function of the initial
$\Gammaeff$ for binaries in the set \fiducial. Shaded areas represent
the probability of binaries of changing their initial circular
eccentricities to values $e>0.1$. Thin black lines show the total
fraction of binaries in each bin for simulations with
$\Sigmacl=0.1\:{\rm g\:cm^{-2}}$ (solid) and $\Sigmacl=1\:{\rm
  g\:cm^{-2}}$ (dashed).  Only binaries with very high initial
$\Gammaeff$ are able to considerably modify their initial
eccentricity, however there are very small number of binaries that
fulfill this condition.}\label{fig:gamma}
\end{figure}

Our modeling includes a full treatment of binaries, so we are able to
examine their effects and evolution in detail. A binary will be
significantly perturbed by an external star (or multiple) if the
potential energy of the encounter is similar to that of the initial
binary, i.e.,
\begin{eqnarray}
      E_{\rm bin} =  - \frac{G m_1m_2}{2a} &\sim & -\frac{G(m_1+m_2)m_{s}}{b},
\end{eqnarray}
where $b$ is the closest approach of the perturber of mass $m_s$, and
$m_1$ and $m_2$ are the primary and secondary masses of the binary,
respectively.  Therefore, defining $\mu\equiv m_1m_2/(m_1+m_2)$ as the
reduced mass of the binary, the closest approach needed to affect
binary properties is:
\begin{eqnarray}
        b &\sim& 2 a \frac{m_{s}}{\mu}.
        \label{eq:b}
\end{eqnarray}

We now estimate 
the perturbation encounter rate of a binary of a given semimajor axis
$a$ in our model star clusters. We first derive the rate assuming
stars move without significant deflection, then include the effects of
gravitational focusing. If we assume that the cluster has only single
stars and binaries, the mean mass per system is $\langle m_{s}
\rangle=(1+\fbin) \langle m_{i} \rangle$. If there exist higher order
multiples, then $\langle m_{s} \rangle$ will be higher, however these
are not included as initial conditions in our models and we will see
that interactions are typically at a relatively low rate so that such
multiples do not form in significant numbers during the dynamical
evolution of the clusters.

The mean rate of interactions that are able to modify the properties
of a binary, $\Gammab$, is proportional to the cross section defined
by $b$, i.e., $\pi b^2$, multipled by the number density of perturbing
systems $n_{s}$
and a typical velocity in the cluster, i.e., the 1D velocity
dispersion, $\sigma$. 
Thus the interaction rate for binaries with a given semi-major axis $a$ is
\begin{eqnarray}
\Gammab&=& 4\pi \left(  \frac{ \langle m_{s} \rangle  }{\mu}a\right)^2 n_s \sigma.
        \label{eq:gamma1}
\end{eqnarray}

As we show in Figure \ref{fig:evol} the effective number density in
our model clusters quickly falls from initial values of $\sim10^3$ to
$10^4$ stars/pc$^3$ (depending on the initial environment mass surface
density) to values similar to $1$ stars/pc$^3$ at 20 Myr, in our
fiducial case. The typical velocity dispersions in the cluster,
however, do not vary too much. 

We can rewrite equation \ref{eq:gamma1} for a given binary of
semi-major axis $a$ and reduced mass $\mu$ in a more convenient way
as:
\begin{eqnarray}
        \Gamma_{\rm b}(a,\mu) & \lesssim& 
        9.67\times10^{-3} \left( \frac{n_{s}}{10^4\:{\rm pc}^{-3}} \right) 
        \left(         \frac{\sigma}{2~{\rm km/s}} \right) \nonumber \\
        \times & &  \left( \frac{a}{40~{\rm AU}} \right)^2 \frac{\langle m_s
        \rangle^2}{\mu^2}\:{\rm Myr}^{-1} \quad
        \label{eq:gamma2}
\end{eqnarray}

The above estimate does not include the effects of gravitational
focusing, which will increase the effective cross section of the
encounter by the factor $\foc=(b_{\rm eff}/b)^2$, where $b_{\rm eff}$
is the effective impact parameter that leads to a closest approach
$b$. Treating the binary and the perturbing system as single point
masses, then conservation of energy and angular momentum and
Eq. \ref{eq:b} imply
\begin{eqnarray}
        \foc&=& 1+ \frac{G\mu}{a\sigma^2}\left( \frac{m_1+m_2}{\meanms} +1 \right),
\end{eqnarray}
where a more convenient way to express the last equation is:
\begin{eqnarray}
        \foc&\approx& 1 + \left[ 5.55 
        \left(\frac{\mu}{{M_\odot}} \right)
        \left( \frac{a}{40\:{\rm AU}} \right)^{-1}
        \left( \frac{\sigma}{2\:{\rm km/s}}\right)^{-2} \right. \nonumber \\  
        & &\times \left.  \left( \frac{m_1+m_2}{m_s} +1 \right)^{\vphantom{1}} \right]
\end{eqnarray}
Then, the effective encounter rate including gravitational focussing
effects is
\begin{eqnarray}
      \Gammaeff &=& \Gammab \foc.
      \label{eq:gammaeff}
\end{eqnarray}

The interaction rates of binaries are mainly determined by the number
density or perturbing systems, which varies by several order of
magnitude over the evolution of the clusters, while none of the other
environmental parameters involved in Eq. \ref{eq:gammaeff} vary that
much. The other crucial parameters that determine the interaction rate
of a binary are its own internal parameters, i.e., the internal
binding energy $E_{\rm bin}$, which determines how resistant the
binary is against perturbations, and the total mass of the binary that
determines the strength of the gravitational focusing effect. These
parameters vary by several orders of magnitude between members in the
binary population. To give a general picture of the different
available interaction rates in the simulations we show the binary
binding energy against binary mass in Figure \ref{fig:massebin} for
binaries in the set \fiducial at the start (black points) and at
20~Myr (rad and green open circles). We show the values of $\Gammaeff$ at the
beginning of the simulation as contour lines where labels are the
corresponding values in units of Myr$^{-1}$.

From Figure \ref{fig:massebin} we can define the typical binary as the
one having total mass of $m_1+m_2 \approx 0.2\:\MSun$ and $E_{\rm bin}
\approx 1 \:\MSun\:{\rm km^2/s^2}$. Such a binary has an interaction
rate of $\Gammaeff\approx0.01\:{\rm Myr^{-1}}$ in the low $\Sigma$
case and $\Gammaeff\approx0.2 \:{\rm Myr^{-1}}$ in the high $\Sigma$
case. These interaction rates are quite small and will fall quickly as
the cluster expands. Massive binaries have higher interaction rates as
they attract other systems more efficiently, however their binding
energies are high and their effective impact parameters become very
small. Only binaries with the smallest binding energies have high
enough interaction rates to be able to interchange energy effectively
with the cluster. These stars are more likely to be low mass stars.

There are several factors that determine how many interactions a
binary will have during the simulations. If the binary is indeed
perturbed its binding energy will change and thus also its interaction
rate. The environment may vary because of several factors, e.g., the
expansion of the cluster, mass segregation, binary fraction, and
therefore it is quite complex to estimate the number of interactions a
binary will have. However, we can use the results of our simulations
and the initial $\Gammaeff$ to calculate the probability that a binary
will suffer at least one important interaction during the
simulation. To do so, we measure the binding energy of all binaries at
$t=0$. At the end of the simulation we calculate the binding energy of
the binaries that have not been disrupted and compare it with their
values at the start. 
We define ``perturbed binaries'' as those with a fractional change in
energy of 1\%.

We also measure $\Gammaeff$ according to Eq. \ref{eq:gammaeff}
assuming density and velocity values measured inside the half mass
radius of the cluster. Figure \ref{fig:gamma} shows the resulting
histogram where the value of each bin has been normalized by the total
amount of (undisrupted) binaries in each bin. We constructed the
histograms shown in Figure \ref{fig:gamma} collecting all binaries
from the 20 simulations performed for the set \fiducial. We can see the
correlation between the initial $\Gammaeff$ and the probability
of being perturbed.  There is an offset in the relation for the
different initial densities. For a given initial $\Gammaeff$ the
probability of suffering an encounter is higher in the low $\Sigma$
clusters.  This seems counter-intuitive, however the rapid initial
expansion of these clusters causes that the initial $\Gammaeff$ to be
less representative as it does not last for long (see the number density
evolution in Figure \ref{fig:evol} for $t<0.3\:\rm Myr$). 

From all the perturbed binaries we have also highlighted the cases for
which the eccentricities suffered some modification ($\Delta e>0.01$)
and displayed this probability as the shaded areas in Figure
\ref{fig:gamma}. The probability of modifying the initial eccentricity
appears to increase linearly with $\Gammaeff$ at first but then it
remains flat for higher binary interaction rates. Even stars with high
initial $\Gammaeff$ have only $\approx 30\%$ chance of modify their
initially circular orbits in a 1\%. We have measured only a few rare
cases where the initial eccentricity increases by a significant
factor.

Even though binaries with high $\Gammaeff$ have a greater chance to
interact and exchange energy with the cluster, these systems are very
rare, as can be seen in the thin black lines of Figure
\ref{fig:gamma} that shows the total fraction of binaries in each
$\Gammaeff$ bin.

The variations in these effects between our considered models is
small. The models with $\fbin=1$ have $\Gammab$ a factor 1.33 higher
than simulations with $\fbin=0.5$, because of the effect on the number
density of perturbing systems ($n_s\propto (1+f_{\rm bin})^{-1}$), and
this difference becomes smaller when considering effects of
gravitational focusing.  Models with initial mass segregation have
central number densities about 10 times lower than the fiducial case,
but it increases as the cluster evolves, until a point where the few
binary interactions that happen cause the cluster to expand relatively
quickly. Note that most of the variables shown in Eq. \ref{eq:gamma2}
tend to increase towards the cluster center. Especially in the case of
the mass segregated cluster, the mean mass per system is higher there,
which according to Eq. \ref{eq:gamma2}, is one of the most important
parameters since it strongly affects gravitational focusing.  However,
as the interactions in the centre become important the cluster expands
faster, therefore is not possible to maintain high number densities.

If we assume that binaries are born as we have modeled them, i.e.,
with initially circular orbits, then to have significant modification
of orbits, e.g., eccentricities, requires a longer high density phase,
e.g., for several Myr.
However, a longer timescale of cluster formation should not only keep
the higher densities for longer. It also would give more time for mass
segregation, which can also boost interaction rates.

\subsection{The effect of stellar evolution}
\label{sec:evolution}

\begin{figure*}
\includegraphics[width=\textwidth]{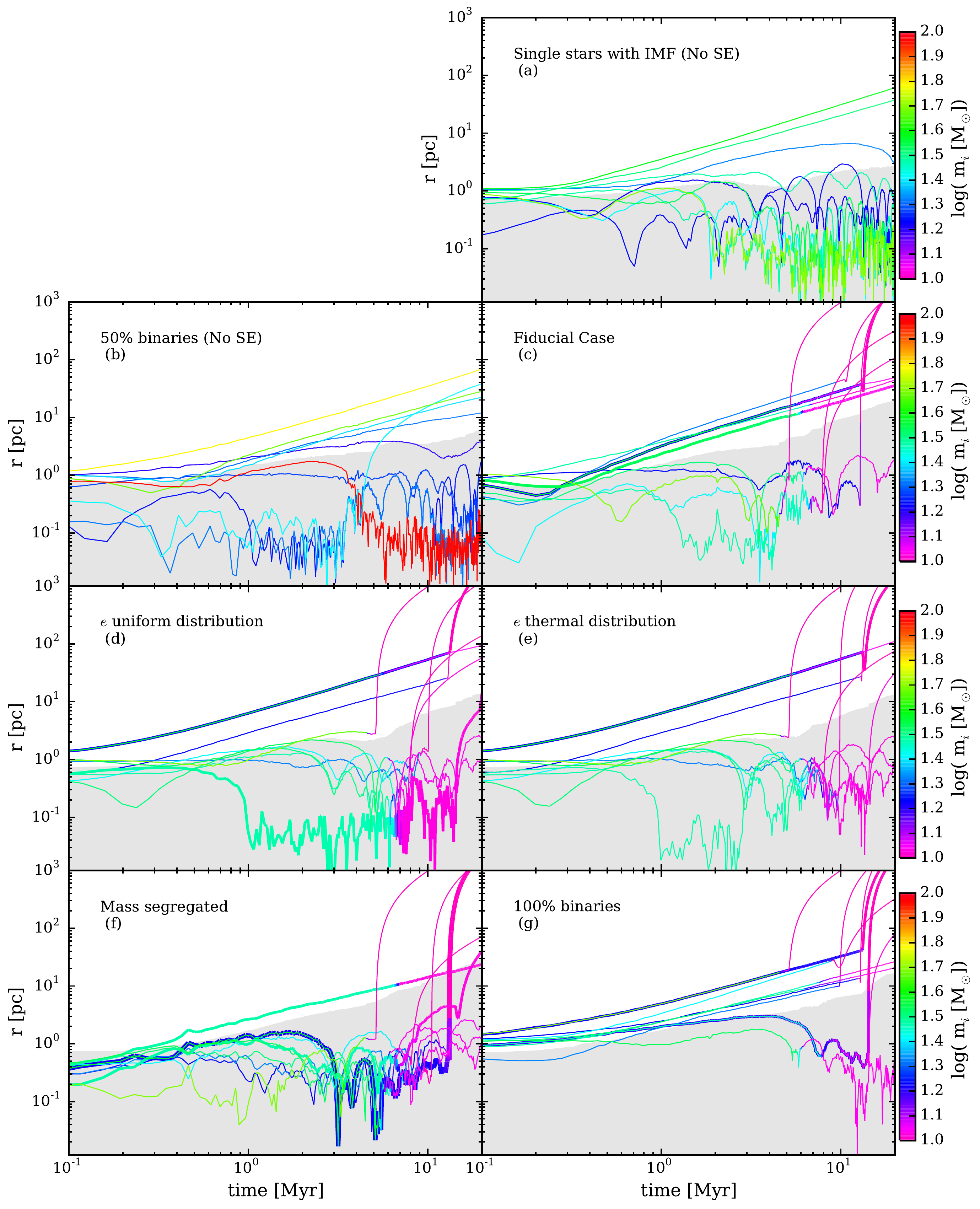}
\caption{
The evolution of the radial location of the ten most massive stars in
example simulations drawn from the model sets that have an IMF and SFE
of 50\%: (a) \nobinaries, (b) \nose, (c) \fiducial, (d) \binariesun,
(e) \binariesth (f) \segregated and (g) \fullbinaries. 
Lines show stellar distances to the density center of the cluster and line
colors the mass of the individual star. Each line represents a single
star, and in the case of a binary there is a thinner line inside the
thicker line representing the trajectory of the companion, only if the
companion is also part of the 10 most massive stars, i.e., no more
than 10 lines are shown on each panel. Shaded area is the region
inside the half mass radius of the cluster measured with respect to
the initial mass.}
        \label{fig:traj}
\end{figure*}

One important feature that can affect the evolution of cluster
dynamics is stellar evolution, especially mass loss from winds and
supernovae. Enroute to constructing our fiducial model set, we
consider two sets with no stellar evolution, i.e., \nobinaries and
\nose, this last one only differing from the \fiducial set by having
stellar evolution turned off. We also show the total mass with respect
to the initial as the shaded areas in the top panels of Figure
\ref{fig:evol}, with decrease by about 12\% on average caused entirely
by stellar evolution effects. Bound fractions at the end of the
simulation in the \fiducial set are $\simeq 0.55$, compared to $\simeq
0.62$ in the \nose runs. Thus we see that in fact the decrease in the
bound mass fraction can be explained entirely by the stellar evolution
mass loss, rather than a significantly increased tendency for
individual stellar members to be lost from the clusters.

Another way of losing mass from the cluster is due to the sudden
ejection of the members of a binary system caused by a supernova
explosion. After the supernovae explodes the binding energy suddenly
drops and the system may be broken \citep{Zwicky1957,Blaauw1961}.
The stars, both the remnant of core collapse and the secondary star of
the binary, may be ejected from the cluster as runaway stars.
In this case it is expected that the models with 100\% binaries
(\fullbinaries) experience a higher loss of members since all
supernova occur in binary systems.
However, this has only a modest effect on the bound mass fraction, as
shown in Fig.~\ref{fig:evol}. The orbital velocity of a $10\:\Msun$ star
in a typical binary is $\sim0.3$ km/s if in the peak of the period
distribution, and can vary from $\sim0.001$ to 50 km/s if we move one
$\sigma$ from the mean period.  We find in our simulations that the
mean escape velocity of the bound cluster varies from $\sim7$ to 0.6
km/s over the course of the simulation. Thus it is not certain that a
binary star will be ejected from the cluster due to binary
disruption. However, we have also included velocity kicks due to
asymmetric SN explosions, with typical values of $\sim100\:$km/s. This
effect is the main factor responsible for ejections of the remnants of
supernova explosions. The concomitant ejection of the secondaries will
depend on the binary properties at time of supernova explosion, which
for the models presented here depends mostly on primordial binary
properties.

Figure \ref{fig:traj} shows radial trajectories of the ten most
massive stars in example simulations drawn from the sets of
investigated models. Moving from panels (a) to (b), we see the effects
of primordial binaries increasing the likelihood of dynamical ejection
from an unstable multiple. Then from (b) to (c), we see the effects of
stellar evolution, especially ejection of stars after supernova
explosions. These types of ejections are more common than fast
ejections resulting from the decay of unstable multiples. Varying the
initial eccentricity distribution has only minor effects compared to
the fiducial model. The fully mass-segregated case leads to the most
extreme concentration of the ten most massive stars in the core of the
cluster. Having 100\% binaries may also lead to greater concentration
of those massive star that are bound in the cluster to its
core. However, note that there is a large degree of variation caused
by stochastic sampling of the IMF in these examples shown in
Fig.~\ref{fig:traj}.

\subsection{Ejected stars and kinematic structure} 
\label{sec:ejections}

\begin{figure*}
        $\begin{array}{cc}
             \includegraphics[width=\columnwidth]{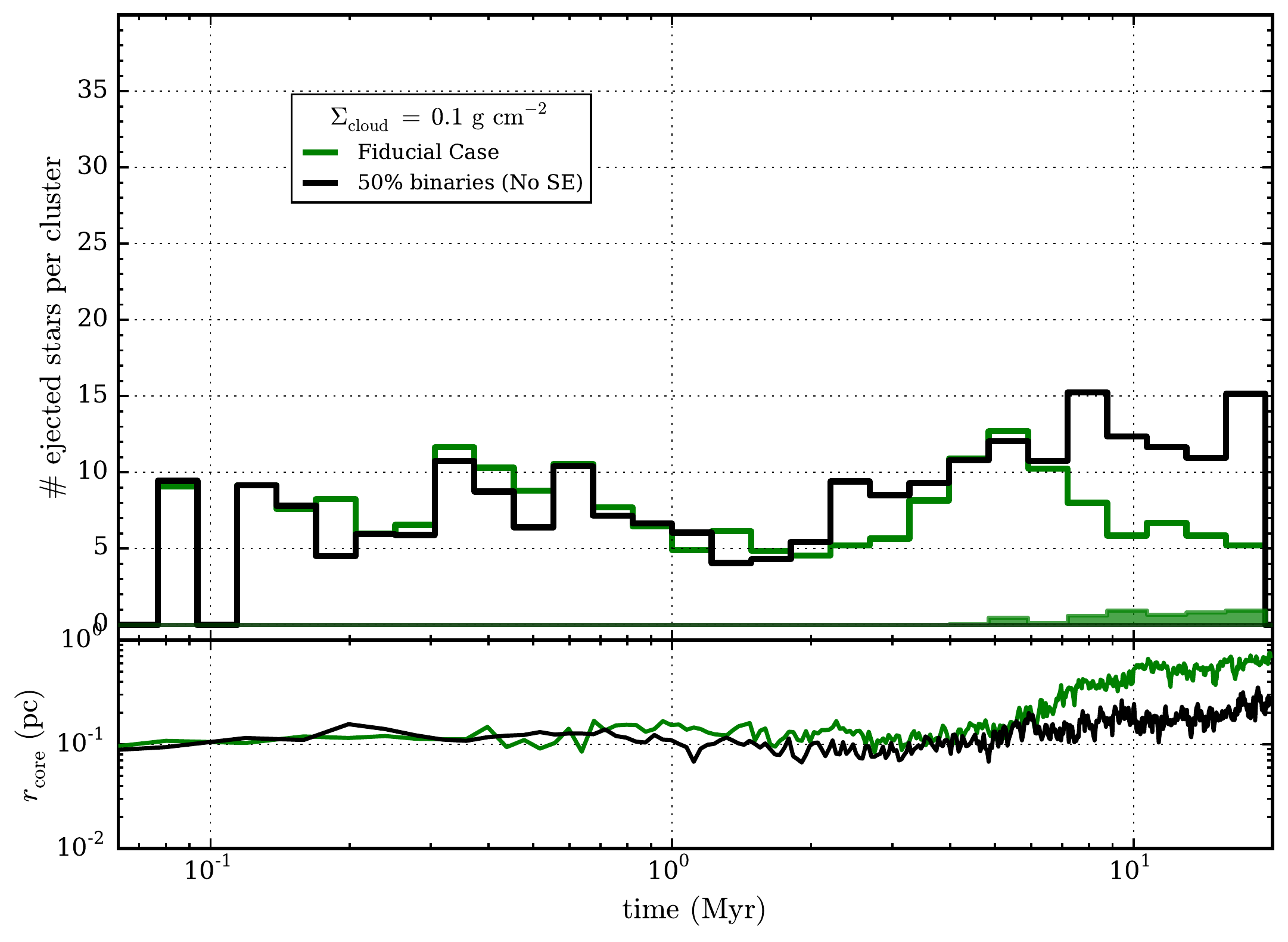} &
             \includegraphics[width=\columnwidth]{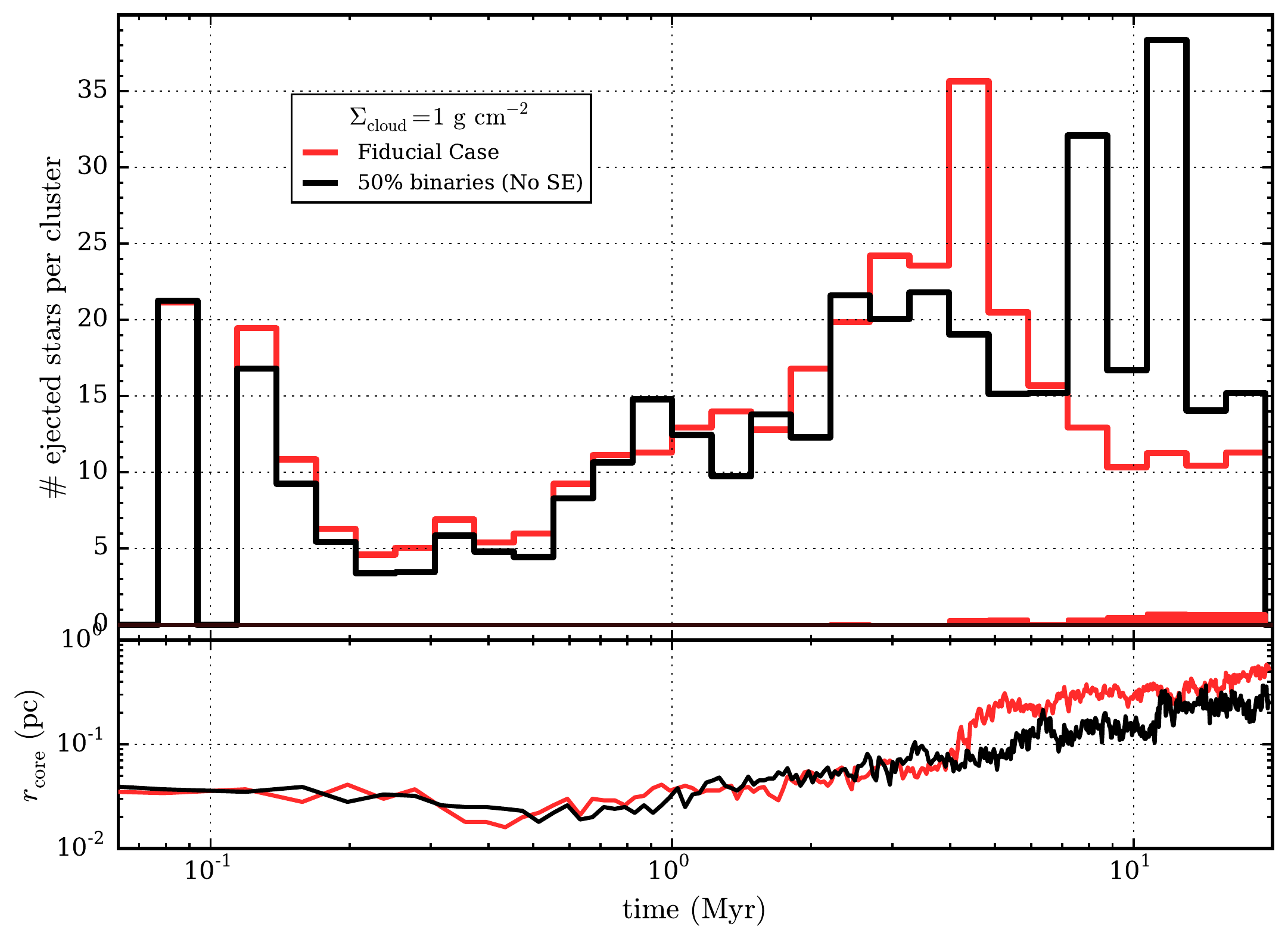}\\
     \end{array}$
\caption{
The number of strong dynamical ejections (see text) per logarithmic
time interval per cluster simulation for the set \fiducial compared
with the set without stellar evolution \nose for the low $\Sigma$
clump (left) and high $\Sigma$ clump (right). Shaded histograms show
ejections caused by supernovae explosions. Bottom panels show the
evolution of the average cluster core radii.}
     \label{fig:ejectiontime}
\end{figure*}

\begin{figure*}
        $\begin{array}{cc}
             \includegraphics[width=\columnwidth]{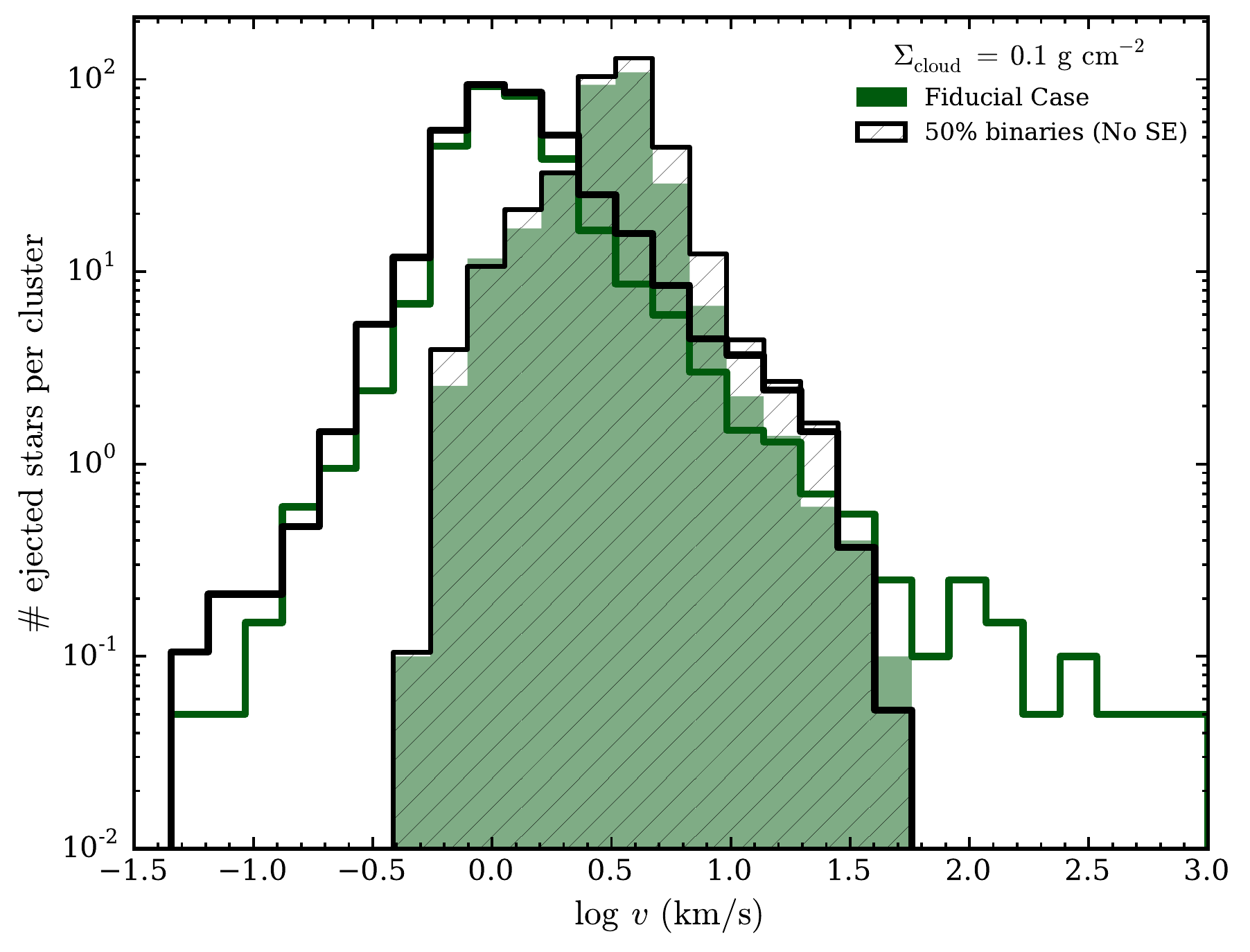} &
             \includegraphics[width=\columnwidth]{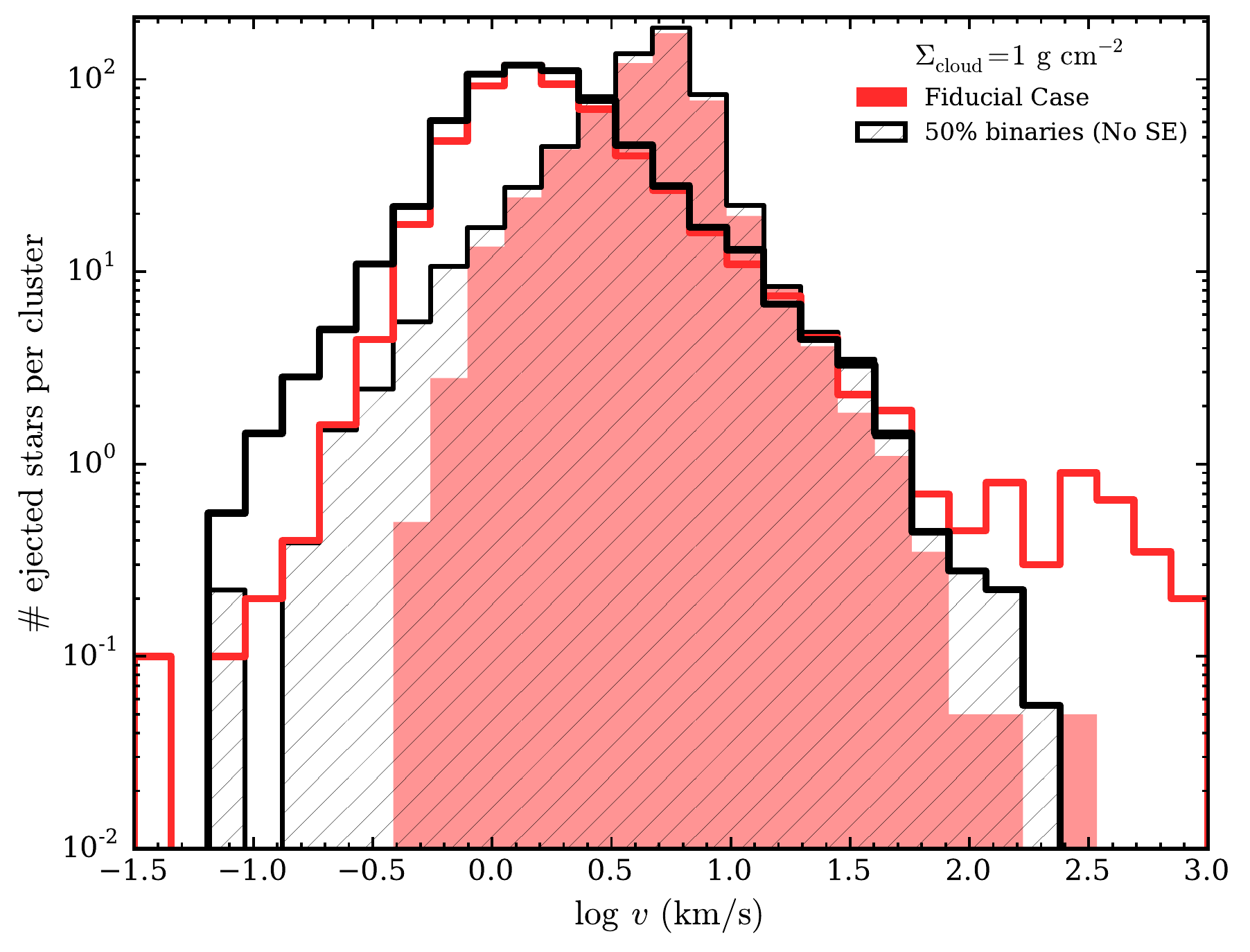}\\
     \end{array}$
\caption{
Velocity distribution of stars ejected dynamically in the \fiducial
set compared with ejected stars in the set with no stellar evolution
\nose.  Left panel show the low $\Sigma$ case and right panel the high
$\Sigma$ simulations. Solid lines shows the velocity distribution at
20 Myr ($v_\infty$) while shaded histograms shows the distribution of
velocities right after ejection ($v_0$). Blue dashed line represents
the best fit to the set \fiducial for velocities at 20 Myr, i.e. when
stars already left the cluster. }
     \label{fig:ejectionvel}
\end{figure*}

We classify the stars in three main groups: (1) unbound stars, which
are born unbound from the cluster because of the initial conditions,
i.e., because of the loss of gravitational potential and confining
pressure due to gas expulsion; (2) bound stars, which are the stars
still bound at the end of the simulation; and (3) ejected stars, which
become unbound during cluster evolution. Amongst ejected stars we
identify three different mechanisms that can lead to ejection: (A)
``supernova ejection'' either by the kick received to the core
collapse remnant and/or the disruption of a binary that contained the
supernova progenitor (see \S\ref{sec:evolution}); (B) ``dynamical
ejection'' due to decay of unstable triple or higher order multiple
systems or due to slingshot super-elastic encounters, which leave
behind a more tightly bound binary or multiple; (C) ``gentle
ejection'' where a star finds itself unbound as a result of the global
evolution of the cluster potential.

In order to obtain detailed information about the ejection events we
identify bound members in the simulations snapshot by snapshot,
recording information about the stars the first time they appear
unbound and comparing with the previous output time. We also record
their positions and velocities at the end of the simulation. 

We are especially interested in ``strong'' dynamical ejections that
lead to relatively fast ejection velocities from the cluster, and
identify such stars as having $\Delta T_i/\Delta \Omega_i \geq 2$ from
the previous time output. Such stars will be easier to identify in
proper motion studies of young clusters.

Figure \ref{fig:ejectiontime} shows the number of such strong ejection
events, including also supernova ejections, per cluster per
logarithmic time interval. We compare simulations with (\fiducial) and
without stellar evolution (\nose) for the two different initial
$\Sigma$ cases. 

As expected, high initial $\Sigmacl$ leads to bound clusters with
smaller core radii and so results in a larger rate of strong dynamical
ejections than the low $\Sigma$ case. Another difference is the number
of ejections after stellar evolution becomes relevant. The number of
massive stars in the cluster decreases significantly after the first
supernova explosions and resulting ejections. Massive stars are likely
to be near the cluster center, therefore supernova explosions lead to
a drop in the central density of the cluster that has a direct effect
on the subsequent number of dynamical ejections, as can be seen in the
anti-correlation with core radius. Cases without stellar evolution
show a constant decrease in their rate of dynamical ejection events
(i.e., a flat distribution in equally logarithmically spaced time
bins), while in cases with stellar evolution the decrease becomes
steeper after the first supernovae.

We also have information about the velocities right after ejection
($v_0$) and at the end of the simulation ($v_{\infty}$), we show
the corresponding distributions in Figure~\ref{fig:ejectionvel} with
shaded areas for $v_0$ and solid lines for $v_{\infty}$. As
expected, low velocity stars show a more significant relative decrease
in their velocities due to their transit out of the cluster
potential. We also notice a high velocity tail of stars appearing in
the fiducial case with stellar evolution: these are the result of
dynamical ejection of massive stars, which later explode as supernovae
resulting in a secondary kick for their remnants and any binary
companions. Such a \emph{two-step ejection} scenario has been proposed
by \cite{Pflamm2010} \citep[see also][]{Gvaramadze2008} and has been
argued to explain some O-type runaway stars and remnants with no
apparent origin cluster (an alternative could be isolated massive star
formation), as well as the rare observed cases of hyperfast runaways
with velocities above 1000 km/s \citep[e.g.\ as in][]{Chatterjee2005}
(an alternative could be interaction with the supermassive black hole
in the Galactic center).

Out of all the ejected O-type stars in our models, i.e., stars more
massive than $16\:\MSun$, 35\% and 22\% of them were ejected
dynamically in the low and high $\Sigma$ cases, respectively.  While
in the same order: 24\% and 33\% of them were ejected because of
supernova explosions; and the remaining 41\% and 45\% were gentle
ejection caused by the drop in the cluster potential or ``weak''
dynamical ejections (i.e., $\Delta T_i/\Delta \Omega_i < 2$). We
expect that these numbers will change when more realistic models of
gradual star cluster formation are considered, since these are likely
to lead to different cluster core densities.

\begin{figure*}
        $\begin{array}{l}
        \includegraphics[width=\textwidth]{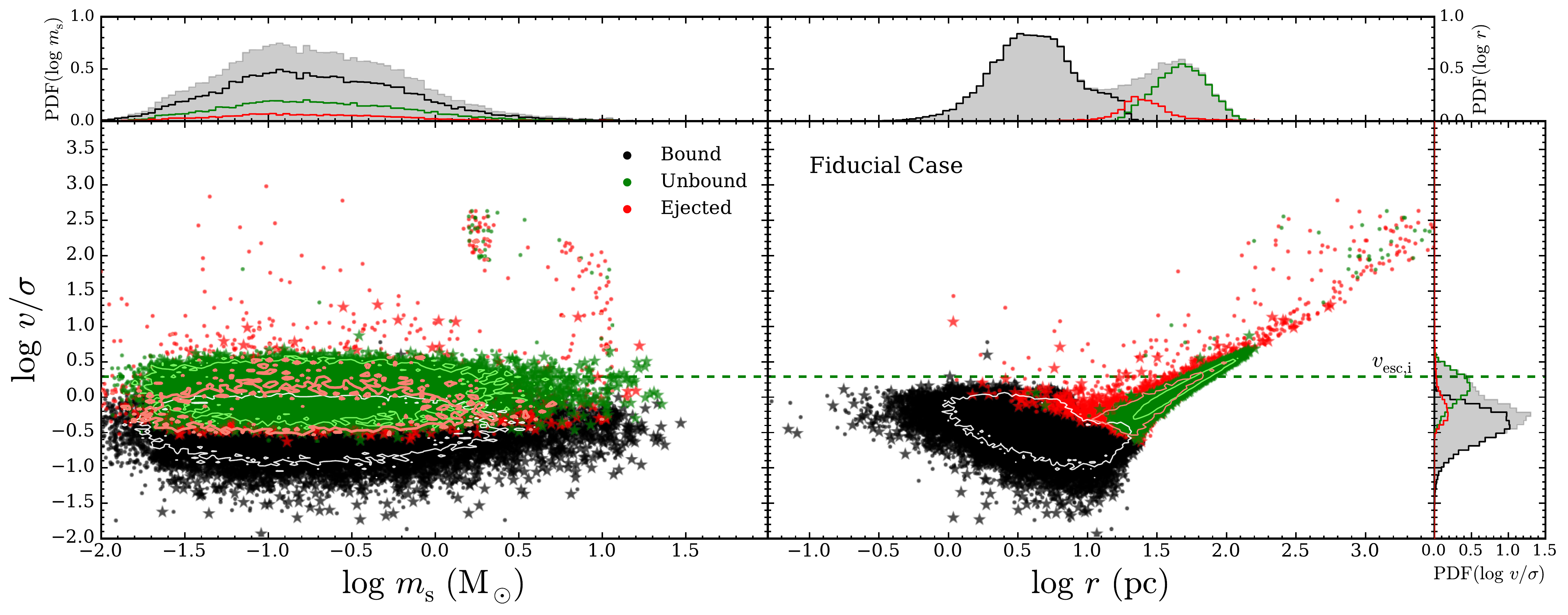} \\
        \includegraphics[width=\textwidth]{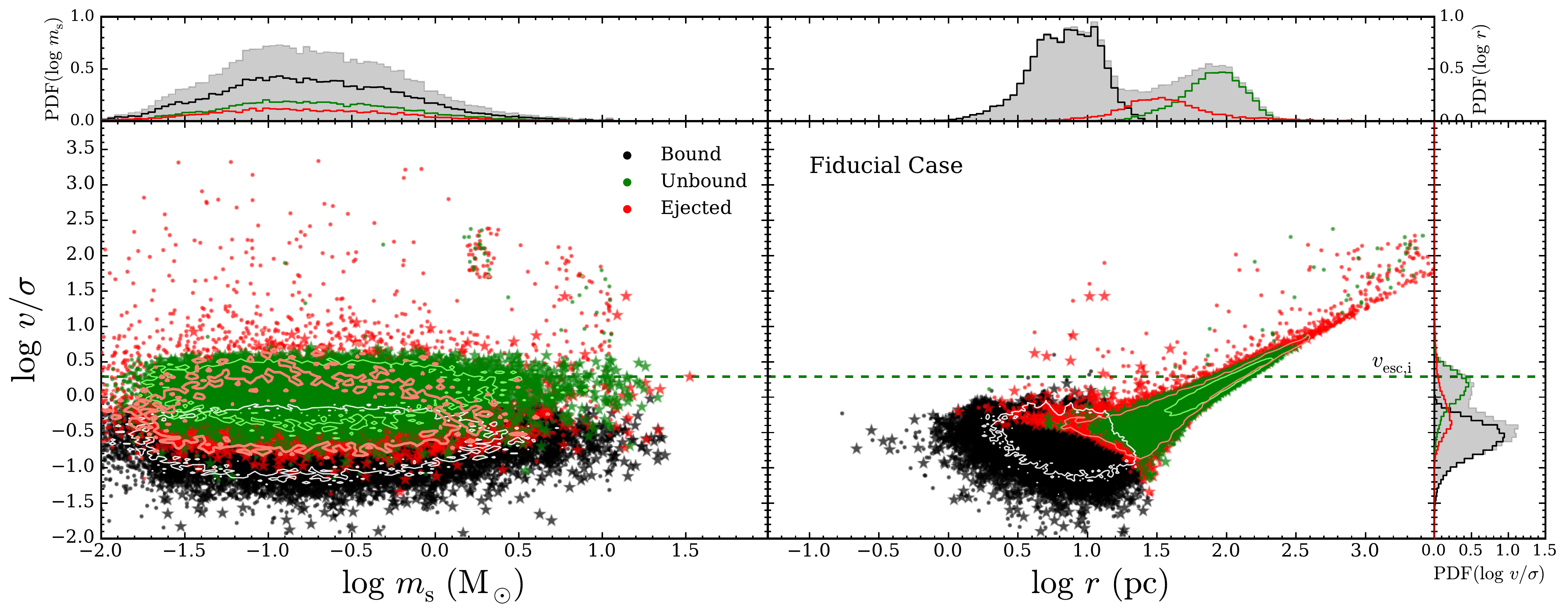} 
        \end{array}$
        \caption{
Normalized velocity versus mass (left panels) and versus distance
(right panels) for all stars in the fiducial case for simulations with
$\Sigmacl=0.1\:{\rm g\:cm}^{-2}$ (top) and $\Sigmacl=1\:{\rm
  g\:cm}^{-2}$ (bottom). Stars are separated in three groups: bound
stars (black); stars born unbound (green); and ejected stars
(red). Different symbols represent either if the star is a single
(circle) or a binary (star). Velocity values are normalized by the
mass average velocity dispersion of the parent clump (see
\ref{tab:ic}). Small top and side panels show the PDFs considering all
stars in the set (gray shaded area), and the fraction of the PDFs that
correspond to each group of stars (lines). The escape velocity from
the stellar cluster at its surface at the start of the simulation is
shown by a green dashed line. In order to show some of the structure
hiding in the cloud of points, contours that contains 90\% of the
stars on each set are shown in a lighter color, i.e., white for bound,
light green for unbound, and light red for ejected stars.}
        \label{fig:mv}
\end{figure*}\vspace{12pt}

Figure \ref{fig:mv} shows the normalized velocity versus mass (left
panels) and normalized velocity versus radial distance from cluster
center (right panels) for all the stars in the simulations of the
\fiducial set (see Figure \ref{fig:mvall} in the Appendix \ref{ap:1}
for same plots for other sets) with low (top panels) and high (bottom
panels) values of $\Sigmacl$. Velocities are normalized by the mass
average velocity dispersion of the parent clump (see $\sigma$ in Table
\ref{tab:ic} and Eq.  \ref{eq:sigma}). Masses show the final mass of
the single stars (filled circles) or combined binary (filled stars).

With exception of massive stars undergoing supernova explosions,
unbound stars (green symbols) do not change their velocities
significantly on leaving the cluster.  
They have a peak on the PDF at $\sigma$, below the initial mean escape
velocity of the cluster, $v_{\rm esc,i}$, shown by a green dashed line
in Figure \ref{fig:mv}. 
Figure \ref{fig:mvall} shows that the width of the velocity PDF of
unbound stars is quite constant over all the different simulation
sets. 

Ejected stars have a velocity PDF that peaks between those of the
unbound and bound stars. This is because the escape velocity decreases
with time as the cluster expands and loses members. As discussed
earlier, a cluster born in a denser state expands more quickly and at
20 Myr its half mass radius is about 10 times larger than the same
cluster born in the lower density state. This leads to a larger
relative decrease in the mean escape speed of the cluster over the
20~Myr of evolution followed by these simulations, compared to the low
$\Sigma$ case.
This in turn causes a broadening of the ejected star velocity
PDF. Otherwise, the widths of the velocity PDFs do not vary much
between simulations sets.

Bound and unbound stars are more clearly distinguished in the velocity
versus radial distance diagram. Bound stars have, in general, higher
velocities at the center of the cluster and lower velocities at the
outskirts. They thus populate different areas from the unbound and
ejected stars: higher velocities of these stars carry them further
from the cluster, modulated also by the time when they were ejected.
Supernova induced velocity kicks also lead to modification of a small
fraction of stars in this diagram. The group of neutron stars seen in
the velocity-mass diagram is another manifestation of such
effects. Their velocities are a direct result of the assumed
Maxwellian distribution for supernova-induced kick velocities with
$\sigma=265\:{\rm km\:s}^{-1}$.

\begin{figure}
        \includegraphics[width=\columnwidth]{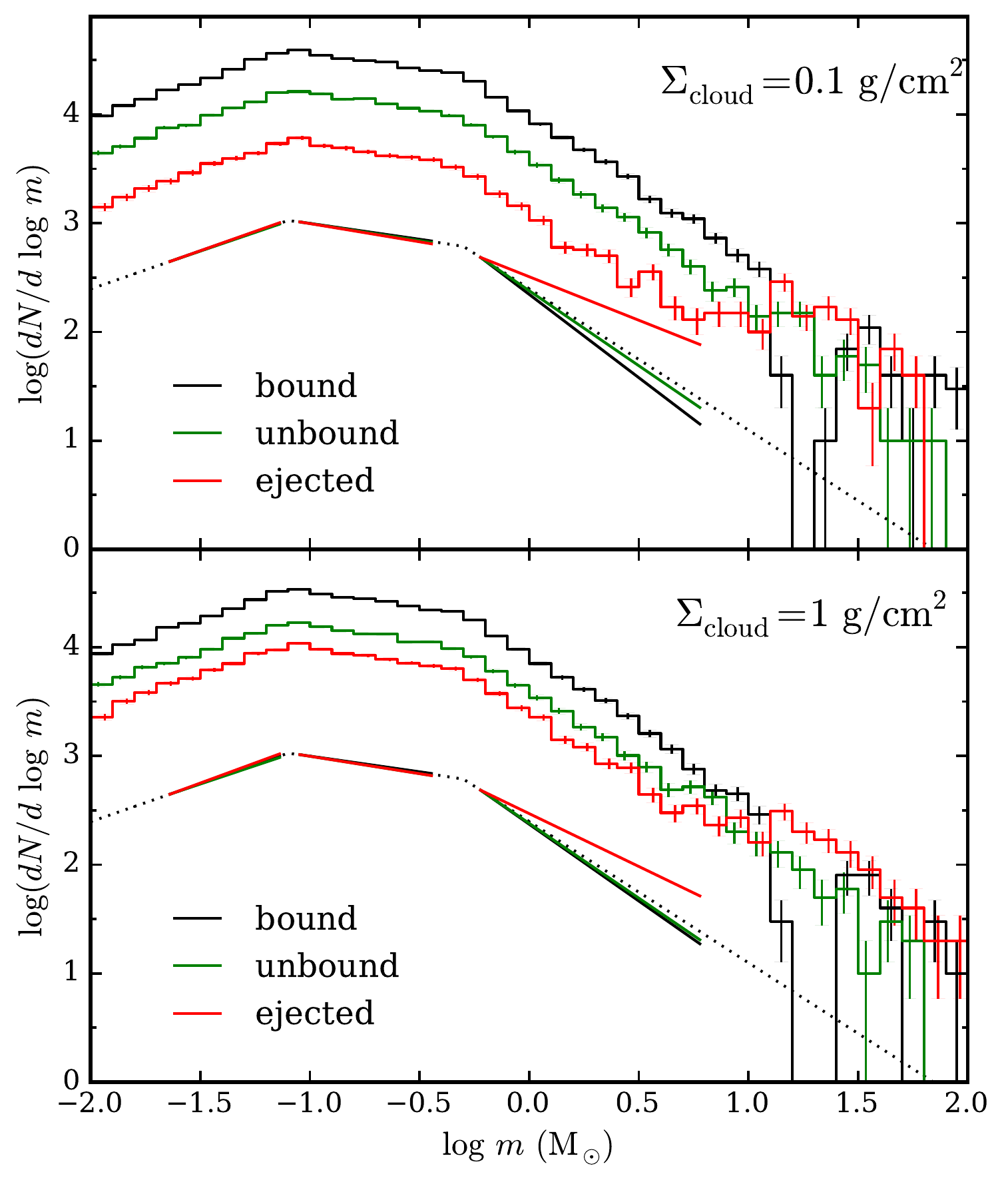} 
        \caption{
Collected IMFs for all the stars in the fiducial case separated in
bound (black steps), unbound (green steps) and ejected (red steps)
stars as in Figure \ref{fig:mv}. Top panel shows simulations with
$\Sigmacl=0.1$ g/cm$^2$ and bottom panel simulations with $\Sigmacl=1$
g/cm$^2$.  Lines below the steps shows the shape of the best fit on
each range of masses for each group of stars over-plotted on the
canonical IMF used as initial condition for comparison. In all groups
the IMF below 0.5 $\Msun$ is the same and do not change
from the original, but the group of ejected stars shows a top heavy
IMF.}
\label{fig:imf}
\end{figure}

Figure \ref{fig:imf} shows the IMF of the three different classes of
stars: bound, unbound and ejected. The IMF of the unbound group (green
histogram) mirrors the assumed primordial distribution; stars from all
masses are initially randomly distributed and are equally likely to be
born unbound. The initially bound cluster shows the same pattern,
however as the cluster evolves and stars are ejected, almost all
massive stars are lost resulting in the black histogram shown in
Figure \ref{fig:imf}. Thus the IMF of the ejected group (red
histogram) shows a clear signature of being top heavy---mostly a
consequence of stellar evolution. Eventually, a large majority of the
stars able to explode as supernovae are ejected from the cluster. When
comparing with other simulation sets (see Figure \ref{fig:imfall}), we
see that even before including stellar evolution the IMF of ejected
stars is already top heavy, which is a result only of dynamical
ejections. However, this effect is not strong enough to significantly
change the shape of the bound cluster IMF. 

Further variations from the set \fiducial do not change these results
significantly, with exception of the \segregated sets. Initial extreme
mass segregation causes a very different evolution in the three groups
of stars. Stars born unbound are preferentially the lowest mass stars,
therefore the initially bound clusters have top heavy IMFs. Later
evolution causes the cluster to lose its massive stars. However, such
extreme mass segregation is a very idealized model that is not
expected to be a very realistic description of observed clusters.

\subsection{Radial structure}\label{sec:rad}

\begin{figure*}
\centering
\includegraphics[width=0.95\textwidth]{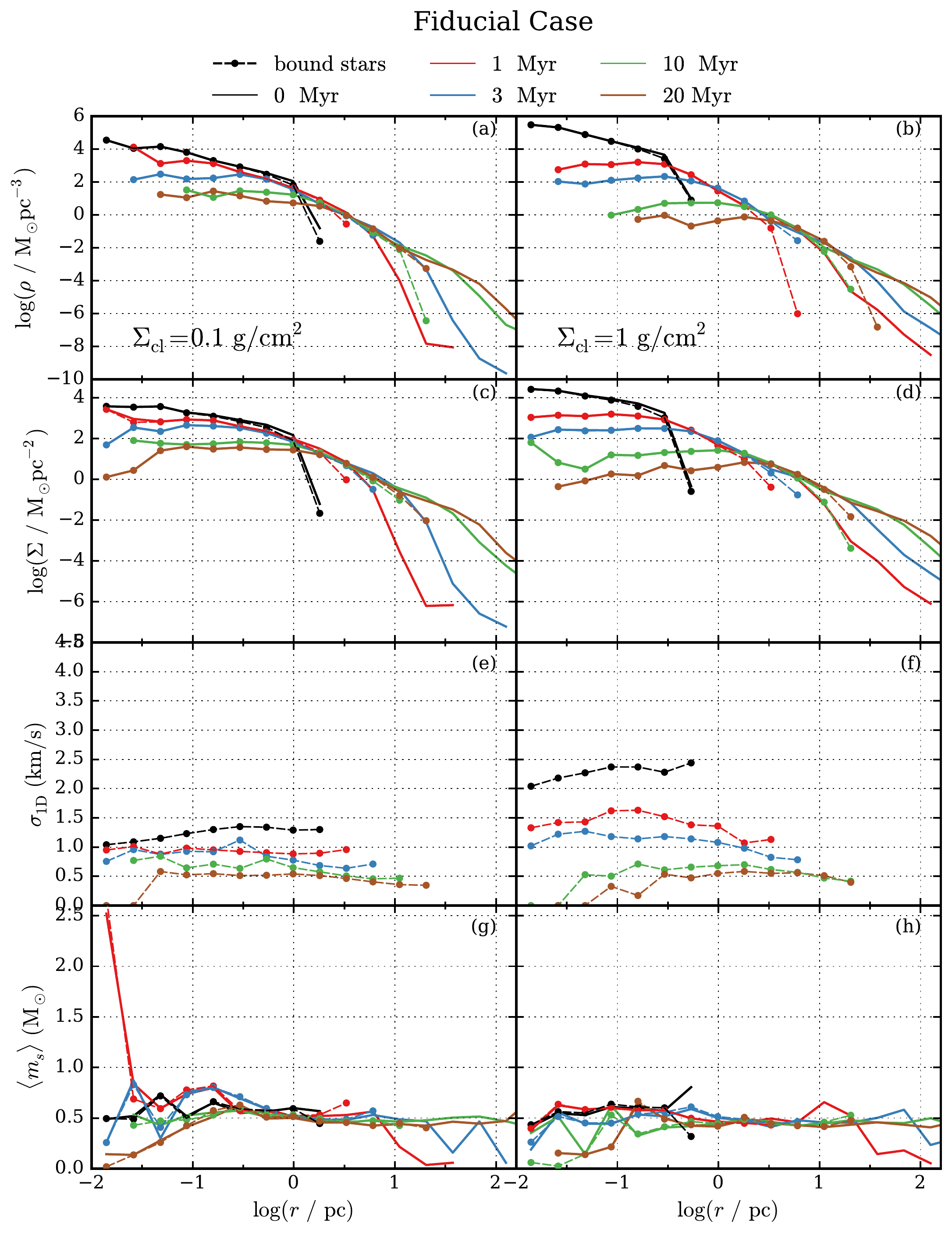}
\caption{
Radial profiles of the fiducial simulations at: 0~Myr (black); 1~Myr
(red); 3~Myr (blue); 10~Myr (green); 20~Myr (brown), with results for
low $\Sigmacl$ shown in the left column and high $\Sigmacl$ in the
right column. Dashed lines and circles show the mean values of the 20
realizations for the bound clusters, while solid lines show these
averages for the total stellar population.
{\it Top row, panels (a) and (b):} stellar volume density radial
profile as a function of spherical radial coordinate from cluster
center. {\it 2nd row, panels (c) and (d):} mass surface density
profile as a function of projected radial coordinated from cluster
center. {\it 3rd row, panels (e) and (f):} 1D velocity dispersion
profiles as a function of projected radius. {\it 4th row, panels (g)
  and (h):} average mass per system profile as a function of projected
radius.
Note, in the 3rd row $\sigma_{\rm 1D}$ for all stars has been omitted
because of the large variations caused by runaway stars.
Note also, binary stars are treated as unresolved systems, i.e., we
use their combined mass, position and velocity to construct each
profile.
}
\label{fig:fiducialrad}
\end{figure*}

We now summarize the evolution of various radial distributions of
stellar properties in the clusters, which, in their projected forms,
are one of the most direct observables of real systems (see Figure
\ref{fig:fiducialrad} for results for the \fiducial set of simulations
and Appendix \ref{ap:1} for the other sets). For the low and high
$\Sigmacl$ cases we show radial profiles for volume density (panels a
and b), projected mass surface density (panels c and d), projected 1D
velocity dispersion $\sigma_{\rm 1D}$ (panels e and f) and projected
mean stellar mass $\langle m_{\rm i} \rangle $ (panels g and h).

These figures show the expansion of the clusters and the flattening of
the initial power law density profile, i.e., development of a constant
density core.  Stars born unbound separate from the cluster as it
evolves and appear as an excess ``halo'' around the bound cluster.

As expected, the velocity dispersion profiles of the bound stars show
a general trend of evolving towards smaller values as the clusters
expand. This can also be seen in Figure~\ref{fig:evol} with a similar
evolution for all the different simulation sets.  
The velocity profiles are relatively flat, but with a modest tendency
to decrease in the outer regions.

The last row of Figure~\ref{fig:fiducialrad} shows the average stellar
mass per system, i.e., masses of binary stars are combined. 
By this metric, we do not see significant signatures of mass
segregation developing in the clusters.
Then at later times, stellar evolution, i.e., wind mass loss and
supernovae, acts to remove massive stars. Stochastic effects due to
IMF sampling are still noticeable, even when averaging over 20
clusters.

The most extreme primordially segregated case we considered (set
\segregated) is able to maintain its mass segregation for much of the
20 Myr evolution (note, stars born unbound are mostly low mass stars),
but it becomes less prominent in the high $\Sigma$ case since massive
stars are ejected more efficiently by dynamical interactions. Then
after 3~Myr, massive star start to be lost due to stellar evolution
(supernova) effects.
\section{Discussion and Conclusions}
\label{sec:discussion}

We have presented a first modeling of the dynamical evolution of star
clusters forming with initial conditions prescribed by the Turbulent
Clump Model of \cite{mt03}.
These initial conditions involve idealized descriptions of
star-forming protocluster clumps as singular polytropic spheres in
virial equilibrium (including effects of large scale magnetic field
support) and pressure equilibrium with a surrounding cloud medium
(i.e., clump radius is set by truncation of the polytropic sphere
where local clump pressure matches ambient cloud pressure, with the
latter assumed to be due to the self-gravitating weight of the
larger-scale cloud of given mass surface density, $\Sigmacl$).
In this first paper we have assumed, for simplicity, that star
clusters are formed instantaneously with a spatially uniform star
formation efficiency. Subsequent papers will build realism to this
model, in particular allowing for the effects of gradual star
formation.

The first consequence of the above assumptions is that stars follow
the same spatial and kinematic distributions of the parent clump,
which sets the main difference between this and previous related
studies in which velocity profiles of the star clusters are
constructed using either isotropic velocity distributions, using
methods described by \cite{Aarseth1974} \citep[e.g.,][]{Goodwin2006}
or assuming the stars are in equilibrium with their natal gas at the
onset of gas expulsion \citep[e.g.,][]{Baumgardt2007}. A system in
equilibrium has a velocity dispersion profile that decreases with
radius. The Turbulent Clump Model, however, involves larger velocity
dispersions at larger scales, so stars born on the outskirts of the
clump are less likely to remain bound to the cluster. Conversely,
these models contain a central, relatively tightly bound central
region.

One crucial parameter that determines the amount of retained mass
after gas is expelled in this model is the contribution of magnetic
fields to the support of the parent clump.  Without magnetic field
support the virial ratios at the onset of gas expulsion are relatively
high even for SFE of 100\%. We can infer from our results that without
magnetic fields 100\% SFE would result in clusters that retain only
$\sim40\%$ of the stars (see Figures \ref{fig:Qsfe} and
\ref{fig:sfe}). However, the models presented here are the worst case
scenario in terms of cluster survivability. It has been shown that
gradual gas expulsion increases the amount of stellar mass retained
\citep[e.g.,][]{Mathieu1983,Baumgardt2007,Smith2013b}. Another factor is the
central star to gas mass ratio at the onset of gas expulsion
\citep[e.g.,][]{Kruijssen2012}. Our models have assumed a spatially
uniform SFE, but the local SFE may be raised either by dynamical
cluster relaxation before gas expulsion \citep{Smith2011,Farias2015}
or by the star formation process itself, which has been argued to be
faster and globally more efficient in the densest regions of the clump
\citep{Kruijssen2012,Kruijssen2012b,Parmentier2013}.

The fiducial clusters that we have simulated show rapid expansion,
even just considering their bound members. For example, the half-mass
radii show dramatic (factors of several) expansion after 1~Myr,
especially in the high $\Sigmacl=1\:{\rm g\:cm}^{-2}$ case. Thus for
these models to explain observed young star clusters of a given age,
mass and mass surface density \citep[see, e.g., Fig. 1 of ][]{Tan2014},
would require initial clumps that have mass surface densities at least
ten times greater. There is limited evidence for such dense starless
clumps (\citealt{Tan2014}; see also \citealt{Walker2014}).
This may indicate that some aspect of the model needs to be modified,
such as the assumption of instantaneous star formation and gas
expulsion.

Another feature of our work has been the full treatment of binaries,
given assumed primordial binary properties. The processing of these
binaries during the early phases of the dynamical evolution of star
clusters can be a diagnostic of the process, i.e., by comparing their
properties with observed field star and embedded cluster binary
properties. We have also seen that binaries affect some aspects of the
dynamical evolution of the cluster, in particular by enhancing the
rate of dynamical ejection of stars. We consider that our main results
on this topic so far are to show the relative importance of binaries
in our model clusters depending on their input assumptions. In the
clusters that we have investigated in this paper there has been
relatively little processing of the average initial binary properties
given the fairly rapid expansion of the clusters from their initial
dynamical states. These results provide a baseline for comparison of
future models of gradual star cluster formation.

Since we have a full treatment of binaries and stellar evolution, we
are also able to make predictions for the properties of ejected stars,
including via dynamical ejection from unstable triples and higher
order multiples and from supernova explosions. The models presented
here are only able to reproduce high velocity runaway stars by
supernovae kicks. Only the densest initial conditions lead to
dynamical ejection of stars at speeds $>100\:{\rm km\:s}^{-1}$ (but
less than one per cluster) and we have not obtained dynamical ejection
of any massive stars at speeds $>20\:{\rm km\:s}^{-1}$. This probably
again indicates a need for more gradual models of star formation that
retain a dense cluster core for a larger number of crossing times.

In the context of our presented models that apply in the limit of
fast, i.e., ``instantaneous,'' star cluster formation, our main
conclusions are:

\begin{enumerate}
\item 
Magnetic fields that partially support the parent clump against
collapse are a key factor that determines the initial velocity and
bound mass of the new born cluster. Star clusters born from clumps
with no magnetic field support would need very high SFE ($>80\%$) to
retain a significant part of the stellar mass after gas is expelled.

\item 
Regardless of the different initial surface densities, the mass of the
bound cluster that emerges from its natal gas is determined by the
SFE. However, the expansion rate of the unbound stellar population is
determined by the typical internal velocity of the parent clump. This
means that clusters born in high mass surface density environments
will produce unbound populations that expand more rapidly
than those produced from clusters formed in low mass surface density
environments. 

\item
The rapid early expansion of the bound clusters imply that at least
$\sim10\times$ higher mass surface density starless clumps are needed
to explain observed young ($\sim1$~Myr old) clusters that have stellar
mass surface densities $\gtrsim0.1\:{\rm g\:cm^{-2}}$, but there is
limited evidence for such clumps. This may indicate a need for more
gradual models of star cluster formation.

\item 
The low interaction rates of binaries in our simulated clusters lead
to only minor modifications of average initial binary properties. If
star cluster formation is rapid, then the present day observed binary
properties have not changed much from their primordial distributions,
except for those effects induced by stellar evolution.

\item
Based on velocity-position diagrams, it is possible to distinguish the
bound cluster from the unbound stars and we expect that this is made
easier if gas is expelled quickly. Such diagrams also give us
information on the initial properties of the parent clump, such as the
initial escape velocity from the cluster.

\item 
High velocity runaway stars in the clusters simulated so far are
mostly due to supernova explosions that disrupt binaries. Such
statistics likely indicate, again, that star cluster formation does
not proceed in such a rapid manner.\\
\end{enumerate}

It remains to be determined how the above metrics will vary for models
of star clusters forming from turbulent clumps that involve gradual
formation of stars and gradual gas expulsion. In addition, correlated
spatial and kinematic substructures associated with turbulence and
potential infalling accretion flows need to be incorporated into these
models. Variations in global geometry, e.g., elongation of the clump
into more filamentary configurations, also need to be explored. A
wider range of initial conditions of clump properties and assumed star
formation efficiencies, primordial mass segregation and primordial
binary properties need to be investigated for this model family. By an
eventual comparison of model results with observed kinematic
properties of young clusters, we hope to constrain star and star
cluster formation theories.

\acknowledgements
We thank Nicola Da Rio and Antonio Ordo\~nez for
helpful discussions. We acknowledge support from Hubble Space Telescope Cycle
23 Theory Grant 14317, {\it The Orion Experiment} (PI: Tan).


\newpage
\appendix
\section{Ancillary results for the full set of simulations}
\label{ap:1}

\begin{figure*}
        $\begin{array}{rl}
        \includegraphics[width=0.49\textwidth]{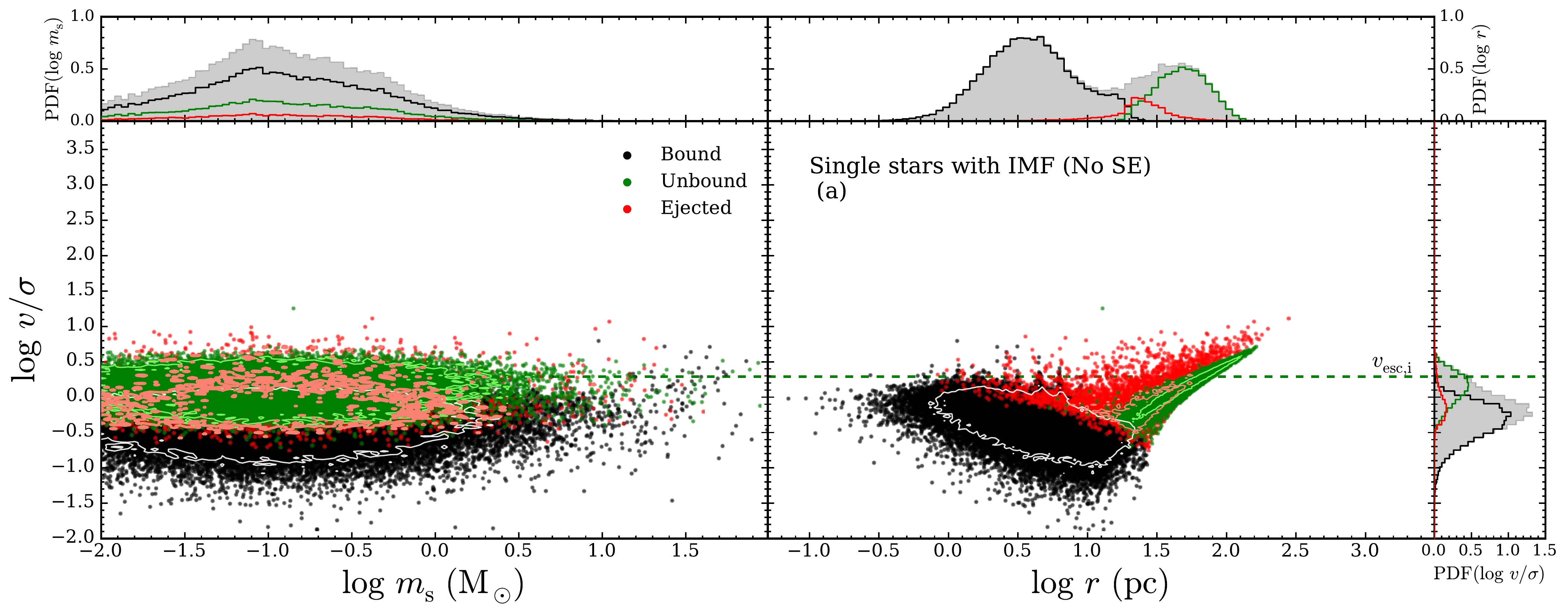} &
        \includegraphics[width=0.49\textwidth]{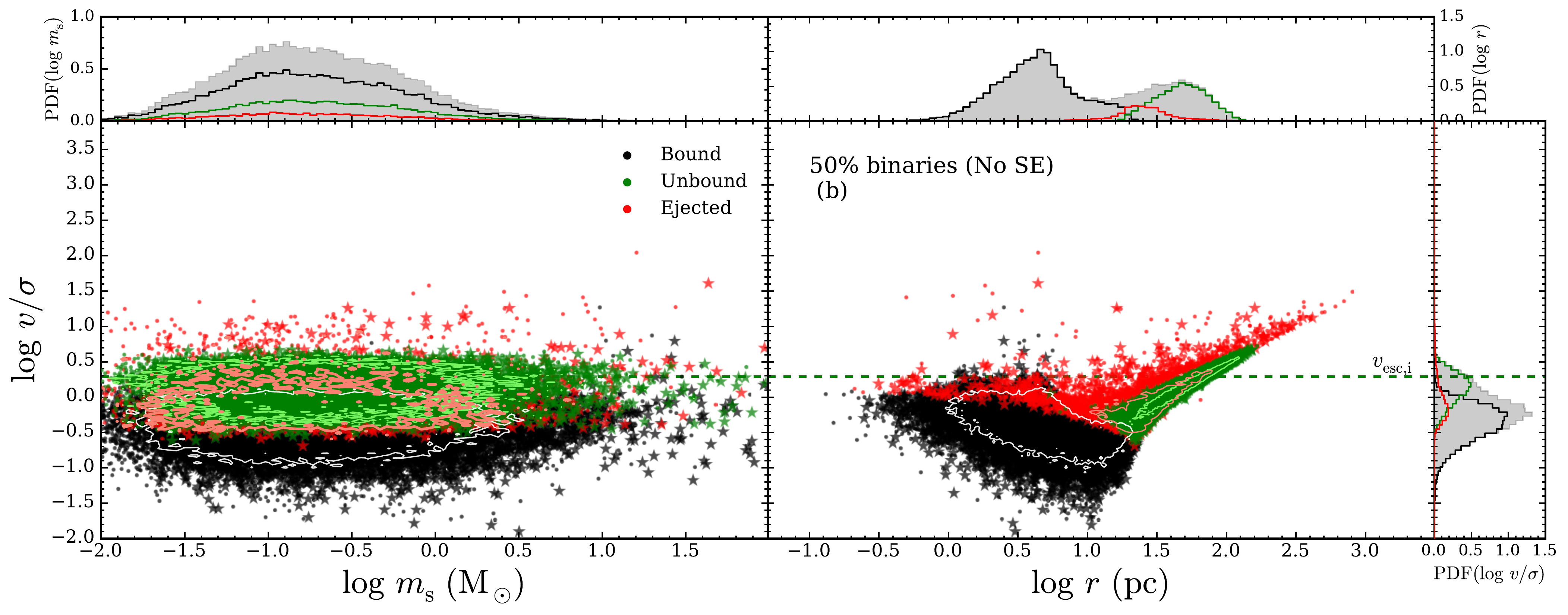} \\
        \includegraphics[width=0.49\textwidth]{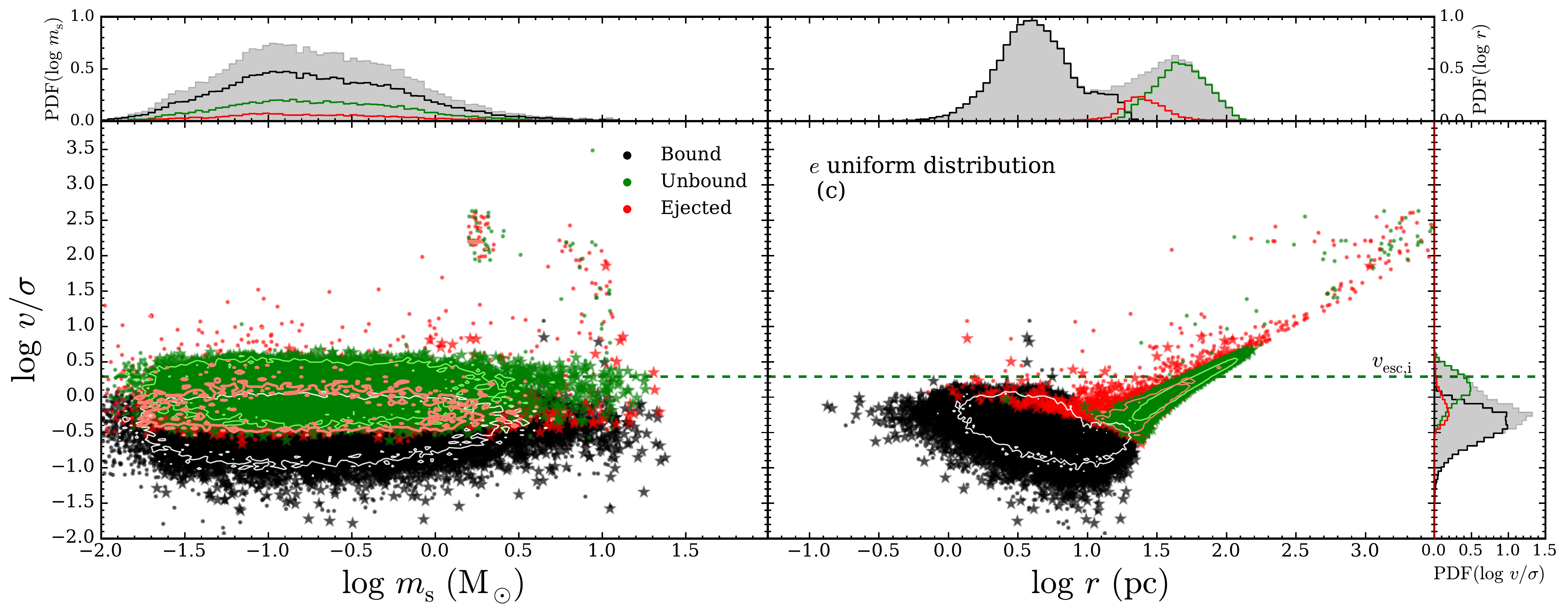} &
        \includegraphics[width=0.49\textwidth]{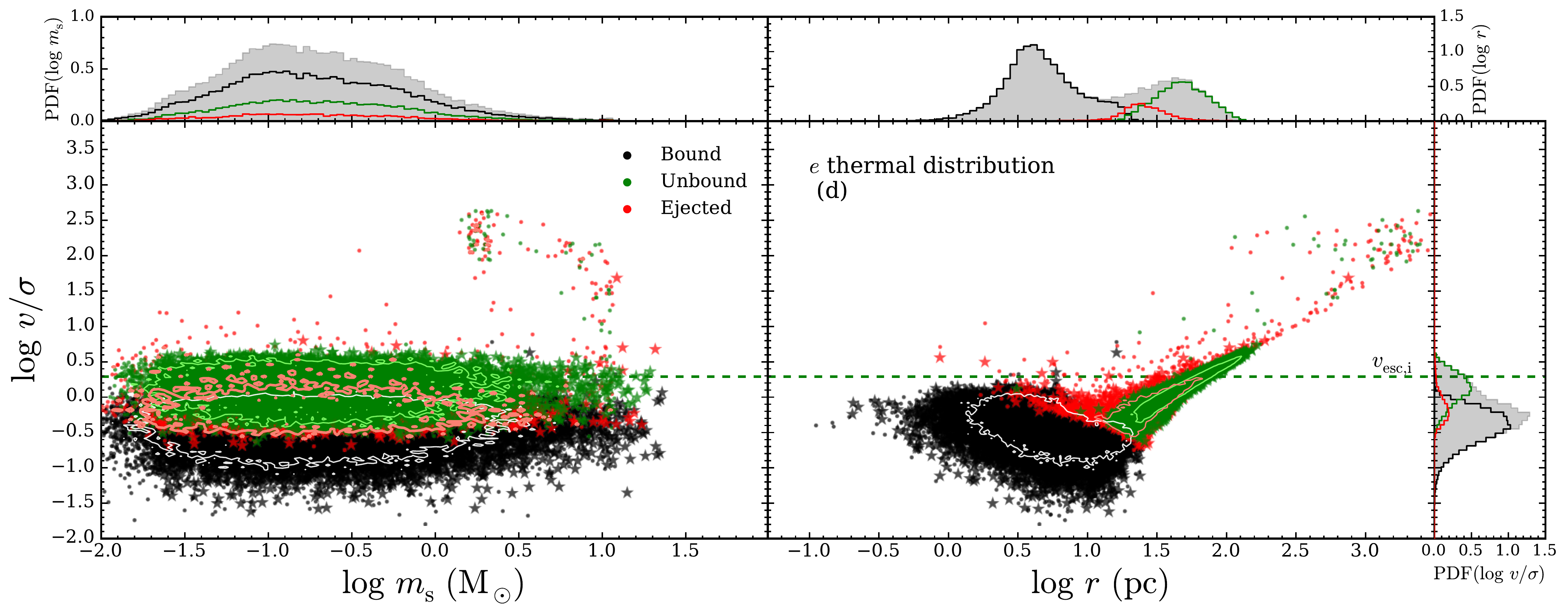} \\
        \includegraphics[width=0.49\textwidth]{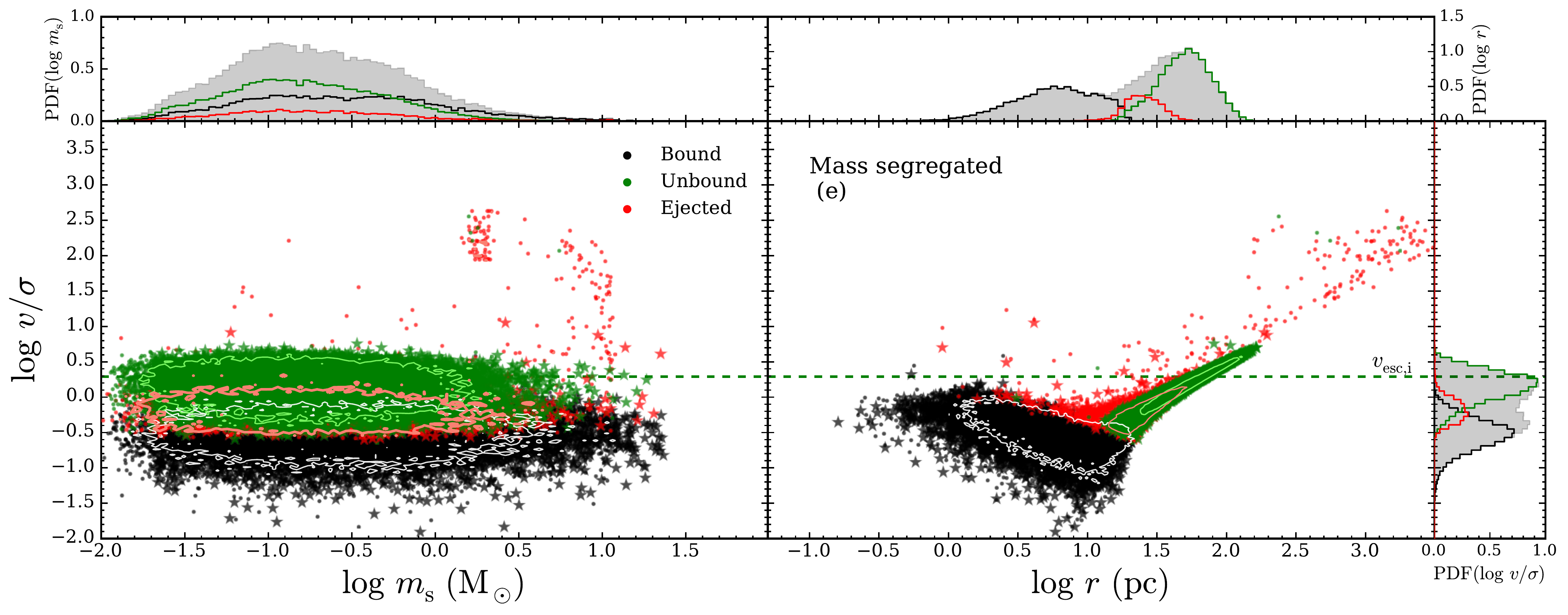} &
        \includegraphics[width=0.49\textwidth]{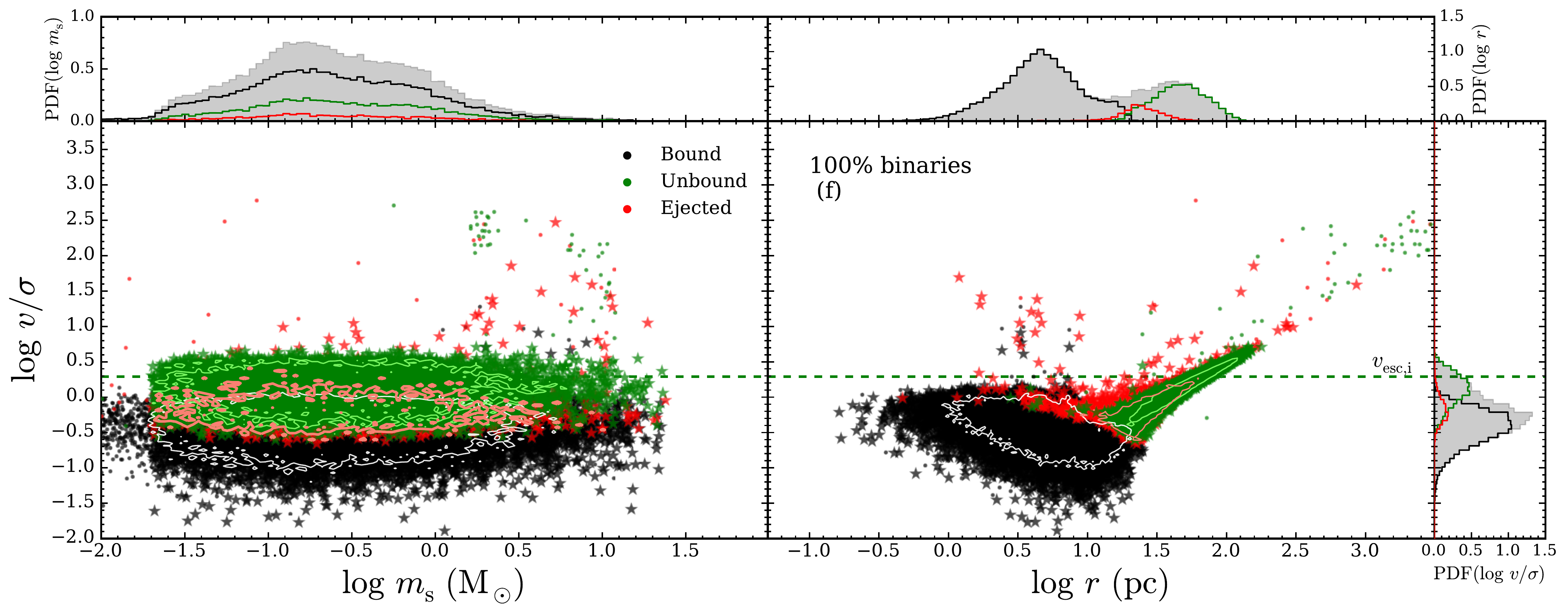} 
        \end{array}$
        \caption{
Same as Figure \ref{fig:mv}, but now for simulations sets: (a)
\nobinaries, (b) \nose, (c) \binariesun, (d) \binariesth, (e)
\segregated and (f) \fullbinaries, all with $\Sigmacl=0.1$ g/cm$^2$.}
        \label{fig:mvall}
\end{figure*}

\begin{figure*}
        $\begin{array}{cc}
        \includegraphics[width=0.49\textwidth]{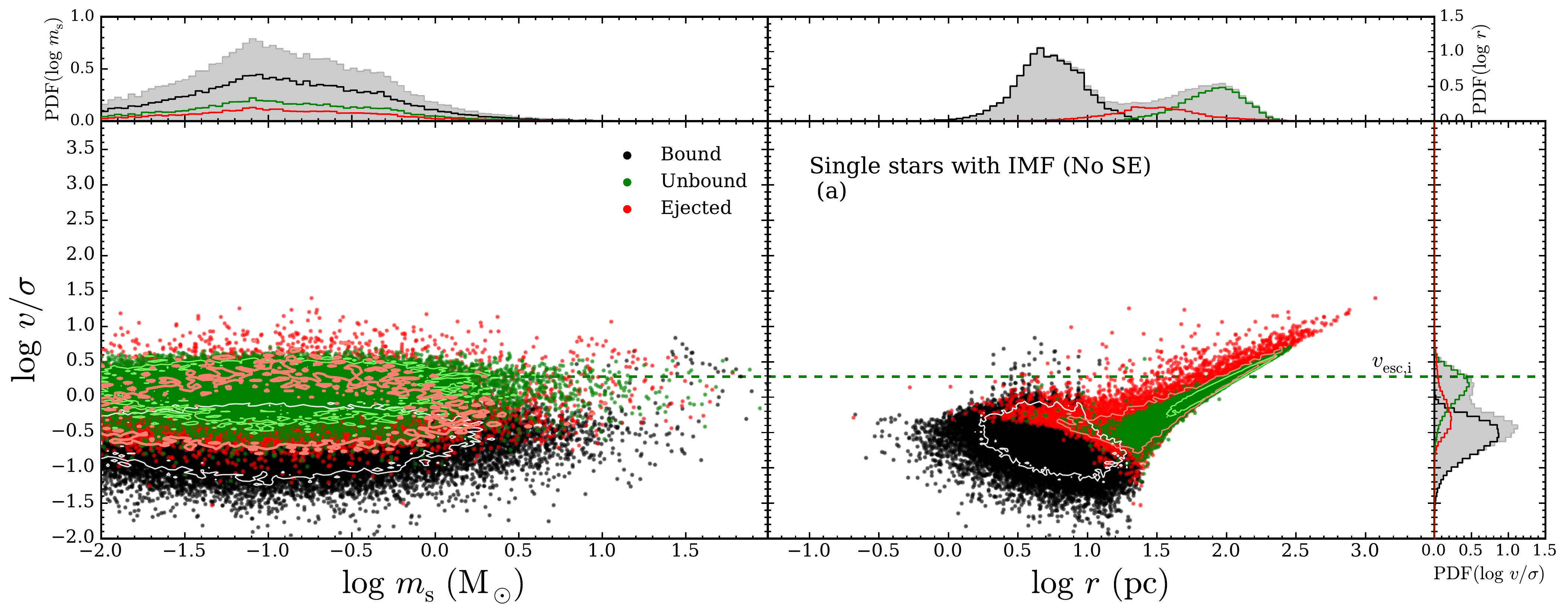} &
        \includegraphics[width=0.49\textwidth]{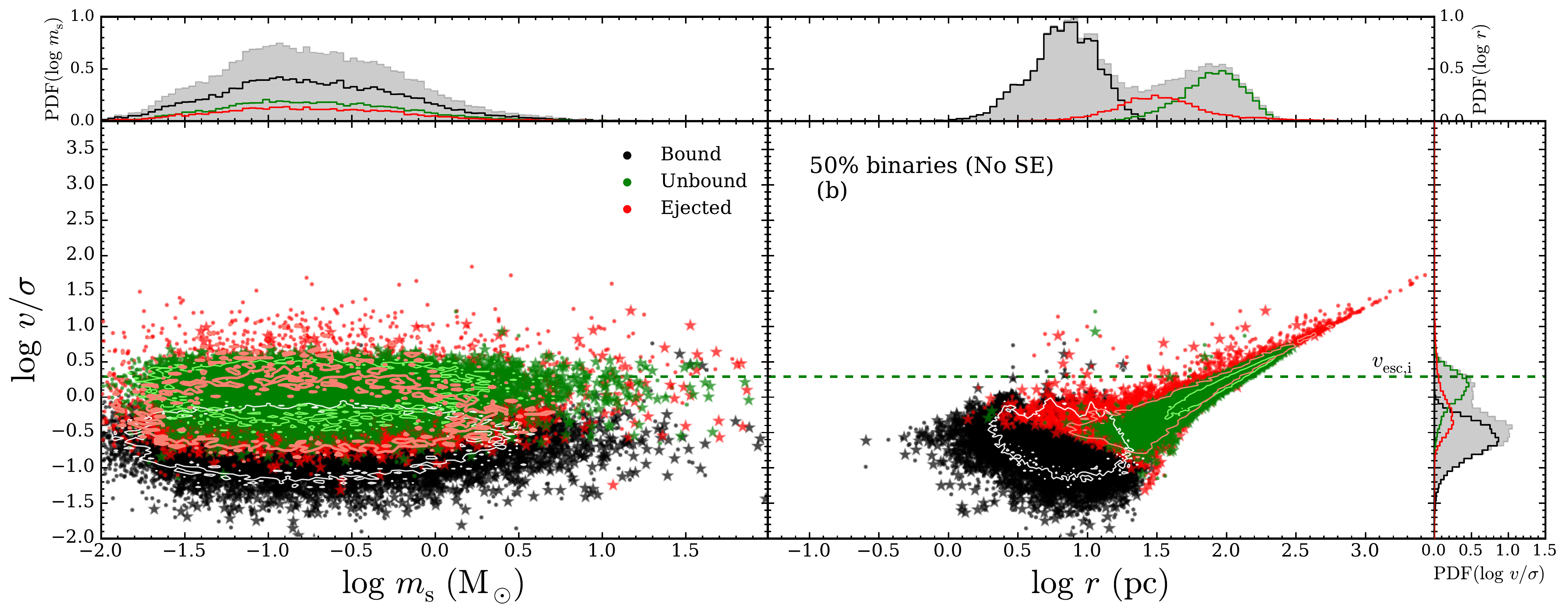} \\
        \includegraphics[width=0.49\textwidth]{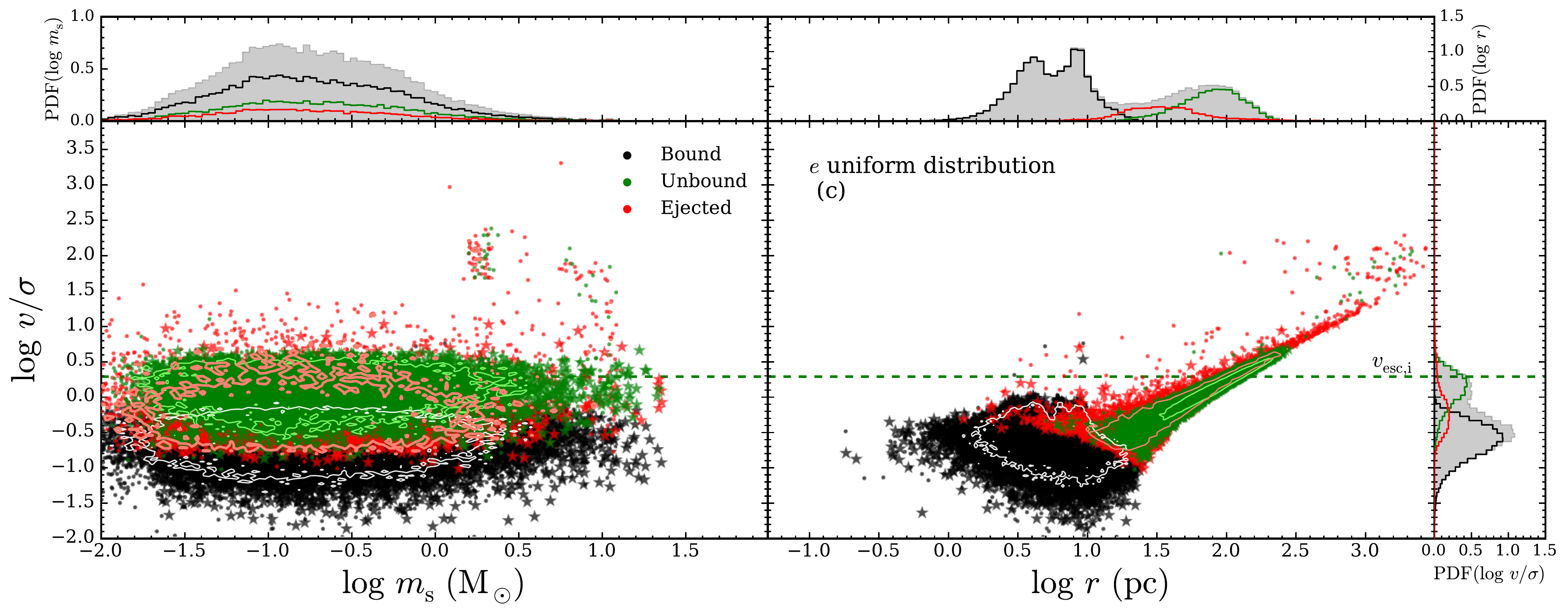} &
        \includegraphics[width=0.49\textwidth]{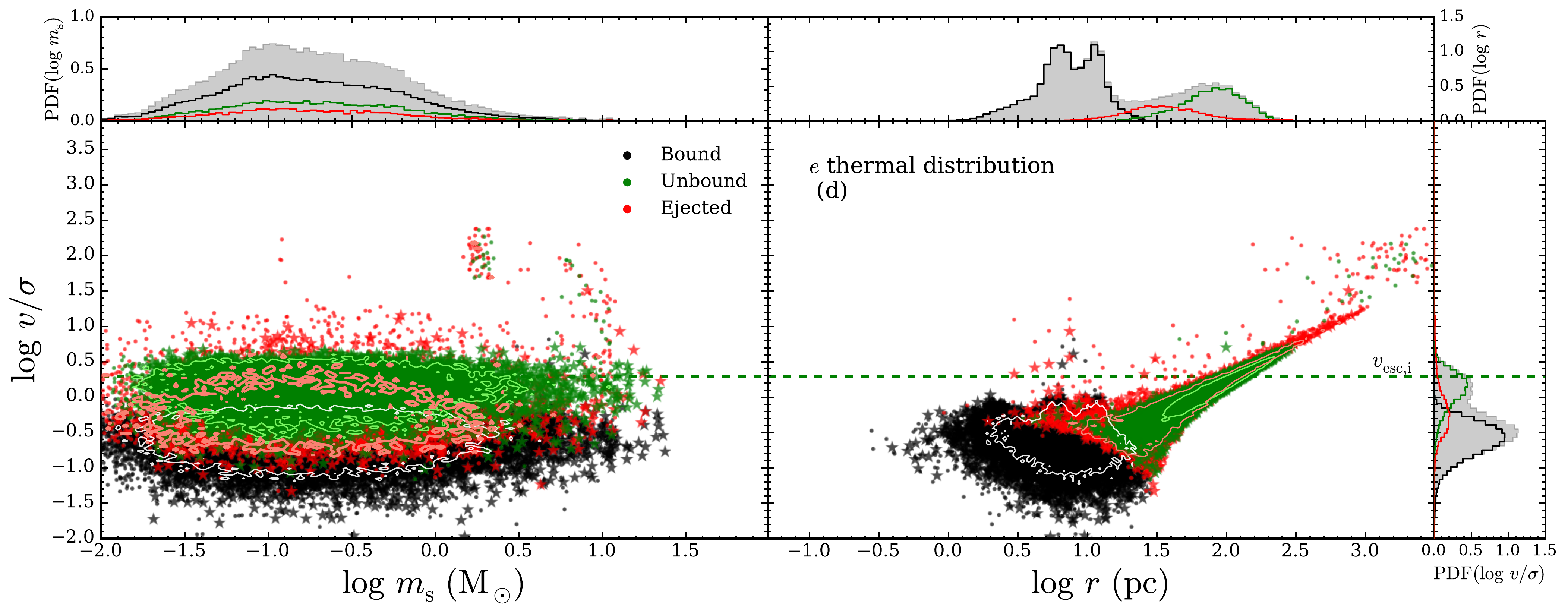} \\
        \includegraphics[width=0.49\textwidth]{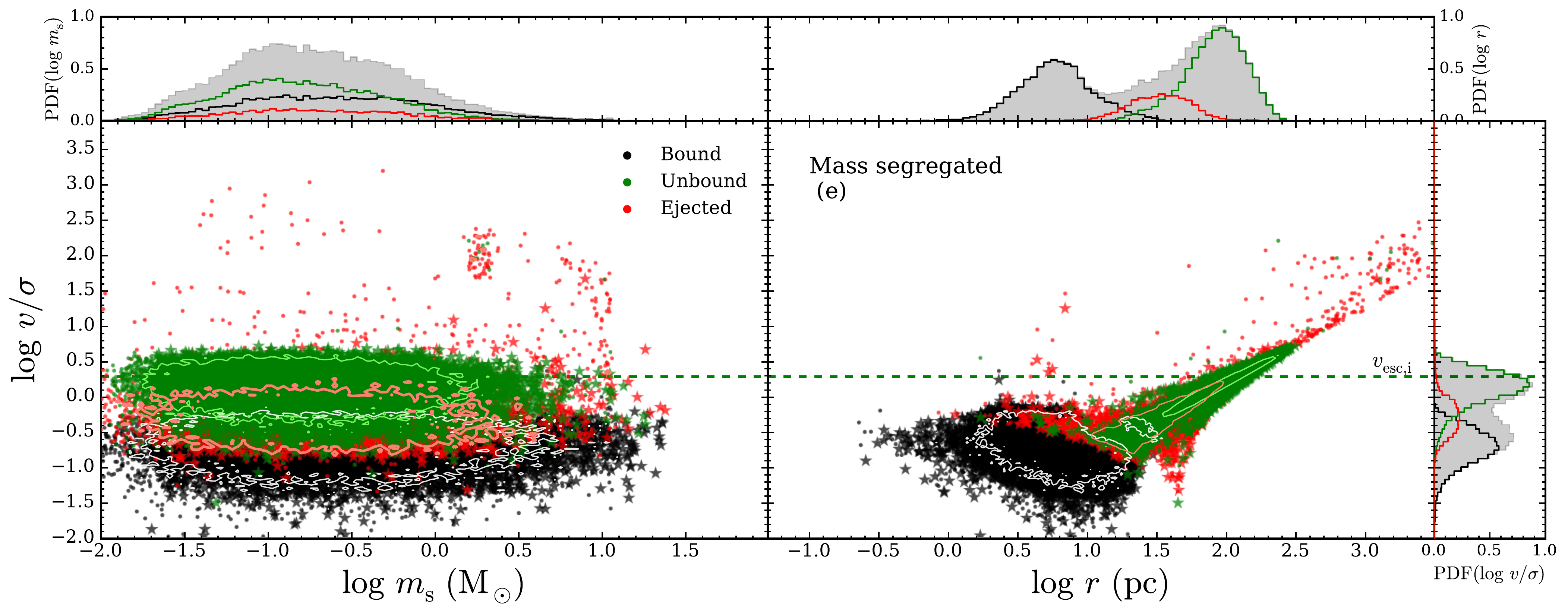} &
        \includegraphics[width=0.49\textwidth]{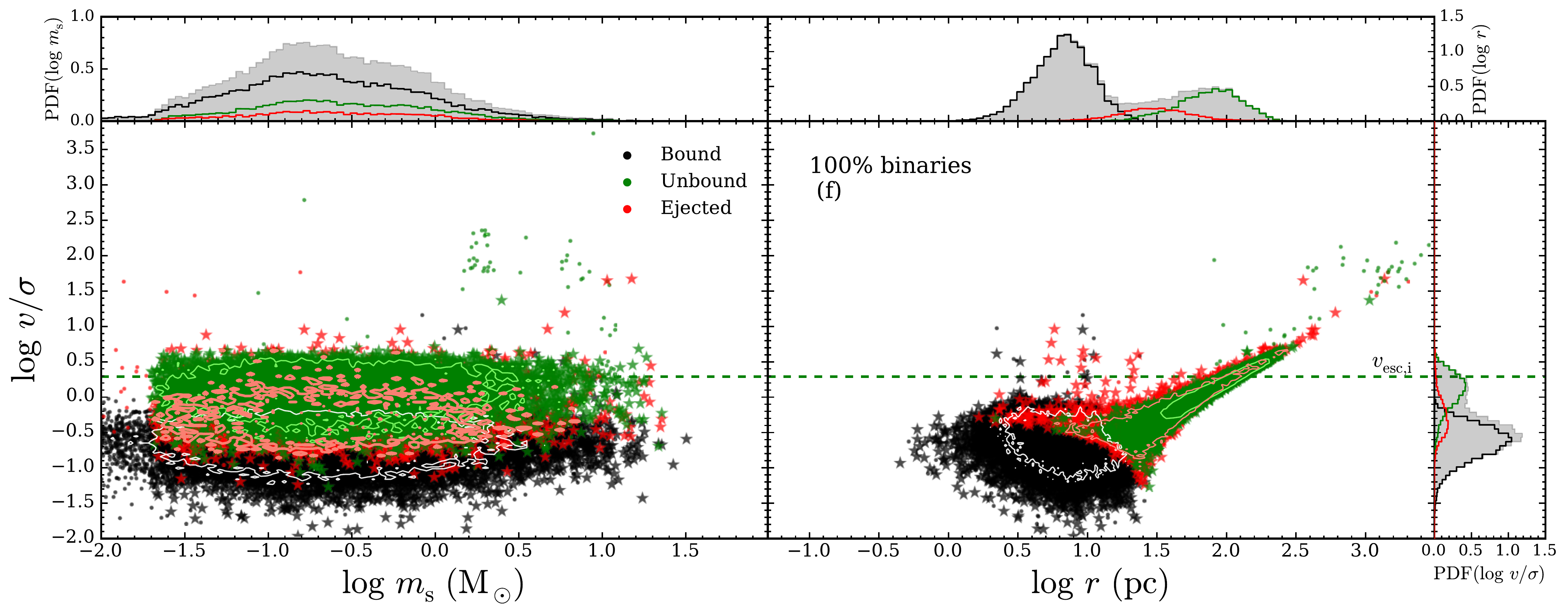} 
        \end{array}$
        \caption{
Same as Figure \ref{fig:mvall}, but now for simulations with
$\Sigmacl=1$ g/cm$^2$.}
        \label{fig:mvallss1}
\end{figure*}

\begin{figure*}
        $\begin{array}{rl}
        \includegraphics[width=0.49\textwidth]{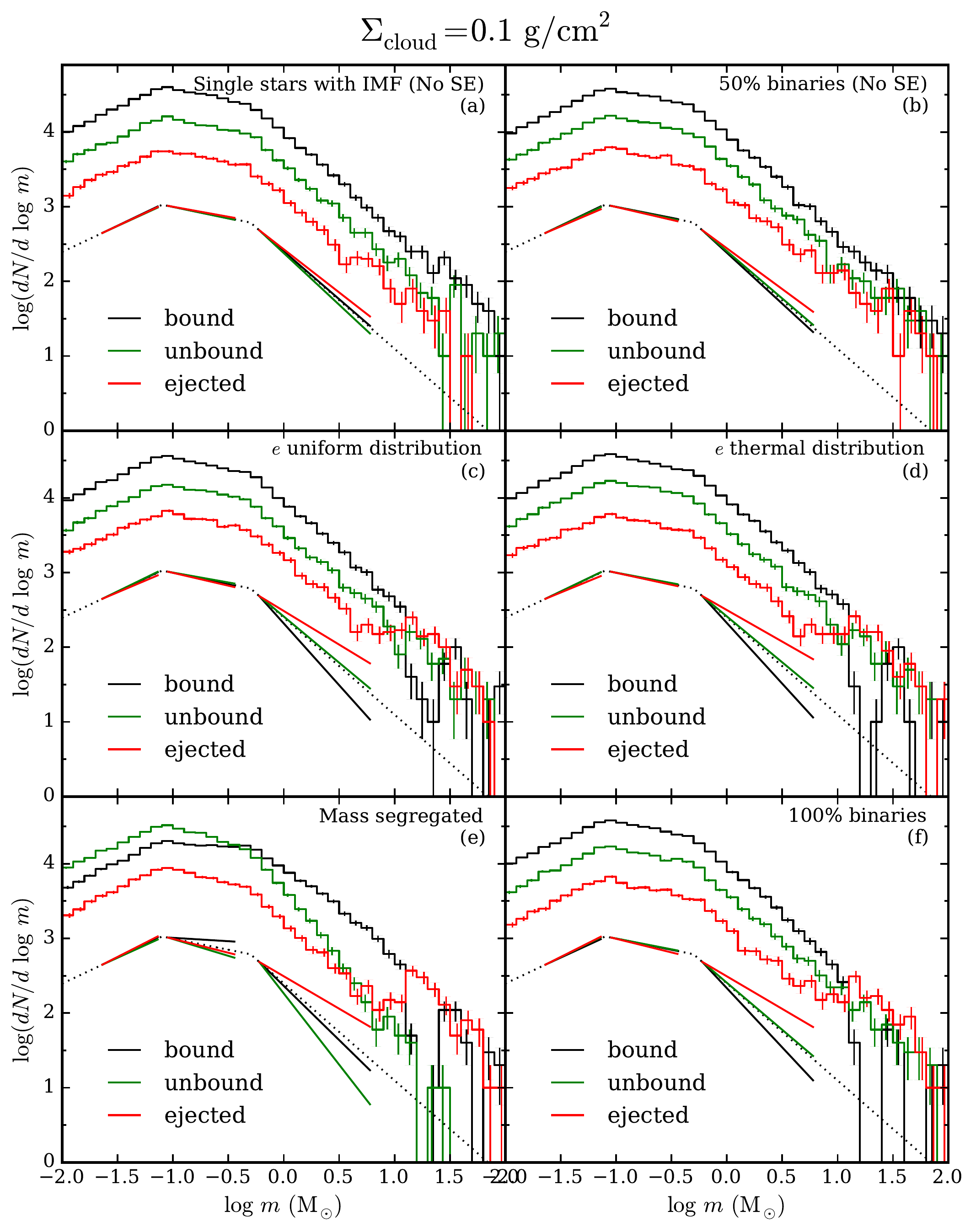} &
        \includegraphics[width=0.49\textwidth]{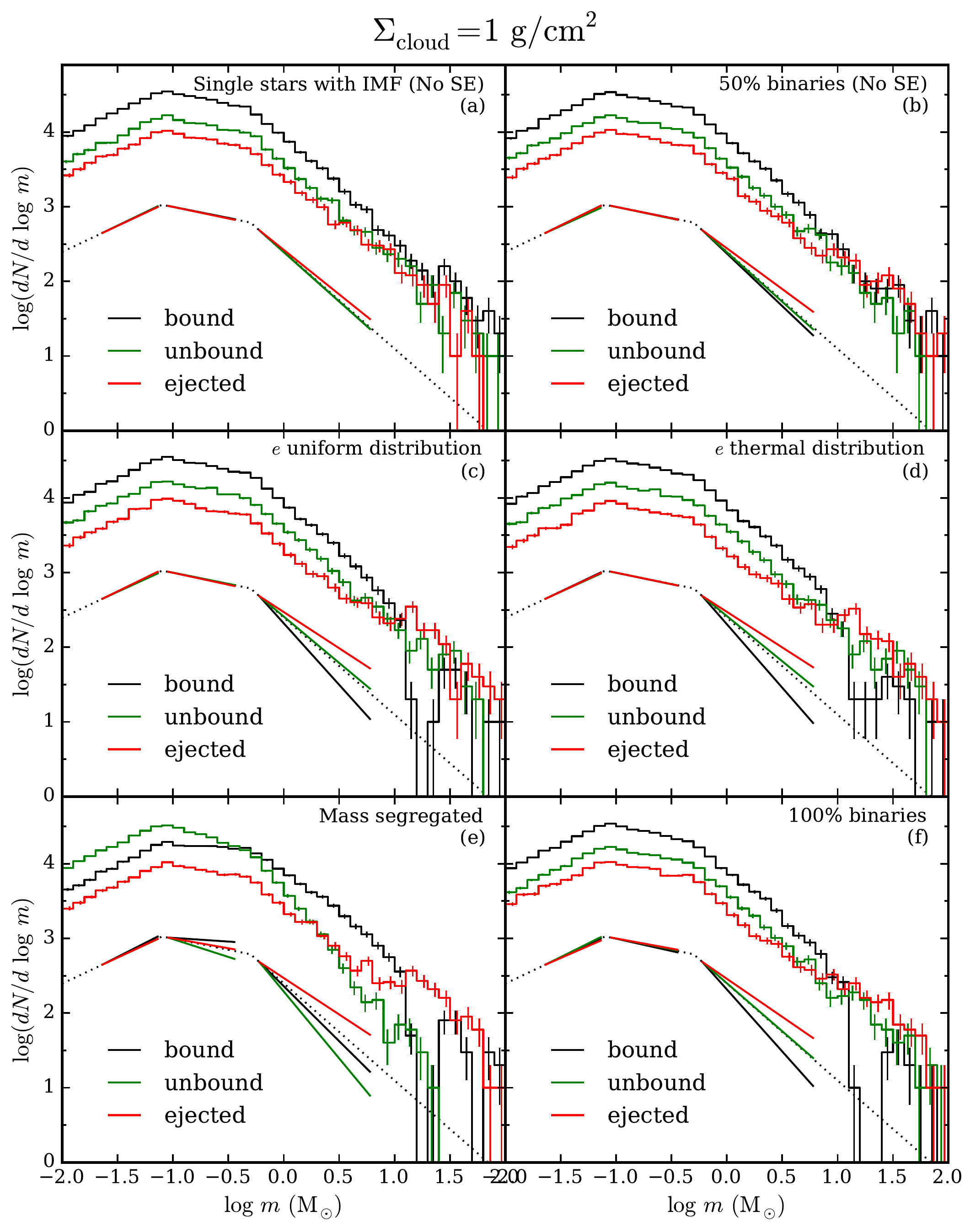}\\
        \end{array}$
        \caption{
Same as Figure \ref{fig:imf}, but now for simulations with $\Sigmacl =
0.1$ g/cm$^2$ (left set of panels) and with $\Sigmacl = 1$ g/cm$^2$
(right set of panels). Labelled panels show the IMFs for the
simulation sets: (a) \nobinaries, (b) \nose, (c) \binariesun, (d)
\binariesth, (e) \segregated~and (f) \fullbinaries.}
        \label{fig:imfall}
\end{figure*}

\begin{figure*}
        $\begin{array}{c}
        \includegraphics[width=0.49\textwidth]{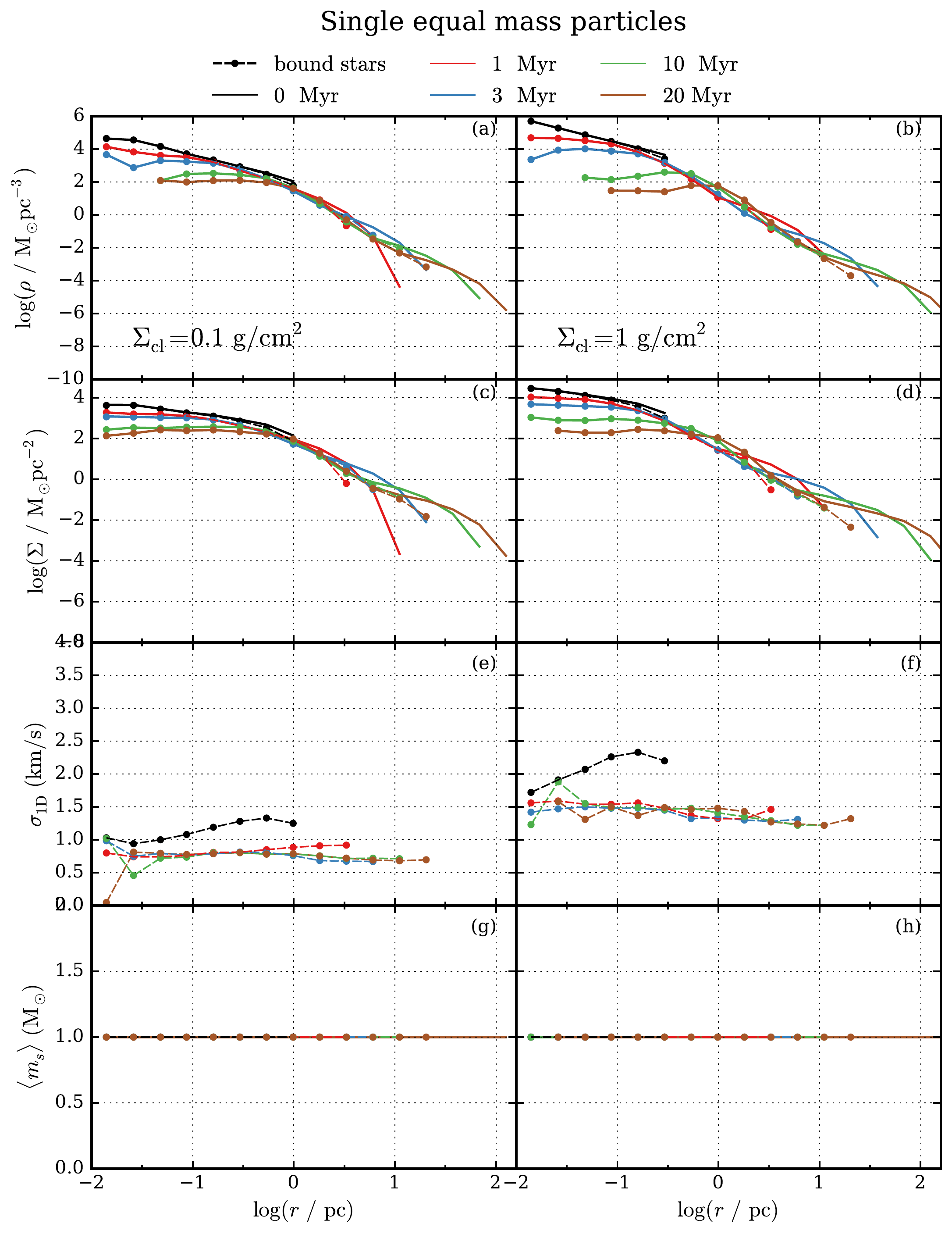} \\
        \begin{array}{rl}
        \includegraphics[width=0.49\textwidth]{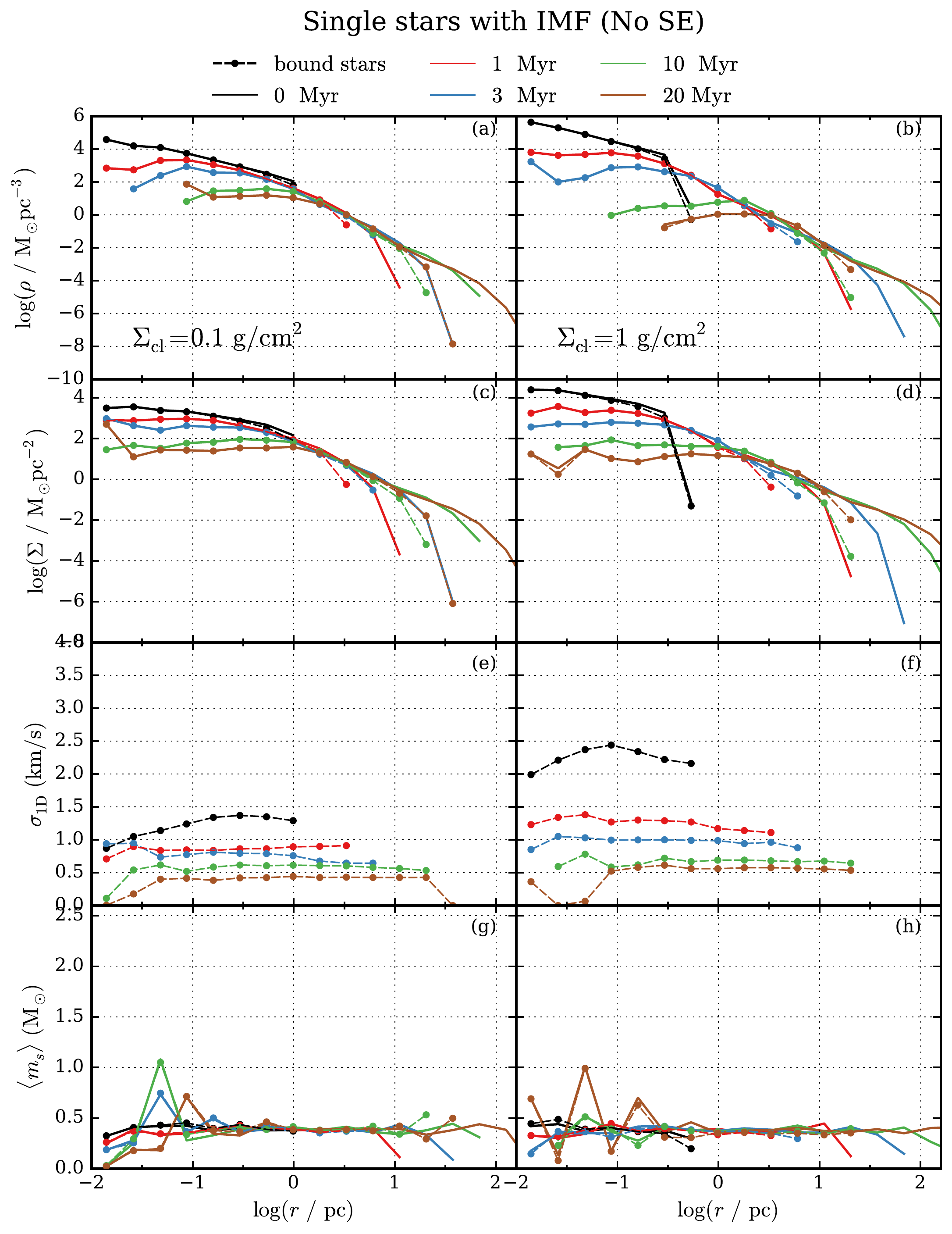} &
        \includegraphics[width=0.49\textwidth]{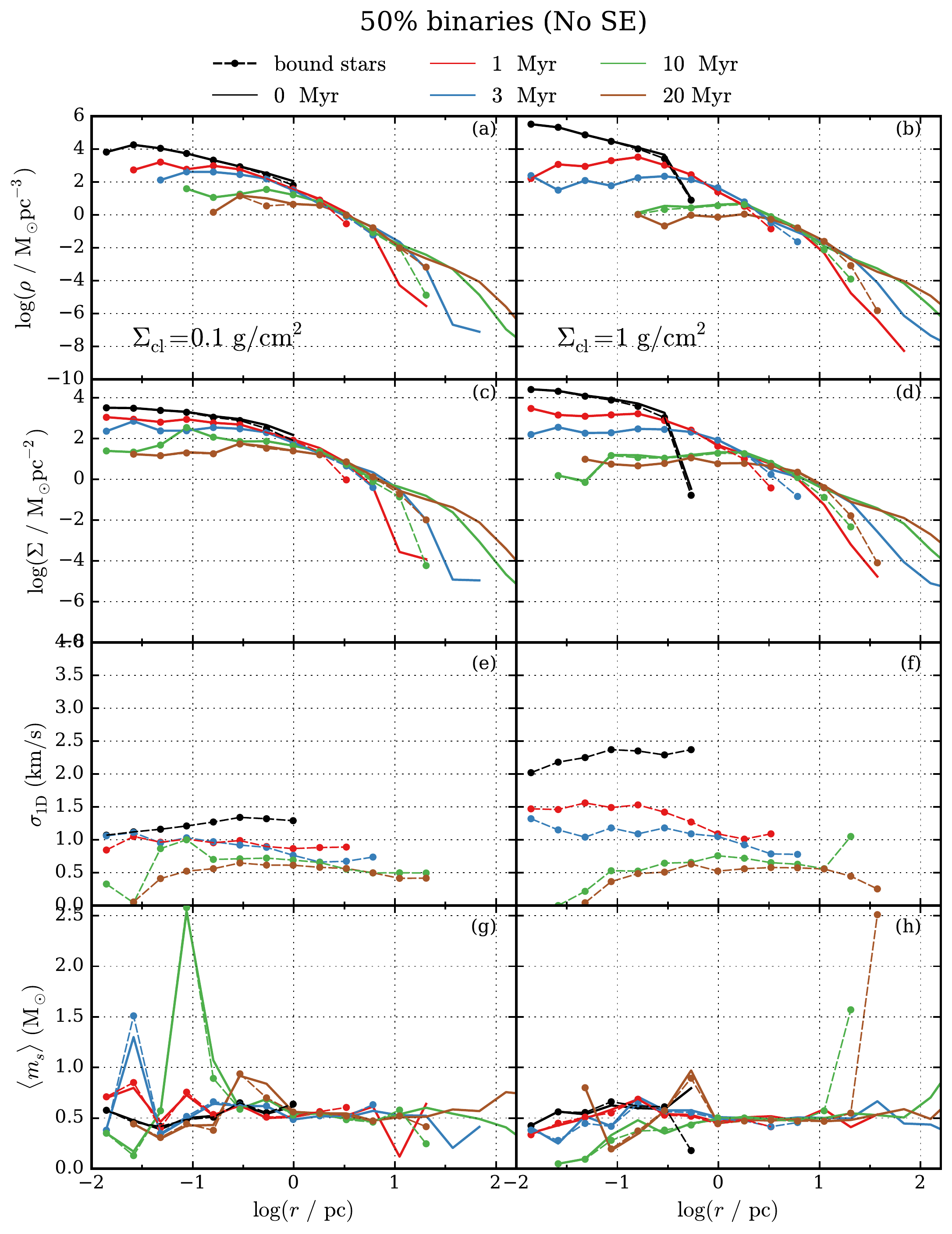}\\
        \end{array}\end{array}$
        \caption{
Same as Figure \ref{fig:fiducialrad}, but now showing radial profiles
for sets \equalmass (top) \nobinaries (bottom left) and \nose (bottom
right), all with no stellar evolution.}
        \label{fig:noserad}
\end{figure*}

\begin{figure*}
        $\begin{array}{rl}
        \includegraphics[width=0.49\textwidth]{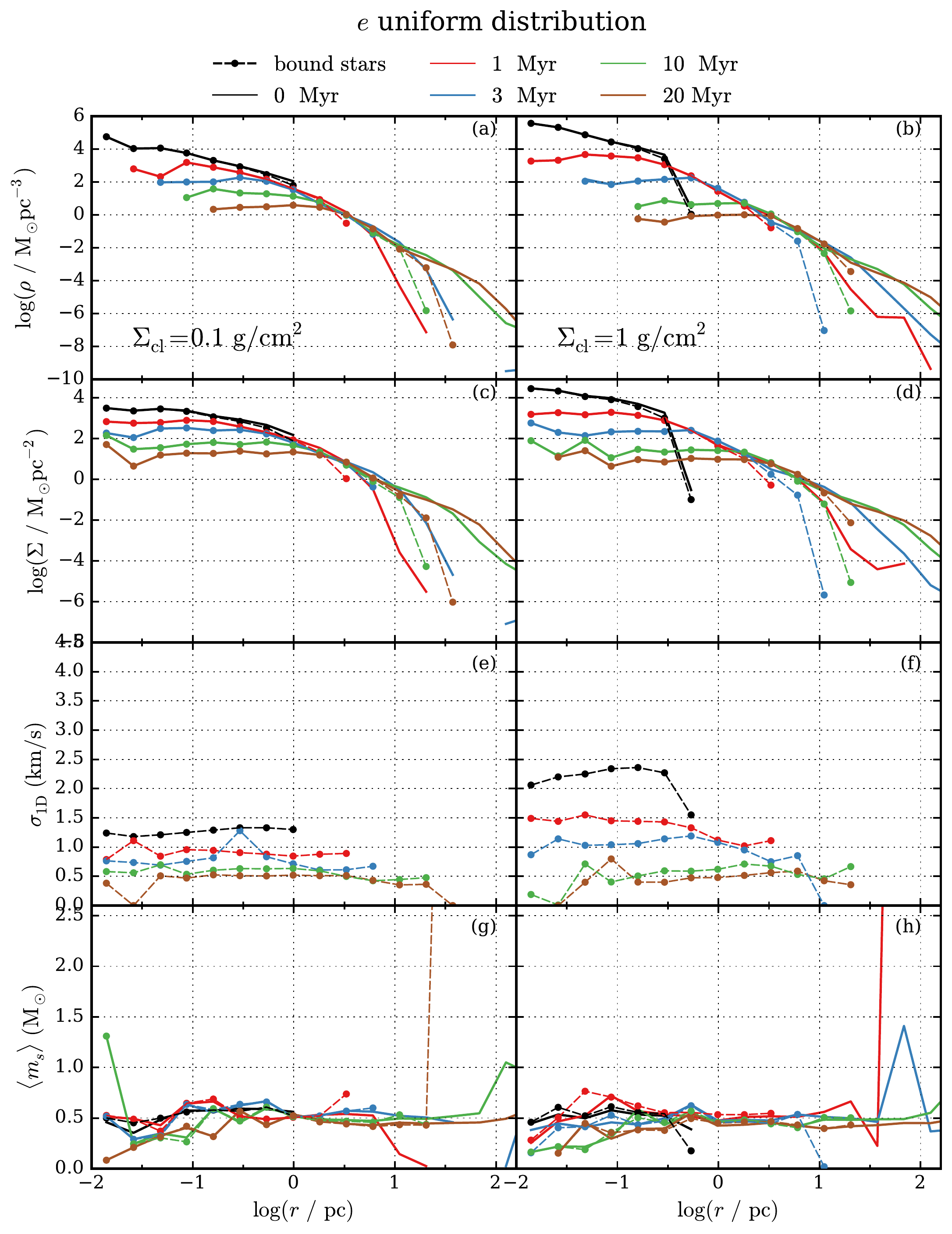} &
        \includegraphics[width=0.49\textwidth]{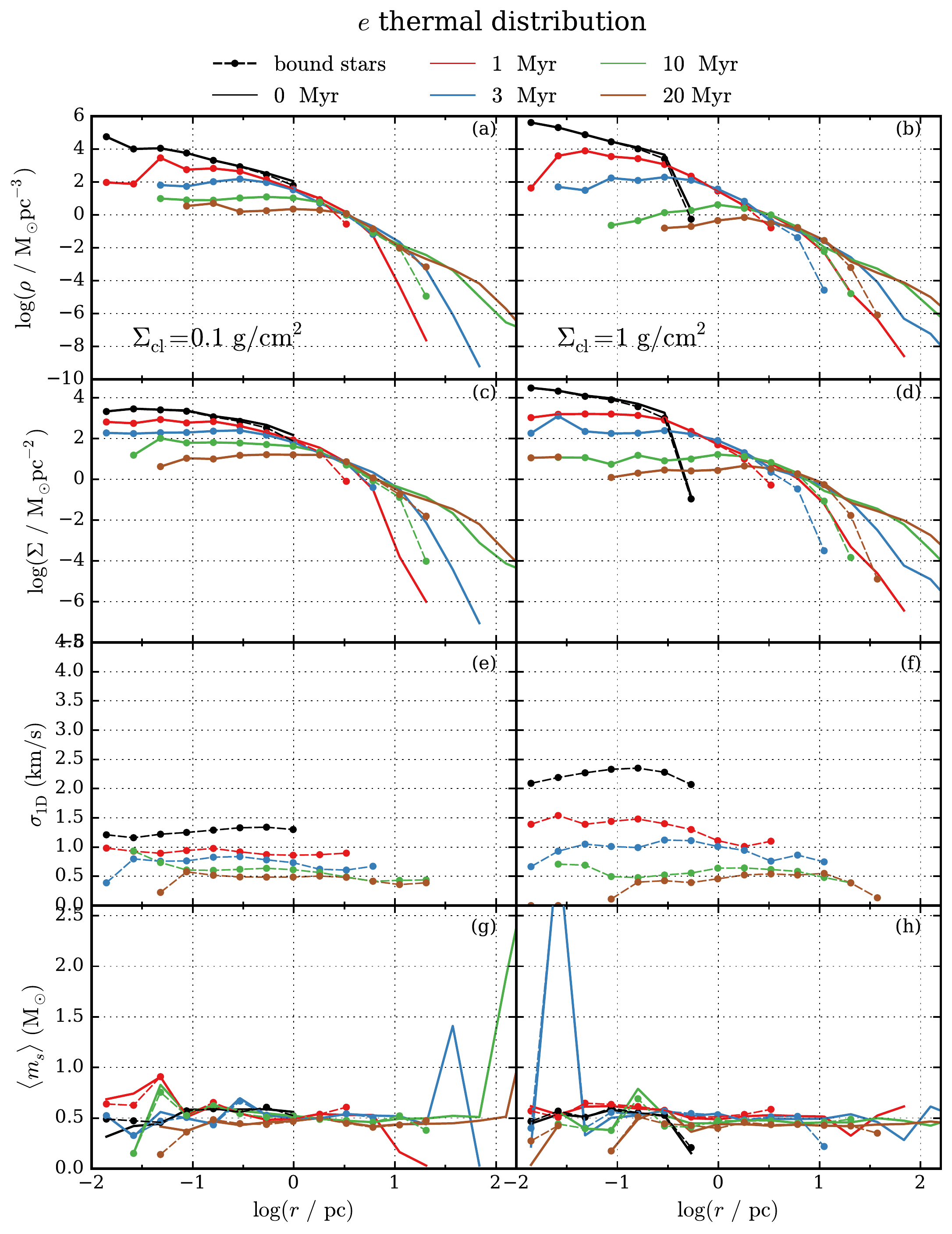} \\
        \includegraphics[width=0.49\textwidth]{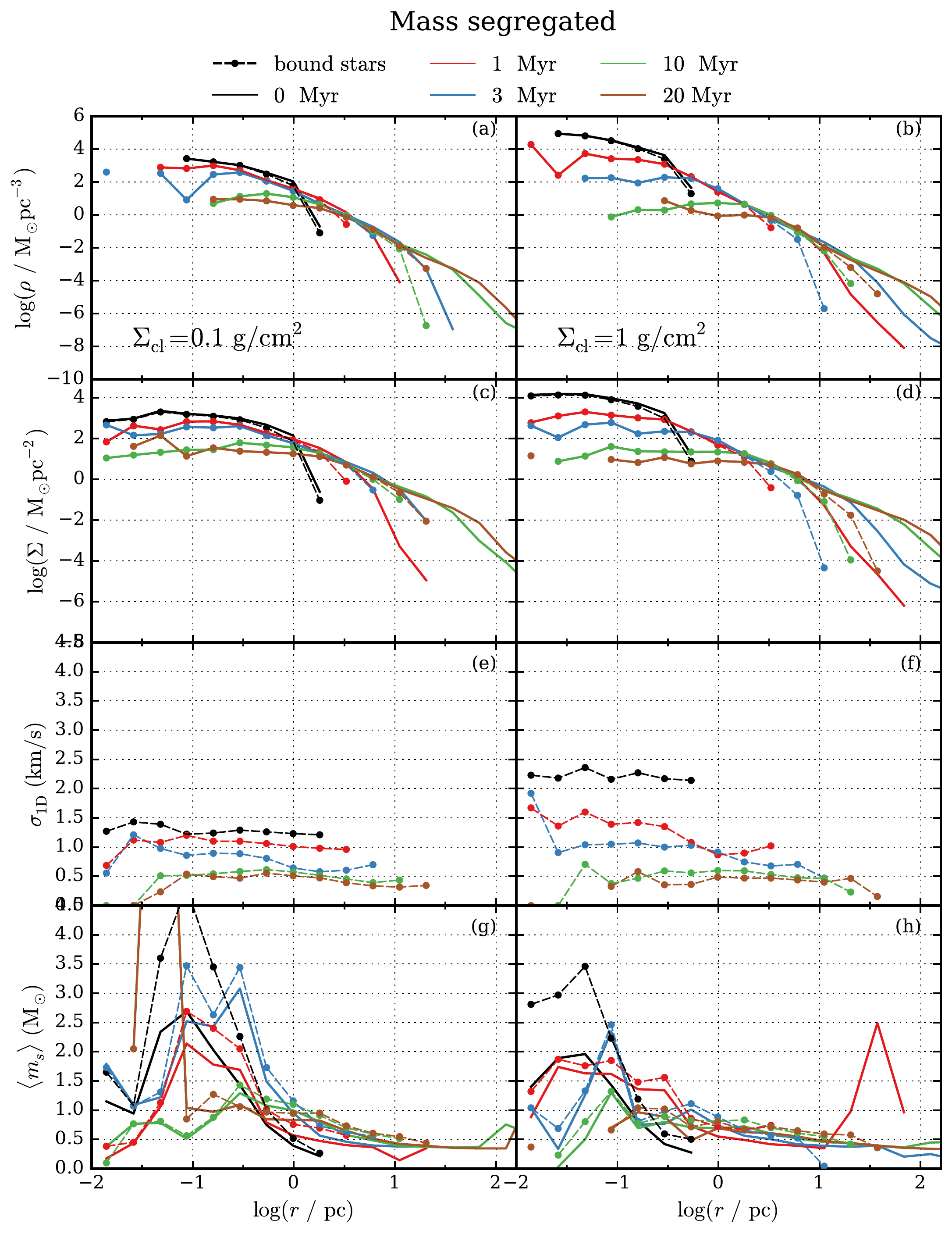} &
        \includegraphics[width=0.49\textwidth]{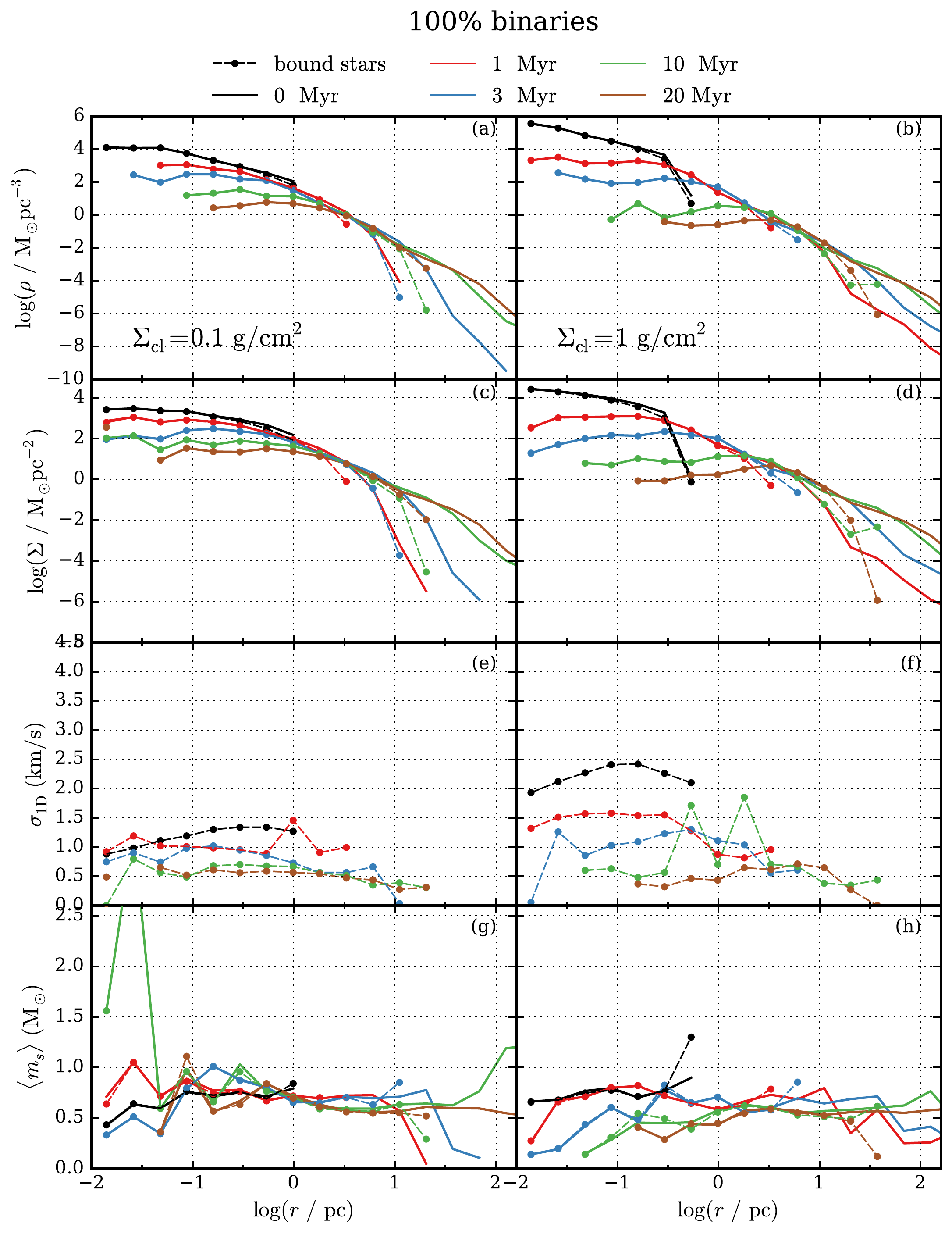}
        \end{array}$
        \caption{
Same as Figure \ref{fig:fiducialrad}, but now for simulations with
stellar evolution turned on. Top panels shows simulations with
different eccentricity distributions, \binariesun (top left) and
\binariesth (top right), and bottom panels shows the most extreme
cases: the set with primordial mass segregation \segregated (bottom
left) and the case with 100 binaries \fullbinaries (bottom right). }
        \label{fig:serad}
\end{figure*}

\end{document}